\documentclass{aa}
\pdfoutput=1

\usepackage{url}
\usepackage{hyperref}
\usepackage[version=3]{mhchem}
\usepackage[varg]{txfonts}
\usepackage{natbib}
\bibpunct{(}{)}{;}{a}{}{,}
\usepackage{pgfplots}
\pgfrealjobname{neat_linear}
\pgfplotsset{compat=newest}
\usetikzlibrary{fadings}
\usepgfplotslibrary{external}
\usepackage{siunitx}
\usepackage{booktabs}
\usepackage{Q4_macros_parent}
\usepackage{T7_macros_parent}
\usepackage{Q4_macros_daughter}
\usepackage{T7_macros_daughter}
\usepackage{T7_HCN_macros}
\usepackage{T7_rms_macros}
\usepackage{Q4_rms_macros}
\usepackage{comp_macros}

\newcommand{\ratran}{\texttt{ratran}}
\newcommand{\herschel}{{\it Herschel}}
\newcommand{\neat}{C/2001 Q4 (NEAT)}
\newcommand{\nea}{C/2001 Q4}
\newcommand{\linear}{C/2002 T7 (LINEAR)}
\newcommand{\lin}{C/2002 T7}
\newcommand{\machholz}{C/2004 Q2 (Machholz)}
\newcommand{\volatiles}{\ce{HCN}, \ce{H2CO}, \ce{CO}, \ce{CS},
			\ce{CH3OH}, and \ce{HNC}}
\newcommand\Kkms{\si{\kelvin\kilo\meter\per\second}}
\newcommand\kms{\si{\kilo\meter\per\second}}
\newcommand\ms{\si{\meter\per\second}}
\newcommand\s{\si{{molec.}.s^{-1}}}
\newcommand\vnea{\SI{0.73}{\km\per\second}}
\newcommand\vlin{\SI{0.90}{\km\per\second}}
\newcommand{\rh}{r_\mathrm{h}}
\newcommand{\xne}{x_{n_\mathrm{e}}}
\newcommand{\vexp}{v_\mathrm{exp}}
\newcommand\rotlinear{\SI{119(34)}{\kelvin}}
\newcommand\rotneat{\SI{54(9)}{\kelvin}}
\newcommand\rotlin{\SI{119}{\kelvin}}
\newcommand\rotnea{\SI{54}{\kelvin}}
\DeclareGraphicsExtensions{.pdf,.png,.jpg}

\newcommand{\rmsbars}[3]{%
  \draw[blue,fill=blue!25] (axis cs:10.1, #1) rectangle
      (axis cs:10.6, -#1);
  \draw[red,fill=red!25] (axis cs:10.7, #2) rectangle
      (axis cs:11.2, -#2);
  \draw[fill=black!25]
      (axis cs:11.3, #3) rectangle (axis cs:11.8, -#3);
  \draw [/pgfplots/every axis grid]
      (axis cs:10, 0) -- (axis cs:11.8, 0);
  \clip (rel axis cs:0,0) rectangle (rel axis cs:1,1);
}
\newcommand{\rmshcnbars}[3]{%
  \draw[blue,fill=blue!25] (axis cs:-0.85, #1) rectangle
      (axis cs:-0.35, -#1);
  \draw[red,fill=red!25] (axis cs:-0.25, #2) rectangle
      (axis cs:0.25, -#2);
  \draw[fill=black!25]
      (axis cs:0.35, #3) rectangle (axis cs:0.85, -#3);
  \draw [/pgfplots/every axis grid]
      (axis cs:-.85, 0) -- (axis cs:0.85, 0);
}
\newcommand{\rmschohbars}[3]{%
  \draw[blue,fill=blue!25] (axis cs:241.9102, #1) rectangle
      (axis cs:241.9110, -#1);
  \draw[red,fill=red!25] (axis cs:241.9112, #2) rectangle
      (axis cs:241.9120, -#2);
  \draw[fill=black!25]
      (axis cs:241.9122, #3) rectangle (axis cs:241.9130, -#3);
  \draw [/pgfplots/every axis grid]
      (axis cs:241.91, 0) -- (axis cs:241.9130, 0);
  \clip (rel axis cs:0,0) rectangle (rel axis cs:1,1);
}
\newcommand{\rmshncbars}[2]{%
  \draw[blue,fill=blue!25] (axis cs:-0.55, #1) rectangle
      (axis cs:-.05, -#1);
  \draw[red, fill=red!25] (axis cs:.05, #2) rectangle
      (axis cs:.55, -#2);
  \draw [/pgfplots/every axis grid]
      (axis cs:-0.55, 0) -- (axis cs:0.55, 0);
}

\pgfplotscreateplotcyclelist{speclist}{%
  black\\%
  blue,semithick\\%
  red,thin\\%
}
\pgfplotscreateplotcyclelist{aoslist}{%
  blue,semithick\\%
  red,thin\\%
}
\pgfplotscreateplotcyclelist{rotdiaglist}{%
  black,mark=triangle,rotdiag style\\%
  blue,mark=o,rotdiag style\\%
  red,mark=square,rotdiag style\\%
  black,mark=triangle*,rotdiag style\\%
  blue,mark=*,rotdiag style\\%
  red,mark=square*,rotdiag style\\%
}
\pgfplotsset{
  ticklabel style={/pgf/number format/.cd,fixed,precision=3},
  spectrum style/.style={
    clip=false,
    const plot mark mid,
    cycle list name=speclist,
    no markers,
    extra y ticks       = 0,
    extra y tick labels = ,
    extra x ticks       = 0,
    extra x tick labels = ,
    extra tick style  = { grid = major },
    xmin=-10, xmax=10,
    xlabel={$v$ [\kms]},
    ylabel={$T_\mathrm{mB}$ [\si{\kelvin}]},
  },
  rotdiag style/.style={only marks,
    error bars/.cd,y explicit,y dir=both
  },
  methanolaos style/.style={name=plot1,
    clip=false, axis on top=true,
    const plot mark mid,
    no markers,
    cycle list name=aoslist,
    extra y ticks       = 0,
    extra y tick labels = ,
    extra y tick style  = { grid = major },
    xlabel={$\nu_\mathrm{LSB}$ [GHz]},
    ylabel={$T_\mathrm{mB}$ [\si{\kelvin}]},
    enlarge x limits=false,
    width=\hsize,
    height=9cm
  },
  methanolcts style/.style={name=plot1,
    at=(plot1.below south), anchor=north,
    const plot mark mid,
    axis on top=true,
    no markers,
    clip=false,
    cycle list name=speclist,
    extra y ticks       = 0,
    extra y tick labels = ,
    extra y tick style  = { grid = major },
    xlabel={$\nu_\mathrm{LSB}$ [GHz]},
    ylabel={$T_\mathrm{mB}$ [\si{\kelvin}]},
    enlarge x limits=false,
    width=\hsize,
    height=9cm
  },
  colormap={blured}{color(0cm)=(blue); color(1cm)=(red)},
  comp style/.style={only marks,
    error bars/.cd,
    x dir=both,x explicit,
    y dir=both,y explicit
  },
}

\tikzfading[name=fade out,
  inner color=transparent!50,
  outer color=transparent!98]

\tikzstyle{every pin}=[pin edge=gray]

\begin{document}

\title{A survey of volatile species in Oort cloud comets \neat{} and
\linear{} at millimeter wavelengths\thanks{Based on observations
acquired with the 10-m Submillimeter Telescope at the Arizona Radio
Observatory, Steward Observatory, Mount Graham, Arizona, USA.}
\fnmsep\thanks{Full Tables~\ref{tbl:neat:prodrates},
  \ref{tbl:linear:prodrates}, \ref{tbl:neat:variability},
  \ref{tbl:linear:variability}, and reduced spectra as FITS files
  are only available at the CDS via anonymous ftp to
  \url{cdsarc.u-strasbg.fr} (130.79.128.5)
  or via \url{http://cdsweb.u-strasbg.fr/cgi-bin/qcat?J/A+A/}
}}

\author{M.~de~Val-Borro\inst{\ref{inst1}}\fnmsep\inst{\ref{inst2}}
	\and M.~K\"uppers\inst{\ref{inst3}}
	\and P.~Hartogh\inst{\ref{inst1}}
	\and L.~Rezac\inst{\ref{inst1}}
	\and N.~Biver\inst{\ref{inst4}}
	\and D.~Bockel\'ee-Morvan\inst{\ref{inst4}}
	\and J.~Crovisier\inst{\ref{inst4}}
	\and C.~Jarchow\inst{\ref{inst1}}
	\and G.~L.~Villanueva\inst{\ref{inst5}}\fnmsep\inst{\ref{inst6}}
  }

\titlerunning{A survey of volatiles in \neat{} and \linear{}}
\authorrunning{M.~de~Val-Borro et al.}

\institute{
  Max Planck Institute for Solar System Research,
  Max-Planck-Str.~2, 37191 Katlenburg-Lindau, Germany\\
    \email{[hartogh;rezac;jarchow]@mps.mpg.de}\label{inst1}
  \and Department of Astrophysical Sciences, Princeton University,
  Princeton, NJ 08544, USA\\
  \email{valborro@princeton.edu}\label{inst2}
  \and Rosetta Science Operations Centre, European Space Astronomy
  Centre, European Space Agency, PO Box 78, 28691 Villanueva de la
  Ca\~nada, Madrid, Spain\\
  \email{michael.kueppers@sciops.esa.int}\label{inst3}
  \and LESIA, Observatoire de Paris, CNRS, UPMC, Universit\'e
  Paris-Diderot, 5 place Jules Janssen, 92195 Meudon, France\\
  \email{[nicolas.biver;dominique.bockelee;jacques.crovisier]@obspm.fr}\label{inst4}
  \and Solar System Exploration Division, NASA Goddard Space Flight
  Center, Greenbelt, MD 20771, USA\\
  \email{Geronimo.Villanueva@nasa.gov}\label{inst5}
  \and Department of Physics, Catholic University of America,
  Washington, DC 20064, USA\label{inst6}
}

\date{Received 15 July 2013 / Accepted 1 August 2013}

\abstract
  {
  The chemical composition of comets can be inferred using spectroscopic
  observations in submillimeter and radio wavelengths.
  }
  {
  We aim to compare the production rates ratio of several volatiles in
  two comets, \neat{} and \linear{}, which are generally regarded as
  dynamically new and likely to originate in the Oort cloud. This type
  of comets is considered to be composed of primitive material that
  has not undergone considerable thermal processing.
  }
  {
  The line emission in the coma was measured in the comets, \neat{} and
  \linear{}, that were observed on five consecutive nights, 7--11 May
  2004, at heliocentric distances of 1.0 and 0.7 AU, respectively, by
  means of high-resolution spectroscopy using the 10-m Submillimeter
  Telescope at the Arizona Radio Observatory.  Both objects became very
  bright and reached naked-eye visibility during their perihelion
  passage in the spring of 2004.
  }
  {
  We present a search for six parent- and product-volatile species
  (\volatiles) in both comets.  Multiline observations of the \ce{CH3OH}
  $J$ = 5--4 series allow us to estimate the rotational temperature
  using the rotation diagram technique. We derive rotational
  temperatures of \rotneat{} for \neat{} and \rotlinear{} for \linear{}.
  The gas production rates are computed using the level distribution
  obtained with a spherically symmetric molecular excitation code that
  includes collisions between neutrals and electrons. The effects of
  radiative pumping of the fundamental vibrational levels by infrared
  photons from the Sun are considered for the case of HCN.  We find an
  HCN production rate of \qqhcn{}~\s{} for comet \neat{}, corresponding
  to a mixing ratio with respect to \ce{H2O} of \qqhcnqho. The mean HCN
  production rate during the observing period is \qthcn{}~\s\ for comet
  \linear{}, which gives a mixing ratio of \qthcnqho.  Relative
  abundances of CO, \ce{CH3OH}, \ce{H2CO}, CS, and HNC with respect to
  \ce{HCN} are \qqcoqhcn, \qqchohqhcn, \qqhcoqhcn, \qqcsqhcn{}, and
  \qqhncqhcn{} in comet \neat{} and \qtcoqhcn, \qtchohqhcn, \qthcoqhcn,
  \qtcsqhcn{}, and \qthncqhcn in comet \linear{}.
  }
  {
  With systematically lower mixing ratios in comet \neat{}, production
  rate ratios of the observed species with respect to \ce{H2O} lie
  within the typical ranges of dynamically new comets in both objects.
  We find a relatively low abundance of \ce{CO} in \neat{} compared to the
  observed range in other comets based on millimeter/submillimeter
  observations, and a significant upper limit on the CO production in
  \linear{} is derived.  Depletion of \ce{CO} suggests partial
  evaporation from the surface layers during previous visits to the
  outer Solar System and agrees with previous measurements of
  dynamically new comets.  Rotational temperatures derived from
  \ce{CH3OH} rotational diagrams in both \neat{} and \linear{} are
  roughly consistent with observations of other comets at similar
  distances from the Sun.
  }


\maketitle

\section{Introduction}\label{sec:intro}

Having spent most of their lifetime in the cold outer regions of the
Solar System, comets contain pristine material that have not evolved very
much since their formation.  Therefore, the composition of cometary
nuclei reflects that of the early solar nebula; characterizing their
chemical conditions can help to constrain planetary formation models.
Spectroscopic observations of cometary atmospheres at various
wavelengths are an efficient tool for investigating the physical and
chemical diversity of comets, and substantial efforts have been made in
the last two decades to develop a chemical classification of comets that
displays a great compositional diversity
\citep{1995Icar..118..223A,2002EM&P...90..323B,2004come.book..391B,2011IAUS..280..261B}.
These observations reveal critical information about the composition of
the primordial material in the solar nebula and the early formation
stages of the Solar System.  In addition, studying the role of the
volatile ice composition in the sublimation of material from the
surface is important in the understanding of cometary activity.

According to the standard dynamical model \citep[see][for a
comprehensive review of cometary dynamics]{2010LNP...790..341R}, Oort
cloud comets are believed to have formed in the giant planet region
between Jupiter and Neptune before being ejected to the outer regions of
the Solar Nebula.  However, it has been speculated that a fraction of
the objects in the Oort cloud have been captured from other stars in the
vicinity of the Sun during its formation \citep{2010Sci...329..187L}.
Objects in the Oort cloud are occasionally perturbed by passing stars or
galactic tides and injected into the inner Solar System.  Although the
present cometary impact rates are low, several collisions of km-sized
bodies have occurred in Jupiter in recent years since the Shoemaker-Levy
9 impact in 1994, which produced large atmospheric disturbances
\citep[see e.g.][]{2010ApJ...715L.155S,2013A&A...553A..21C}.

Conversely, Jupiter-family comets (JFC) form by accretion in a different
region, the Edgeworth-Kuiper belt, beyond Neptune's orbit up to about
100 AU from the Sun.  Although Oort cloud comets were probably assembled
on average closer to the Sun, observations of the abundance of cometary
volatiles suggest that there could have been mixing of material and an
overlap of the regions, where both kind of objects formed
\citep{2004come.book..391B,2011IAUS..280..261B,2012ApJ...758...29A}.
Dynamical simulations also indicate that formation of a fraction of the
JFC and Oort cloud comet populations can occur in the transneptunian
scattered disk region, where they could be gravitationally scattered by
the giant planets \citep{2010LNP...790..341R}.

In addition to several radicals and ions formed by photodissociation in
the coma, more than 20 different volatile species have been detected via
ground-based spectroscopic surveys at infrared to millimeter wavelengths
and measured in situ in comets belonging to different dynamical
families.  Many of these molecules are parent species that originate
directly in the nucleus ices and play an important role in organic
chemistry (e.g.\ \ce{H2O}, HCN, and \ce{CH3OH}).  The composition of
some cometary ices show strong evidence of processing in the solar
nebula and can provide clues about their place of formation and
subsequent evolution.  Submillimeter observations using heterodyne
receivers can be used to resolve the line shape and enable the
determination of accurate production rates with the aid of a molecular
excitation code \citep[e.g.,][]{2002EM&P...90..323B}.  Moreover, the
coma structure and outgassing velocity can be derived by fitting the
observed line shapes, and mixing ratios of volatile species can be
compared to observed chemical abundances in protoplanetary disks to
improve our understanding of the formation and evolution of the Solar
System.  However, there is no evident systematic correlation between the
observed relative abundances and the dynamical class of the comets
\citep[see e.g.][]{2012ApJ...758...29A}; therefore studying a larger
sample of comets is needed.

We have derived the production rates for the observed molecular lines in
the bright comets \neat{} and \linear{} during their 2004 apparitions
and compared them to the mixing ratios measured in other comets.  In
Sect.~\ref{sec:observations}, the observations of several volatile
species, namely \volatiles{}, that were first reported by
\citet{2004DPS....36.2505K} and \citet{2005DPS....37.1108V}, and the
reduction method are summarized.  Section~\ref{sec:results} presents the
data analysis of the observations using the molecular excitation
calculated with a code based on \ratran{}
\citep{2000A&A...362..697H,2004ApJ...615..531B} and discusses the
derived production rates for the detected molecules and the short-term
time variability of the HCN production rates.  Finally, we summarize the
obtained results and discuss the main conclusions in
Sect.~\ref{sec:discussion}.

\section{Observations}\label{sec:observations}

A spectral line survey of primary and daughter volatile species in
comets \linear{} and \neat{} was made in 7--11 May 2004 using the
Submillimeter Telescope \citep[SMT;][formerly known as the Heinrich
  Hertz Submillimeter
Telescope]{1996RvMA....9..111B,1999PASP..111..627B} located at the Mount
Graham International Observatory (MGIO), a division of Steward
Observatory on Mount Graham, Arizona (latitude
32\degr42\arcmin05\farcs8~N, longitude 109\degr53\arcmin28\farcs5~W and
elevation \SI{3186}{\meter}).  The SMT telescope has a parabolic 10-m
primary dish and a hyperbolic nutating secondary reflector.  Several
sensitive heterodyne receivers and backends are available that allow
deep line searches with a broad frequency coverage.  The observations
presented here were performed in remote observing mode using the 1.3 mm
double sideband superconductor-insulator-superconductor (SIS)
receiver system using various spectrometers.  This receiver allows us to
observe horizontal and vertical polarizations simultaneously and to
improve the signal-to-noise ratio (S/N) by averaging data from both
polarization channels.

The main goal of our observing program was to obtain the relative
abundances of the dynamically new comets \neat{} and \linear{} and to
confront them with the composition of dynamically old ones, such as
comet \machholz{} that was also observed at the SMT
\citep{2012A&A...545A...2D}.  An additional purpose of these
observations was to test the performance and stability of the newly
installed high-resolution chirp transform spectrometer (CTSB) built at
the Max Planck Institute for Solar System Research
\citep{1990MeScT...1..592H,1997SPIE.3220..115H,1997SPIE.3221..328H,2004ExA....18...77V,2006ITMTT..54.1415V}
and to compare these observations to the results from the
acousto-optical spectrometers.  High spectral resolution is crucial for
resolving the shape of rotational lines in comets and for studying the
gas velocity and asymmetries related to non-isotropic outgassing and
self-absorption effects.  Therefore, the CTSB spectrometer has been used
in parallel with other spectrometers in several cometary observation
programs at the SMT, since its installation in early 2004 \citep[see
e.g.][]{2004DPS....36.2505K,2005DPS....37.1108V,2010A&A...510A..55D,2010ApJ...715.1258P,2011Icar..215..153J,2011AOGS,2012A&A...545A...2D}.

We observed the comet simultaneously with the Forbes filterbanks (F1.M
and F250) with bandwidths of 2~GHz and 64~MHz, the CTSB with a bandwidth
of 215~MHz, and the acousto-optical spectrometers (AOSA, AOSB, and AOSC)
with total bandwidths of 1~GHz, 970~MHz, and 250~MHz, respectively.  The
spectral resolution provided was 1~MHz and 250~kHZ for the F1.M and F250
filterbanks; 40~kHz for the CTSB; and \SIlist{934;913;250}{\kilo\hertz}
for the AOSA, AOSB, and AOSC, respectively.  A typical system
temperature of \SI{1200}{\kelvin} was attained with $\sim$ 40\%
fluctuations during the observing period.  We focus on the observations
acquired with the AOSA, AOSB, and CTSB backends.

We performed pointing and focus calibration on Saturn, Mars, and Uranus,
because they were close to the comets during the course of our
observations.  Flux density reference observations of bright standard
sources, like the evolved star IRC +10216, the protobinary system W3(OH),
and the protoplanetary nebula CRL 2688, were obtained interleaved with
the comet observations for calibration purposes.

We used the latest osculating orbital elements provided by JPL's
HORIZONS online Solar System data
service\footnote{\url{http://ssd.jpl.nasa.gov/?horizons}} during the
observations to track the position and relative motion of the comet with
respect to the observer \citep{1996DPS....28.2504G}.  Typical telescope
pointing errors were $\sim 2\arcsec$ during our observing run and were
included in the derivation of the production rates.

\subsection{Comet \neat}

Comet \object{\neat} (hereafter referred to as \nea{}) is generally
considered a dynamically new comet, which is likely to originate in the
Oort cloud and is visiting the inner Solar System for the first time.
It was discovered on 24 August 2001 as a 20\textsuperscript{th}
magnitude object at a heliocentric distance of 10.1 AU by the Near-Earth
Asteroid Tracking (NEAT) program using the 1.2-m Samuel Oschin telescope
at Palomar observatory \citep{2001IAUC.7695....1P}.  Comet \nea{}
reached perihelion on 16 May 2004 at a distance of 0.962~AU from the Sun
and had its closest approach to Earth on 7 May 2004 at a distance of
0.32 AU.  Spectral observations of \nea{} were carried out shortly
before perihelion from 7 to 11 May 2004 using the SMT. We searched for
several rotational lines of molecular species (namely \volatiles{})
using the position-switching observing mode with a 0\fdg5 offset in
azimuth.  Water was first detected on 6 March 2004 with a production
rate of $Q_\ce{H2O} = \num{1.79\pm.55e29}\ \s{}$ as measured by the {\it
Odin} satellite \citep{2004IAUC.8304....2L,2009A&A...501..359B}.  From
the periodic variations in the outgassing rate of the 557~GHz \ce{H2O}
line, \citet{2009A&A...501..359B} derived a rotation period of
\SI{0.816\pm0.004}{days}.  The comet showed an asymmetry in the overall
outgassing activity with a peak about two weeks before perihelion
\citep{2009AJ....137.4734C,2009A&A...501..359B}.

\subsection{Comet \linear}

Comet \object{\linear} (hereafter \lin{}) was discovered on 14 October
2002 at 6.9 AU by the Lincoln Near-Earth Asteroid Research (LINEAR)
project with a visual magnitude of 17.5 \citep{2002IAUC.8003....1B}.
This comet is also regarded as a dynamically new Oort cloud object with
a slightly hyperbolic orbit at perihelion. It passed perihelion on 23
April 2004 at a heliocentric distance of 0.61~AU and reached perigee on
19 May 2004 at 0.27 AU from the Earth.  We observed \lin{} on 8--11
May 2004 at the SMT when the comet was at heliocentric distance of
$\sim$ 0.72 AU and 0.45 from the Earth.  Due to its favorable geometry
for observations from Earth, a large number of volatile species have
been found in this comet from infrared, submillimeter and radio
observational campaigns
\citep{2005ApJ...630..623F,2006ApJ...650..470D,2006ApJ...649.1169M,2008ApJ...689..613R,2009AJ....137.4837H}.

\subsection{Data reduction}

\begin{table*}
  \caption{
    Log of the observed rotational transitions in comets \neat{} and
    \linear{} in our SMT observing campaign from 7 to 11 May 2004.
  }
  \label{tbl:log}
  \centering
  \begin{tabular}{c c c S
      S[table-number-alignment = center,
      output-decimal-marker = {\text{--}},
      parse-numbers = false]
      c S[table-number-alignment = center,
	  table-format=3]
      c c
      S[output-decimal-marker = {\fdg}]
      S[output-decimal-marker = {\farcs}]
    }
    \toprule
    Target & Molecule\tablefootmark{\emph{a}} & Frequency &
    {Date\tablefootmark{\emph{b}}} & {Transition} &
    {$N$\tablefootmark{\emph{c}}} & {Exp\tablefootmark{\emph{d}}} &
    {$\langle\rh\rangle$\tablefootmark{\emph{e}}} &
    $\langle\Delta\rangle$\tablefootmark{\emph{f}} &
    {$\langle\phi\rangle$\tablefootmark{\emph{g}}} & {Beam FWHM}\\
    & & {(GHz)} & {(UT)} & & & {(\si{\minute})} & {(AU)} &
    {(AU)} & {(\si{\degree})} & {(\si{\arcsecond})} \\
    \midrule
    Q4  & \ce{HCN} & 265.886180 & 7.98 & 3.2 & 20 & 120 & 0.972 & 0.323 & 87.18 & 27.91\\
Q4  & \ce{H2CO} & 225.697775 & 8.12 & 3_{12}.2_{11} & 15 & 90 & 0.972 & 0.324 & 87.23 & 32.88\\
T7  & \ce{HCN} & 265.886180 & 8.68 & 3.2 & 30 & 120 & 0.702 & 0.522 & 110.10 & 27.91\\
T7  & \ce{H2CO} & 225.697775 & 8.82 & 3_{12}.2_{11} & 23 & 92 & 0.704 & 0.517 & 110.54 & 32.88\\
Q4  & \ce{CO} & 230.538000 & 9.01 & 2.1 & 43 & 240 & 0.970 & 0.329 & 87.36 & 32.19\\
T7  & \ce{CO} & 230.538000 & 9.67 & 2.1 & 24 & 96 & 0.713 & 0.487 & 113.25 & 32.19\\
T7  & \ce{HCN} & 265.886180 & 9.70 & 3.2 & 15 & 60 & 0.713 & 0.486 & 113.35 & 27.91\\
T7  & \ce{CS} & 244.935643 & 10.74 & 5.4 & 60 & 240 & 0.724 & 0.450 & 116.70 & 30.30\\
T7  & \ce{CH3OH} & 241.700159 & 10.74 & 5_0.4_0~\text{E} & 60 & 240 & 0.724 & 0.450 & 116.70 & 30.70\\
T7  & \ce{CH3OH} & 241.767234 & 10.74 & 5_{-1}.4_{-1}~\text{E} & 60 & 240 & 0.724 & 0.450 & 116.70 & 30.69\\
T7  & \ce{CH3OH} & 241.791352 & 10.74 & 5_0.4_0~\text{A}^+ & 60 & 240 & 0.724 & 0.450 & 116.70 & 30.69\\
T7  & \ce{CH3OH} & 241.832718 & 10.74 & 5_3.4_3~\text{A}^+ & 60 & 240 & 0.724 & 0.450 & 116.70 & 30.68\\
T7  & \ce{CH3OH} & 241.833106 & 10.74 & 5_3.4_3~\text{A}^- & 60 & 240 & 0.724 & 0.450 & 116.70 & 30.68\\
T7  & \ce{CH3OH} & 241.842284 & 10.74 & 5_2.4_2~\text{A}^- & 60 & 240 & 0.724 & 0.450 & 116.70 & 30.68\\
T7  & \ce{CH3OH} & 241.843604 & 10.74 & 5_3.4_3~\text{E} & 60 & 240 & 0.724 & 0.450 & 116.70 & 30.68\\
T7  & \ce{CH3OH} & 241.879025 & 10.74 & 5_1.4_1~\text{E} & 60 & 240 & 0.724 & 0.450 & 116.70 & 30.68\\
T7  & \ce{CH3OH} & 241.887674 & 10.74 & 5_2.4_2~\text{A}^+ & 60 & 240 & 0.724 & 0.450 & 116.70 & 30.68\\
T7  & \ce{CH3OH} & 241.904147 & 10.74 & 5_{-2}.4_{-2}~\text{E} & 60 & 240 & 0.724 & 0.450 & 116.70 & 30.67\\
T7  & \ce{CH3OH} & 241.904643 & 10.74 & 5_2.4_2~\text{E} & 60 & 240 & 0.724 & 0.450 & 116.70 & 30.67\\
Q4  & \ce{CS} & 244.935643 & 11.03 & 5.4 & 78 & 312 & 0.966 & 0.349 & 87.07 & 30.30\\
Q4  & \ce{CH3OH} & 241.700159 & 11.03 & 5_0.4_0~\text{E} & 78 & 312 & 0.966 & 0.349 & 87.07 & 30.70\\
Q4  & \ce{CH3OH} & 241.767234 & 11.03 & 5_{-1}.4_{-1}~\text{E} & 78 & 312 & 0.966 & 0.349 & 87.07 & 30.69\\
Q4  & \ce{CH3OH} & 241.791352 & 11.03 & 5_0.4_0~\text{A}^+ & 78 & 312 & 0.966 & 0.349 & 87.07 & 30.69\\
Q4  & \ce{CH3OH} & 241.832718 & 11.03 & 5_3.4_3~\text{A}^+ & 78 & 312 & 0.966 & 0.349 & 87.07 & 30.68\\
Q4  & \ce{CH3OH} & 241.833106 & 11.03 & 5_3.4_3~\text{A}^- & 78 & 312 & 0.966 & 0.349 & 87.07 & 30.68\\
Q4  & \ce{CH3OH} & 241.842284 & 11.03 & 5_2.4_2~\text{A}^- & 78 & 312 & 0.966 & 0.349 & 87.07 & 30.68\\
Q4  & \ce{CH3OH} & 241.843604 & 11.03 & 5_3.4_3~\text{E} & 78 & 312 & 0.966 & 0.349 & 87.07 & 30.68\\
Q4  & \ce{CH3OH} & 241.879025 & 11.03 & 5_1.4_1~\text{E} & 78 & 312 & 0.966 & 0.349 & 87.07 & 30.68\\
Q4  & \ce{CH3OH} & 241.904147 & 11.03 & 5_{-2}.4_{-2}~\text{E} & 78 & 312 & 0.966 & 0.349 & 87.07 & 30.67\\
Q4  & \ce{CH3OH} & 241.904643 & 11.03 & 5_2.4_2~\text{E} & 78 & 312 & 0.966 & 0.349 & 87.07 & 30.67\\
T7  & \ce{HCN} & 265.886180 & 11.60 & 3.2 & 7 & 28 & 0.734 & 0.421 & 119.50 & 27.91\\
T7  & \ce{HNC} & 271.981142 & 11.69 & 3.2 & 30 & 120 & 0.735 & 0.418 & 119.79 & 27.28\\
Q4  & \ce{HNC} & 271.981142 & 11.95 & 3.2 & 45 & 180 & 0.964 & 0.362 & 86.67 & 27.28\\

    \bottomrule
  \end{tabular}
  \tablefoot{
  \tablefoottext{\emph{a}}{Observations of the \ce{CH3OH} and \ce{CS}
  lines were obtained simultaneously in the lower and upper sidebands
  respectively.}
  \tablefoottext{\emph{b}}{Observing times in UT decimal days in May 2004.}
  \tablefoottext{\emph{c}}{Number of individual scans.}
  \tablefoottext{\emph{d}}{Total integration time.}
  \tablefoottext{\emph{e}}{Heliocentric distance.}
  \tablefoottext{\emph{f}}{Geocentric distance.}
  \tablefoottext{\emph{g}}{Solar phase angle.}
  }
\end{table*}

The observed antenna temperature scale, $T_\mathrm{A}^{*}$, was
calibrated using the chopper-wheel method with the on-site data
processing software \citep[see][]{1976ApJS...30..247U}, and the combined
dataset was stored using the Continuum and Line Analysis Single-dish
Software (CLASS) file type.  Moreover, we have corrected the temperature
scale for the average beam efficiency of the telescope, which is
estimated to be $\eta_\mathrm{B} = \num{0.74(2)}$ from observations of
Mars and Saturn in the fall 2007--spring 2008%
\footnote{\url{http://kp12m.as.arizona.edu/smt_docs/smt_beam_eff.htm}}.
This gives $T_\mathrm{mB} = T_\mathrm{A}^{\prime}/\eta_\mathrm{mB}$,
where $T_\mathrm{A}^{\prime}$ is the antenna temperature corrected for
atmospheric attenuation, radiative loss, rearward scattering, and
spillover losses.  The reduction of the SMT spectra was performed using
the open-source \texttt{pyspeckit} spectroscopic toolkit written in
Python, which depends on the NumPy library \citep{oliphant-2006-guide}
for manipulating multidimensional arrays efficiently
\citep{2011ascl.soft09001G}.  The \texttt{pyspeckit} package supports
many file types including some versions of the non-standard CLASS file
format\footnote{The source code repository is hosted at
  \url{http://pyspeckit.bitbucket.org.}}.

Table~\ref{tbl:log} summarizes the frequencies of the molecular
species transitions, observing dates, comet distances, and beam sizes in
comets \neat{} and \linear{} during our observing run at the SMT.

The data analysis was carried out as described by
\citet{2012A&A...545A...2D}.  A standing wave appears as a baseline
ripple in the observed spectra.  Using the Lomb-Scargle periodogram
technique, the baseline was removed by fitting a linear combination of
sine waves to the emission-free background and subtracted from the
original spectrum.  This method has the advantage that it can be applied
to unevenly spaced series, as in the case when the narrow emission
features are masked in the spectra before fitting the baseline. Using
alternative methods to fit standing wave, such as the Hilbert-Huang
transform (HHT), give similar results (Rezac et al.\ in preparation).
After removal of the baseline ripple, the individual scans were averaged
to increase the S/N.  Line intensities were calculated from the weighted
averages of the spectra, where the statistical weights of each individual
spectrum are proportional to the exposure time and inversely
proportional to the system temperature squared.

The statistical uncertainty in the integrated line intensity was
calculated using the root mean square (rms) noise in the background
after subtracting the fitted baseline.  Detected emission lines may
have a different gain response, depending upon which sideband the lines
were observed in.  Thus, the sideband gain ratio between the upper and
lower sidebands ($G_\mathrm{USB}/G_\mathrm{LSB}$) deviates from unity
and introduces an additional uncertainty of $\sim$ 10\% in the absolute
brightness temperature calibration \citep{2012A&A...545A...2D}.

\section{Results}\label{sec:results}

\subsection{Excitation model}
\label{sec:model}

To compute the production rates, we adopted a radiative transfer code
based on \ratran{} \citep{2000A&A...362..697H,2009AJ....137.4837H} that
includes collisions between neutrals and electrons and radiation
trapping effects \citep[see][and references
therein]{2010A&A...521L..50D}.  The radiative pumping of the fundamental
vibrational levels, which are induced by solar infrared radiation and
subsequently decay to rotational levels in the ground vibrational state,
is considered only for the case of HCN.  We used the one-dimensional
spherically symmetric version of the code with a constant outflow
velocity by following the description outlined in
\citet{2004ApJ...615..531B}, which has been used to model water
excitation to interpret \herschel{} cometary observations \citep[see
e.g.][]{2010A&A...518L.149B,2012A&A...544L..15B,2012A&A...546L...4D,2013ApJ...774L..13O}.

Although the outgassing rate may be variable on short timescales due to
the rotation of the nucleus as discussed in Sect.~\ref{sec:variability}, an
isotropic steady-state radial gas density profile for parent molecules was
assumed using the standard Haser spherically symmetric
distribution \citep{1957BSRSL..43..740H}:
\begin{equation}
  n_\mathrm{p}(r)= \frac{Q}{4\pi r^2v_\mathrm{exp}}\,
  \exp\left(-\frac{r\beta}{v_\mathrm{exp}}\right),
\end{equation}
where $Q$ is the total production rate in molecules s$^{-1}$, $v_\mathrm{exp}$
is the expansion velocity, and $r$ is the nucleocentric distance.  Volatiles
that sublimate off the surface of the comet can be photodissociated when they
are exposed to solar UV radiation; the photodissociation rate
$\beta$ considers the dissociation and ionization of molecules by the
radiation from the Sun and determines the spatial distribution of the
species.

\begin{table*}
  \caption{Photodissociation rates from the quiet-Sun reference spectrum
  at 1 AU for the observed molecules and corresponding scale-lengths in
  \nea{} and \lin{} at $\rh$ = 0.97 and 0.72 AU, respectively.
  }
  \label{tbl:beta}
  \centering
  \begin{tabular}{c
    S[table-format = 1.1e+1]
    S[table-format = 7]
    S[table-format = 7]
    S[table-format = 1.1e+1]
    S[table-format = 4]
    S[table-format = 4]
    }
    \toprule
    Molecule
    & {$\beta_0$\tablefootmark{\emph{a}}}
    & \multicolumn{2}{c}{$L$\tablefootmark{\emph{a}}}
    & {$\beta_\mathrm{0,p}$\tablefootmark{\emph{b}}}
    & \multicolumn{2}{c}{$L_\mathrm{p}$\tablefootmark{\emph{b}}}\\
    & {(\si{\per\second})} & \multicolumn{2}{c}{(\si{\kilo\meter})}
    & {(\si{\per\second})} & \multicolumn{2}{c}{(\si{\kilo\meter})}\\
    & & {\nea{}} & {\lin{}} & & {\nea{}} & {\lin{}} \\
    \midrule
    \ce{CO} & 6.5e-07 & 1100100 & 717800\\
\ce{CH3OH} & 1.3e-05 & 55000 & 35900\\
\ce{CS} & 2.5e-05 & 28600 & 18700 & 1.7e-03 & 400 & 300\\
\ce{H2CO} & 2.0e-04 & 3600 & 2300 & 1.0e-04 & 7200 & 4700\\
\ce{HCN} & 1.5e-05 & 47700 & 31100\\
\ce{HNC} & 1.9e-05 & 36700 & 23900\\

    \bottomrule
  \end{tabular}
  \tablefoot{\\
  \tablefoottext{\emph{a}}{Photodissociation rates and scale-lengths for
  release from the nucleus surface.}
  \tablefoottext{\emph{b}}{Photodissociation rates and scale-lengths
  from a parent distributed source.}
  }
\end{table*}

We obtained the photodissociative lifetimes for \ce{CH3OH}, HCN,
\ce{H2CO}, and CO at 1 AU from \citet{1994JGR....99.3777C} assuming the
quiet-Sun reference spectrum, and the HNC and CS radical lifetimes were
taken from \citet{2011A&A...528A.142B}.  Photodissociation rates and
scale-lengths for the observed species are summarized in
Table~\ref{tbl:beta}.  The photodissociation rates were scaled by the
heliocentric distance of the comets as $\rh^{-2}$ at the time of the
observations.  Nevertheless, some daughter molecules, such as CS, HNC, and
\ce{H2CO}, are not correctly described by the Haser parent molecule
profile.  The density profile of daughter species originating from the
photodissociation of an extended parent source in the coma is taken from
\citet{2004come.book..523C}.  We use a distributed source in the coma
with parent species photodissociation rates at 1 AU of
$\beta_\mathrm{0,p} = \SI{1e-4}{\per\second}$ and
\SI{1.7e-3}{\per\second} for \ce{H2CO} and CS, respectively
\citep[see][and references therein]{2012A&A...545A...2D}.  There are no
constraints on the scale-length and distribution of HNC in the comae of
comets \nea{} and \lin{}. Thus, we assumed a parent molecule profile for
HNC to compute the production rates.

The excitation conditions of molecules in cometary atmospheres have been
studied initially by \citet{1983A&A...126..170C} and
\citet{1984A&A...130..361C}.  Neutrals are excited by collisions with
mainly water molecules and electrons, which are the dominant
effects in the inner coma. For HCN, the radiative pumping of the
fundamental vibrational levels by the solar infrared flux has been
included.  Infrared pumping of vibrational bands by solar radiation
becomes the dominant excitation mechanism in the outer coma, where the
gas and electron densities are low and their collision rates become
negligible \citep{1987A&A...181..169B}.  For a beam size of about
30\arcsec{} at the observed frequencies, we are probing a region where
excitation is mainly dominated by collisions and the level populations
can be described approximately by a Boltzmann distribution at the gas
kinetic temperature.

Water is the main volatile component of cometary nuclei and is typically
used to determine relative abundances.  Although water is not directly
accessible from the ground at submillimeter wavelengths, it has been
observed from space using the {\it Odin} and \herschel{} satellites
\citep[see][]{2003A&A...402L..55L,2010A&A...518L.150H,2011Natur.478..218H,2012A&A...539A..68B}.
The water production rate of comet \nea{} derived from {\it Odin}
observations between 2 April and 16 May 2004 has a long-term variation
that depends on the heliocentric distance and seasonal effects. A
Gaussian fit of the daily average of the outgassing rates is given by
\citep{2009A&A...501..359B}:
\begin{equation}
  Q_\ce{H2O} = \num{2.74e29}
  \times \exp\left(- \frac{(t/\mathrm{days}+17)^2}{45.7^2}\right)\ \s,
\end{equation}
where $t$ is the time in days relative to \nea{}'s
perihelion passage, UT 15.97 May 2004. The values obtained from this
formula at the time of the SMT observations have been used in our
excitation model and to calculate the mixing ratios with respect to
\ce{H2O}.  We used a water production rate of $Q_\ce{H2O} = 3 \times
10^{29}\ \s$ for comet \lin{} at the time of our observations, which are
derived from the observation of the fundamental \ce{H2O} transition at
557~GHz with the {\it Odin} satellite at different heliocentric
distances \citep{2007P&SS...55.1058B}.

The outgassing velocity in the coma is assumed to be constant in our
model with values of \vnea{} for comet \nea{} and \vlin{} for comet
\lin{}.  These values were obtained from the half-width-at-half maximum
(HWHM) of a single Gaussian function fitted to the HCN line observed by
the CTSB, whose line profiles are relatively symmetric and have the
highest S/N of the detected transitions.  The HWHM was further reduced
by 10\% to consider the increase in line width due to thermal Doppler
broadening \citep[e.g.][]{1997EM&P...78....5B}.  The HCN line width
decreases by $\sim$ 5\% from UT 8.68 to 11.60 May in comet \lin{}, and
the value for the outflow velocity above has been derived from the
weighted average of the HCN spectra on three different dates. We also
provide for reference the values of the expansion velocities of the
other lines in Tables~\ref{tbl:neat:prodrates}
and~\ref{tbl:linear:prodrates} for both objects.

The distribution of rotational levels in the collisional region is given
by the gas neutral temperature that is assumed to be equal to the
rotational temperature. We estimated the rotational temperatures in both
comets from the relative intensities of the measured \ce{CH3OH}
transitions as described in Sect.~\ref{sec:rottemp}.

We chose an electron density scaling factor of $\xne = 0.2$ with respect
to the reference profile deduced from in situ measurements of comet
1P/Halley as defined by \citet{1997PhDT........51B} and
\citet{1999AJ....118.1850B,2007P&SS...55.1058B}.  The electron density
distribution in comet \nea{} is constrained by the brightness
distribution of the 557~GHz line of \ce{H2O} that was mapped by the {\it
Odin} satellite with high spatial resolution on 16 May 2004
\citep{2007P&SS...55.1058B}. This scaling factor minimizes the radial
variation between production rates derived from the intensity at various
offset points. For \lin{}, we used the same value for the electron
density scaling factor, since its density distribution has not been
estimated by mapping observations.

\subsection{Rotational temperatures}\label{sec:rottemp}

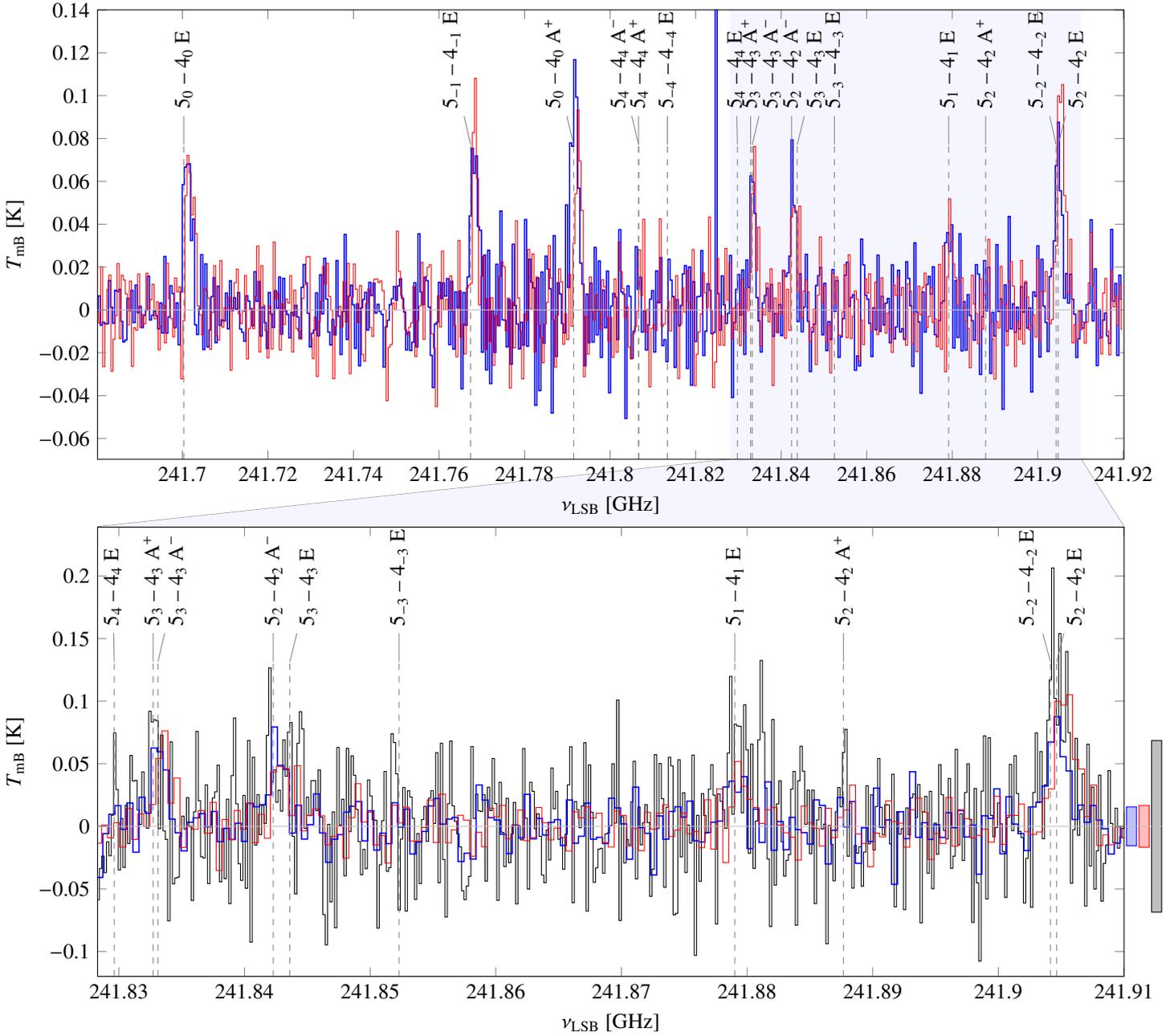
\begin{figure*}
  \centering
  \begin{tikzpicture}
    \begin{axis}[
	methanolaos style,
	xmin=241.68, xmax=241.92,
	ymax=0.14
      ]
      \fill [blue!4] ({axis cs: 241.828,0}|-{rel axis cs:0,0}) rectangle
			({axis cs: 241.91,0}|-{rel axis cs:0,1});
      \draw [gray] ({axis cs: 241.828,0}|-{rel axis cs:0,0}) --
			(rel axis cs:0,-0.15);
      \draw [gray] ({axis cs: 241.91,0}|-{rel axis cs:0,0}) --
			(rel axis cs:1,-0.15);
      \fill [blue!4] ({axis cs: 241.828,0}|-{rel axis cs:0,0}) --
			(rel axis cs:0,-0.15) --
			(rel axis cs:1,-0.15) --
			({axis cs: 241.91,0}|-{rel axis cs:0,0});
      \clip (rel axis cs:0,0) rectangle (rel axis cs:1,1);
      \addplot table [x index=0,y index=1] {Q4_CH3OH_AOSA.dat};
      \addplot table [x index=0,y index=1] {Q4_CH3OH_AOSB.dat};
      \foreach \freq/\angle/\trans in {
	241.7001590/right/$5_0$ -- $4_0$ E,
	241.7672340/5/$5_{-1}$ -- $4_{-1}$ E,
	241.7913520/5/$5_0$ -- $4_0$ A$^+$,
	241.8065240/5/$5_4$ -- $4_4$ A$^-$,
	241.8065250/right/$5_4$ -- $4_4$ A$^+$,
	241.8132550/right/$5_{-4}$ -- $4_{-4}$ E,
	241.8296290/3/$5_4$ -- $4_4$ E,
	241.8327180/right/$5_3$ -- $4_3$ A$^+$,
	241.8331060/-4/$5_3$ -- $4_3$ A$^-$,
	241.8422840/right/$5_2$ -- $4_2$ A$^-$,
	241.8436040/-4/$5_3$ -- $4_3$ E,
	241.8522990/right/$5_{-3}$ -- $4_{-3}$ E,
	241.8790250/right/$5_1$ -- $4_1$ E,
	241.8876740/right/$5_2$ -- $4_2$ A$^+$,
	241.9041470/5/$5_{-2}$ -- $4_{-2}$ E,
	241.9046430/-5/$5_2$ -- $4_2$ E
	}
	{ \edef\temp{\noexpand
	  \draw[gray,dashed]
	    ({axis cs:\freq,0}|-{rel axis cs:0,0}) --
	    ({axis cs:\freq,0}|-{rel axis cs:0,0.7});
	  }
	  \temp
	  \edef\temp{\noexpand
	  \node[
	    coordinate,
	    pin = {[rotate=90]\angle:{\trans}}
	  ] at ({axis cs:\freq,0}|-{rel axis cs:0,0.7}) {};
	  }
	  \temp
	}
    \end{axis}
    \begin{axis}[methanolcts style,
	xmin=241.8283, xmax=241.91,
	ymin=-0.12,
	xtick={241.83,241.84,241.85,241.86,241.87,241.88,241.89,241.90,241.91,241.92},
      ]
      \rmschohbars{\qchohAOSArms}{\qchohAOSBrms}{\qchohCTSBrms}
      \addplot table [x index=0,y index=1] {Q4_CH3OH_CTSB.dat};
      \addplot table [x index=0,y index=1] {Q4_CH3OH_AOSA.dat};
      \addplot table [x index=0,y index=1] {Q4_CH3OH_AOSB.dat};
      \foreach \freq/\angle/\trans in {
	241.8296290/3/$5_4$ -- $4_4$ E,
	241.8327180/right/$5_3$ -- $4_3$ A$^+$,
	241.8331060/-4/$5_3$ -- $4_3$ A$^-$,
	241.8422840/right/$5_2$ -- $4_2$ A$^-$,
	241.8436040/-4/$5_3$ -- $4_3$ E,
	241.8522990/right/$5_{-3}$ -- $4_{-3}$ E,
	241.8790250/right/$5_1$ -- $4_1$ E,
	241.8876740/right/$5_2$ -- $4_2$ A$^+$,
	241.9041470/12/$5_{-2}$ -- $4_{-2}$ E,
	241.9046430/-5/$5_2$ -- $4_2$ E
	}
	{ \edef\temp{\noexpand
	  \draw[gray,dashed]
	    ({axis cs:\freq,0}|-{rel axis cs:0,0}) --
	    ({axis cs:\freq,0}|-{rel axis cs:0,0.7});
	  }
	  \temp
	  \edef\temp{\noexpand
	  \node[
	    coordinate,
	    pin = {[rotate=90]\angle:{\trans}}
	  ] at ({axis cs:\freq,0}|-{rel axis cs:0,0.7}) {};
	  }
	  \temp
	}
    \end{axis}
  \end{tikzpicture}
  \caption{\emph{Upper panel}: shows the \ce{CH3OH} averaged spectrum
  toward comet \neat{} acquired on UT 11.03 May 2004 with a
  \SI{312}{\minute} total integration time using the AOSA (thick blue
  line) and AOSB (thin red line).  \emph{Lower panel}: shows the
  frequency range covered by the CTSB (black line) with the AOSA (thick
  blue line) and AOSB (thin red line) spectra.  The lower sideband
  frequency is shown in the horizontal axes  and the calibrated main
  beam brightness temperature in the vertical axes.  The CTSB spectrum
  was resampled to a 190~kHz resolution per channel with a rectangular
  window function to increase the S/N ratio of the detected lines.  Note
  that a spurious emission feature is present around 241.825~GHz in the
  AOSA spectrum.
  The rms noise level in the main brightness temperature is calculated
  from the background level and is shown by the blue, red, and gray
  bars on the right of the \emph{lower panel} for AOSA, AOSB, and CTSB,
  respectively.  Note that the rms in the CTSB spectrum was calculated
  for the native frequency resolution of 40 kHz.
  }
  \label{fig:neat:ch3oh}
\end{figure*}

Methanol rotational transitions appear in several multiplets at
millimeter wavelengths that are well suited to estimate the
rotational temperature and excitation conditions in the coma in the
optically thin limit.  The rotational levels of \ce{CH3OH} listed in
Table~\ref{tbl:log} are described with three quantum numbers ($J_K\
T_\mathrm{s}$) following the notation by \citet{1999JMoSp.194..171M},
where $J$ is the total angular momentum, $K$ its projection along the
symmetry axis, and $T_\mathrm{s}$ is the torsional symmetry state
(A\textsuperscript{+}, A\textsuperscript{-}, E$_1$ or E$_2$).  The
difference between the E$_1$ and E$_2$ states is indicated by the sign
of the quantum number $K$ with a positive sign corresponding to E$_1$
levels and a negative sign to E$_2$ levels.  All the observed transitions
in comets \nea{} and \lin{} are in the first torsional state (quantum
number $\nu_\mathrm{t}$ = 0).

The rotational temperature was derived using the rotational diagram
(also known as Boltzmann diagram) technique, which is frequently used in
studies of the interstellar medium, from the relative intensities of the
individual \ce{CH3OH} lines between levels with quantum numbers $J =
5$--4. We assume that the population distribution of the levels sampled
by the emission lines is in local thermodynamical equilibrium (LTE),
or described by a Boltzmann distribution characterized by a single
temperature.  Then the column density of the upper transition level
within the beam, $\langle N_\mathrm{u} \rangle$, can be expressed as
\begin{equation}
  \langle N_\mathrm{u} \rangle =  \langle N \rangle
  \frac{g_\mathrm{u}}{Z(T_\mathrm{rot})}
  \exp\left(-\frac{E_\mathrm{u}}{k_\mathrm{B}T_\mathrm{rot}}\right),
\end{equation} 
where $g_\mathrm{u}$ is the degeneracy of the upper
level, $Z$ denotes the rotational partition function, which is a
function of temperature, $T_\mathrm{rot}$ is the rotational temperature,
$E_\mathrm{u}$ is the energy of the upper state, $k_\mathrm{B}$
represents the Boltzmann constant, and $\langle N \rangle$ is the total
column density averaged over the beam \citep[see][]{1994A&A...287..647B}.

For optically thin conditions that normally apply to cometary lines of
most volatile species, the line intensity, $\int T_\mathrm{mB}\,dv$, is
proportional to the averaged column density in the upper level of the
observed transition within the beam:
\begin{equation}
\langle N_\mathrm{u} \rangle =  \frac{8\pi
k_\mathrm{B}\nu_\mathrm{ul}^2}{hc^3A_\mathrm{ul}} \int
T_\mathrm{mB}\,dv,
\end{equation}
where $\nu_\mathrm{ul}$ is the transition frequency, $h$ denotes the
Planck constant, $c$ is the speed of light and $A_\mathrm{ul}$ the
Einstein coefficient for spontaneous emission.  This relation assumes
that the beam sizes of the observed transitions are approximately the
same. Although the observed transitions in \nea{} and \lin{} are at
nearby frequencies, the excitation energy of the upper level varies
between 35 and 85 K, which allows us to retrieve the rotational
temperature using the rotational diagram method.
As noted by \citet{1994A&A...287..647B}, \citep[see
also][]{2002EM&P...90..323B}, it is best to use \ce{CH3OH} lines 
in which $\Delta J = 0$ and $\Delta K \neq 0$; in this case we have
$\Delta J \neq 0$ and $\Delta K = 0$, which makes these lines sensitive to
deviations from LTE.

\subsubsection{Comet \neat}

Figure~\ref{fig:neat:ch3oh} shows the \ce{CH3OH} rotational transitions
($J$ = 5--4) observed in the lower sideband in comet \nea. Ten
transitions are detected by the AOSA and AOSB backends, including three
blended lines around 241.833, 241.843, and 241.904~GHz that are
designated by the dashed lines. Seven of those transitions are also
detected in the frequency range covered by the CTSB (see lower panel in
Fig.~\ref{fig:neat:ch3oh}).  To increase the S/N of the CTSB spectra, we
resampled over four adjacent channels with a rectangular window function
corresponding to an effective spectral resolution of 190~kHz.
Additionally, the CTSB spectrum shows tentative detections of the $5_4$
-- $4_4$ E, $5_{-3}$ -- $4_{-3}$ E and $5_2$ -- $4_2$ A$^+$ transitions
at 241.829, 241.852 and 241.887~GHz, respectively, that are also
indicated by vertical dashed lines. The $5_4$ -- $4_4$ E transition is
marginally detected in both AOSA and AOSB backends as well, but since
the line is very close to a blended pair, the baseline subtraction
technique cannot be performed reliably.

\begin{figure}
  \centering
  \begin{tikzpicture}
    \begin{semilogyaxis}[
      cycle list name=rotdiaglist,
      xlabel={$E_\mathrm{u}/k_\mathrm{B}$ [K]},
      ylabel={$\langle N_\mathrm{u} \rangle/g_\mathrm{u}$ [\si{cm^{-2}}]},
      width=\hsize, xmin=30, xmax=90,
      ]
      \addplot+ table [x index=0,y index=1,y error index=2]
	{rot_ch3oh_Q4_CTSB.txt};
      \addplot+ table [x index=0,y index=1,y error index=2]
	{rot_ch3oh_Q4_AOSA.txt};
      \addplot+ table [x index=0,y index=1,y error index=2]
	{rot_ch3oh_Q4_AOSB.txt};
      \addplot[mark=none] table [x index=0,y index=1] {rotdiag_Q4___S_.txt};
    \end{semilogyaxis}
  \end{tikzpicture}
  \caption{
    Rotation diagram for \ce{CH3OH} lines in comet \neat{} which
    includes 1-$\sigma$ uncertainties.  The natural logarithm of the
    column density of the upper level divided by its degeneracy is
    plotted against the energy of the upper level for the AOSA (blue
    circles), AOSB (red squares), and CTSB (black triangles).  The solid
    line shows the best linear fit with a derived rotational temperature
    of \rotneat{} to the lines detected by all the spectrometers with
    S/N greater than three.
  }
  \label{fig:neat:rotdiagram}
\end{figure}
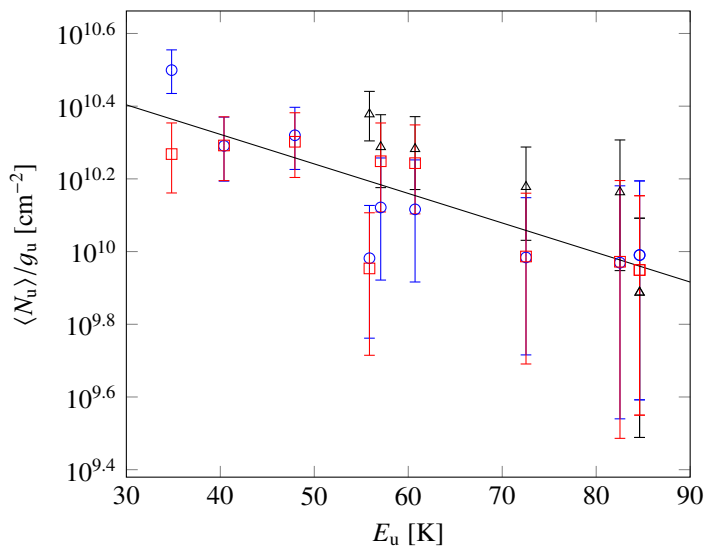

We show the best fit of the rotational diagram for all lines observed in
comet \nea{} with the AOSA, AOSB, and CTSB spectrometers in
Fig.~\ref{fig:neat:rotdiagram}.  The rotation diagram fits all detected
\ce{CH3OH} lines using a weighted linear least-squares method, where the
weights are equal to the reciprocal of the variance of each measurement.
A rotational temperature of \rotneat\ is obtained from the best linear
fit to the transitions detected above the 3-$\sigma$ detection limit
with all the spectrometers.  In comparison, a rotational temperature of
$T_\mathrm{rot} = 70 \times (\rh/\si{AU})^{-1}$~K is derived from the
relative line intensities of \ce{CH3OH} lines observed at the Institut
de Radioastronomie Millim\'etrique (IRAM) 30-m telescope between 7 and
11 May 2004 (\citealt{2009A&A...501..359B}; Biver et al.\ in
preparation), which corresponds to 72 K at the time of our \ce{CH3OH}
observations.  The rotational and spin temperature of methane in \nea{}
are determined to be \SI{104(2)}{\kelvin} and $33^{+3}_{-2}$
\si{\kelvin}, respectively, from near-infrared observations with the
Subaru telescope \citep{2005ApJ...623L..49K}. This results suggest that
the rotational temperature decreases as the sampled region in the coma
increases, which is consistent with model predictions when the
photolytic heating is not very efficient
\citep[e.g.][]{2004come.book..523C}.

\subsubsection{Comet \linear}

\begin{figure*}
  \centering
  \begin{tikzpicture}
    \begin{axis}[methanolaos style,
	xmin=241.68, xmax=241.92,
	ymin=-0.1, ymax=0.26,
      ]
      \fill [blue!4] (axis cs:241.828,-0.1) rectangle
		      (axis cs:241.91,0.26);
      \draw [gray] ({axis cs: 241.828,0}|-{rel axis cs:0,0}) --
			(rel axis cs:0,-0.15);
      \draw [gray] ({axis cs: 241.91,0}|-{rel axis cs:0,0}) --
			(rel axis cs:1,-0.15);
      \fill [blue!4] ({axis cs: 241.828,0}|-{rel axis cs:0,0}) --
			(rel axis cs:0,-0.15) --
			(rel axis cs:1,-0.15) --
			({axis cs: 241.91,0}|-{rel axis cs:0,0});
      \clip (rel axis cs:0,0) rectangle (rel axis cs:1,1);
      \addplot table [x index=0,y index=1] {T7_CH3OH_AOSA.dat};
      \addplot table [x index=0,y index=1] {T7_CH3OH_AOSB.dat};
      \foreach \freq/\angle/\trans in {
	241.7001590/right/$5_0$ -- $4_0$ E,
	241.7672340/right/$5_{-1}$ -- $4_{-1}$ E,
	241.7913520/right/$5_0$ -- $4_0$ A$^+$,
	241.8065240/5/$5_4$ -- $4_4$ A$^-$,
	241.8065250/right/$5_4$ -- $4_4$ A$^+$,
	241.8132550/right/$5_{-4}$ -- $4_{-4}$ E,
	241.8296290/3/$5_4$ -- $4_4$ E,
	241.8327180/right/$5_3$ -- $4_3$ A$^+$,
	241.8331060/-4/$5_3$ -- $4_3$ A$^-$,
	241.8422840/right/$5_2$ -- $4_2$ A$^-$,
	241.8436040/-4/$5_3$ -- $4_3$ E,
	241.8522990/right/$5_{-3}$ -- $4_{-3}$ E,
	241.8790250/right/$5_1$ -- $4_1$ E,
	241.8876740/right/$5_2$ -- $4_2$ A$^+$,
	241.9041470/5/$5_{-2}$ -- $4_{-2}$ E,
	241.9046430/-5/$5_2$ -- $4_2$ E
	}
	{ \edef\temp{\noexpand
	  \draw[gray,dashed]
	    ({axis cs:\freq,0}|-{rel axis cs:0,0}) --
	    ({axis cs:\freq,0}|-{rel axis cs:0,0.7});
	  }
	  \temp
	  \edef\temp{\noexpand
	  \node[
	    coordinate,
	    pin = {[rotate=90]\angle:{\trans}}
	  ] at ({axis cs:\freq,0}|-{rel axis cs:0,0.7}) {};
	  }
	  \temp
	}
    \end{axis}
    \begin{axis}[methanolcts style,
	xmin=241.828, xmax=241.91,
	xtick={241.83,241.84,241.85,241.86,241.87,241.88,241.89,241.90,241.91,241.92},
      ]
      \rmschohbars{\tchohAOSArms}{\tchohAOSBrms}{\tchohCTSBrms}
      \addplot table [x index=0,y index=1] {T7_CH3OH_CTSB.dat};
      \addplot table [x index=0,y index=1] {T7_CH3OH_AOSA.dat};
      \addplot table [x index=0,y index=1] {T7_CH3OH_AOSB.dat};
      \foreach \freq/\angle/\trans in {
	241.8296290/3/$5_4$ -- $4_4$ E,
	241.8327180/right/$5_3$ -- $4_3$ A$^+$,
	241.8331060/-4/$5_3$ -- $4_3$ A$^-$,
	241.8422840/right/$5_2$ -- $4_2$ A$^-$,
	241.8436040/-4/$5_3$ -- $4_3$ E,
	241.8522990/right/$5_{-3}$ -- $4_{-3}$ E,
	241.8790250/-5/$5_1$ -- $4_1$ E,
	241.8876740/right/$5_2$ -- $4_2$ A$^+$,
	241.9041470/12/$5_{-2}$ -- $4_{-2}$ E,
	241.9046430/-5/$5_2$ -- $4_2$ E
	}
	{ \edef\temp{\noexpand
	  \draw[gray,dashed]
	    ({axis cs:\freq,0}|-{rel axis cs:0,0}) --
	    ({axis cs:\freq,0}|-{rel axis cs:0,0.7});
	  }
	  \temp
	  \edef\temp{\noexpand
	  \node[
	    coordinate,
	    pin = {[rotate=90]\angle:{\trans}}
	  ] at ({axis cs:\freq,0}|-{rel axis cs:0,0.7}) {};
	  }
	  \temp
	}
    \end{axis}
  \end{tikzpicture}
  \caption{\ce{CH3OH} averaged spectrum toward comet \linear{} acquired
  on 10.74~UT May 2004 with a total integration time of
  \SI{240}{\minute}. The \emph{upper panel} shows all the lines detected
  with the AOSA (thick blue line) and AOSB (thin red line) within a
  $\sim \SI{240}{MHz}$ range. The \emph{lower panel} shows a close-up
  view corresponding to the frequency range of the shaded region in the
  \emph{upper panel} covered with the AOSA (thick blue line), AOSB (thin red
  line), and the CTSB (black line). The CTSB spectrum was resampled to a
  190~kHz resolution per channel with a rectangular window function to
  increase the S/N ratio of the detected lines.  The vertical axes are
  the calibrated main beam brightness temperatures and the lower
  sideband frequency is shown in the horizontal axes.  Note that the
  AOSA backend has a channel with a spurious emission feature around
  241.825~GHz in the \emph{upper panel}.  The rms noise level measured
  with respect to the zero reference level is shown by the blue, red,
  and gray bars on the right of the \emph{lower panel} for AOSA, AOSB, and
  CTSB, respectively.  Note that the rms in the CTSB spectrum was
  calculated for the native frequency resolution of 40 kHz.
  }
  \label{fig:linear:ch3oh}
\end{figure*}
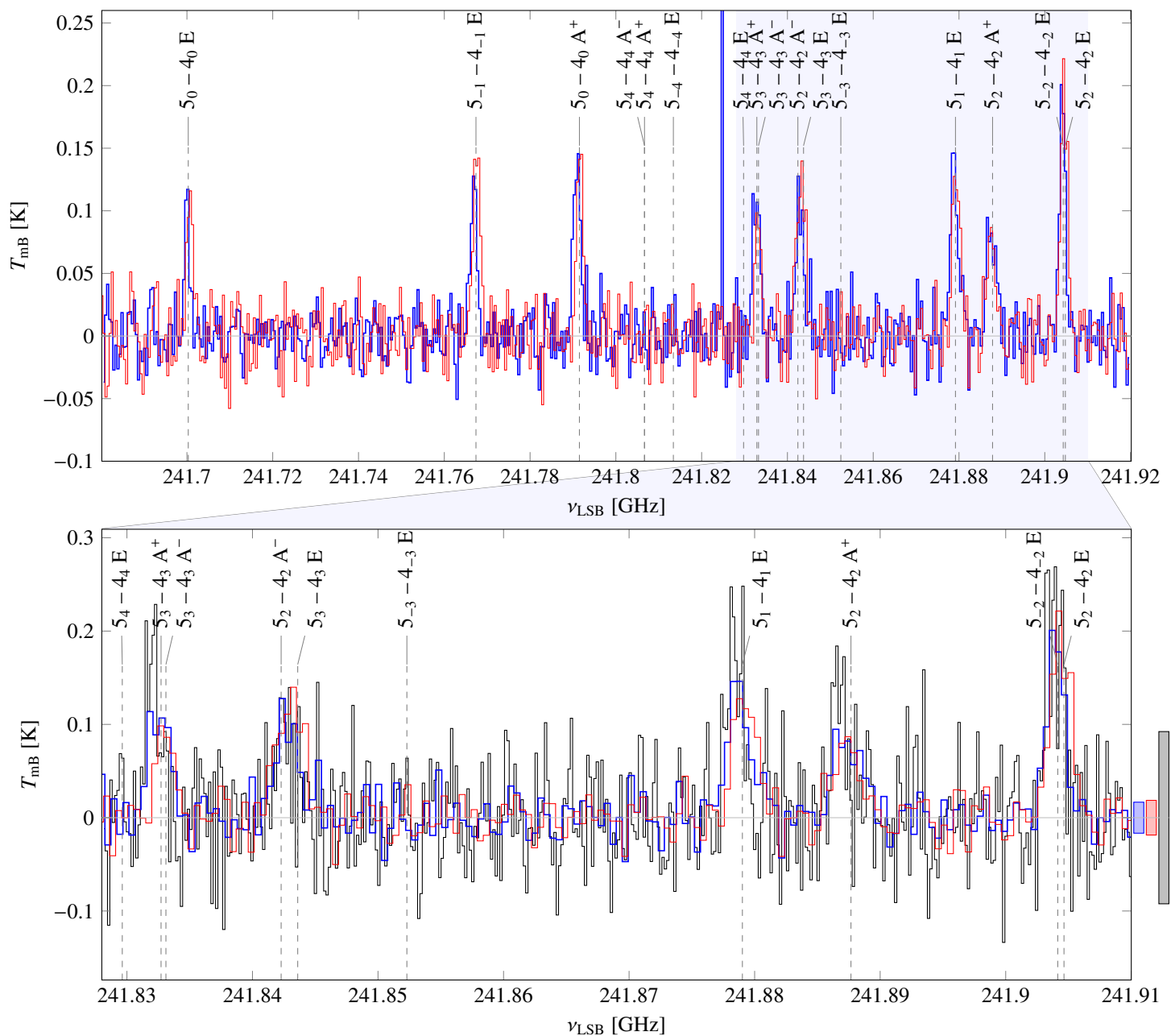

Figure~\ref{fig:linear:ch3oh} shows the observed \ce{CH3OH} rotational
transitions ($J$ = 5--4) with the AOSA, AOSB, and CTSB in comet \lin.
Eleven transitions are detected by the AOSA and
AOSB backends, including three blended lines around 241.833, 241.843, and
241.904~GHz. Eight of those transitions are also observed in the
frequency range covered by the CTSB (see lower panel in
Fig.~\ref{fig:linear:ch3oh}).  There appears to be an emission feature
at the frequency of the $5_4$ -- $4_4$ E transition in the AOSA and AOSB
spectra. However, it has been fitted as part of the continuum baseline,
since we are not confident that it is real due to the proximity of
the stronger blended pair consisting of the $5_3$ -- $4_3$ A$^{+}$ and
$5_3$ -- $4_3$ A$^{-}$ transitions. Although the noise level is about
the same as in the \ce{CH3OH} \nea{} observations, the emission lines
are stronger in \linear{} with larger line intensities by about
20--40\%.

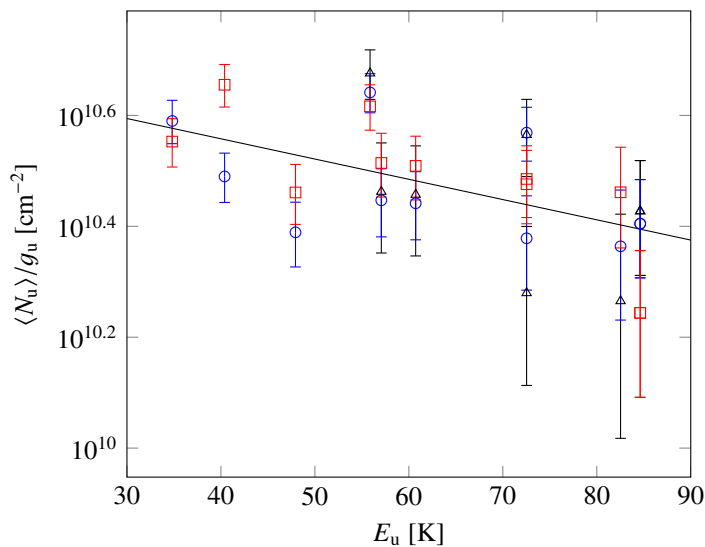
\begin{figure}
  \centering
  \begin{tikzpicture}
    \begin{semilogyaxis}[
      cycle list name=rotdiaglist,
      xlabel={$E_\mathrm{u}$ [K]},
      ylabel={$\langle N_\mathrm{u} \rangle/g_\mathrm{u}$ [\si{cm^{-2}}]},
      width=\hsize, xmin=30, xmax=90,
      ]
      \addplot+ table [x index=0,y index=1,y error index=2]
	{rot_ch3oh_T7_CTSB.txt};
      \addplot+ table [x index=0,y index=1,y error index=2]
	{rot_ch3oh_T7_AOSA.txt};
      \addplot+ table [x index=0,y index=1,y error index=2]
	{rot_ch3oh_T7_AOSB.txt};
      \addplot[mark=none] table [x index=0,y index=1] {rotdiag_T7___S_.txt};
    \end{semilogyaxis}
  \end{tikzpicture}
  \caption{Rotation diagram for \ce{CH3OH} lines in comet \linear{}
  which includes 1-$\sigma$ uncertainties.  The natural logarithm of the
  column density of the upper level divided by its degeneracy is plotted
  against the energy of the upper level of each transition for the AOSA
  (blue circles), AOSB (red squares) and CTSB (black triangles).  The
  solid line shows the best linear fit with a derived rotational
  temperature given by the inverse of the slope of \rotlinear\ to the
  lines detected by all the spectrometers with S/N greater than three.
  }
  \label{fig:linear:rotdiagram}
\end{figure}

The rotation diagram for multiple \ce{CH3OH} lines that fits
A\textsuperscript{+}, A\textsuperscript{-}, and E states simultaneously
is determined using the same weighted linear least-squares method.
Figure~\ref{fig:linear:rotdiagram} shows the best fit for all lines
observed in \lin{} with the AOSA, AOSB and CTSB spectrometers.  A
rotational temperature of \rotlinear{} is derived from the inverse of
the slope of the best linear fit to the observed transitions above the
3-$\sigma$ detection limit with all the spectrometers.
\citet{2008ApJ...689..613R} derived a lower rotational temperature of
\SI{35(5)}{\kelvin} in \lin{} from observations around
\SI{157}{\giga\hertz} at the 12-m Arizona Radio Observatory (ARO)
telescope on UT 21.80 May.  The derived rotational temperatures in
\nea{} and \lin{} are used in the excitation model of all the observed
molecules (see Sect.~\ref{sec:model}).

\subsection{Production rates}

Table~\ref{tbl:neat:prodrates} and \ref{tbl:linear:prodrates} show the
molecules and transitions detected with 1-$\sigma$ uncertainties or with
the upper limits of the integrated brightness temperature and derived
model-dependent production rates in comets \nea{} and \lin{}.  Line
areas are obtained by numerically integrating the signal over the fitted
baseline and velocity offsets in the comet frame that are calculated as
first moments of the velocity over the emission line range ($\sum_i
{T_\mathrm{mB}}_i v_i / \sum_i {T_\mathrm{mB}}_i$ where $v$ is the
Doppler velocity and the index $i$ refers to the channel number).  A
typical telescope pointing error of $\sim 2\arcsec$ was assumed to
compute the production rates in both comets.  Standard deviations for
the uncertainties are shown in brackets after the main part of the
number and are applicable to the last significant digits.  Line
frequencies were obtained from the latest online edition of the JPL
Molecular Spectroscopy Catalog \citep{1998JQSRT..60..883P}.

Most transitions were observed simultaneously in both objects with the
AOSA, AOSB and the higher resolution CTSB, which resolves the
asymmetries in the line shape, except for three methanol lines that are
outside the frequency range of the CTSB ($5_0$ -- $4_0$ E at
241.700~GHz, $5_{-1}$ -- $4_{-1}$ E at 241.767~GHz and $5_0$ -- $4_0$ E
at 241.791~GHz).  Blended \ce{CH3OH} transitions can be distinguished by
missing line velocity offset and production rate values in
Tables~\ref{tbl:neat:prodrates} and \ref{tbl:linear:prodrates}.  Line
areas and standard deviations are shown for the AOSA and AOSB, which are
generally the most sensitive of the AOS backends, and CTSB data.
Production rates shown in Tables~\ref{tbl:neat:prodrates} and
\ref{tbl:linear:prodrates} were derived assuming a parent molecule Haser
profile.  The difference with the production rates obtained using a
density distribution for daughter molecules is discussed briefly for
\ce{H2CO} and CS in Sects.~\ref{sec:h2co} and \ref{sec:cs}.

\begin{table*}
  \caption{
    Observed flux densities integrated over velocity, Doppler
  velocity shifts and production rates in comet \neat{} measured with
  the SMT in 7--11 May 2004.
  Statistical uncertainties are shown in brackets
  after the respective values in units of the last significant digits.
  }
  \label{tbl:neat:prodrates}
  \centering
  \begin{tabular}{c S
		S[table-number-alignment = center,
		output-decimal-marker = {\text{--}},
		parse-numbers = false]
		c c
		S[table-number-alignment = center,
		table-format = 1.3]
		S[table-number-alignment = center,
		table-figures-integer = 1,
		table-figures-uncertainty = 3,
		table-comparator = true]
		S[table-number-alignment = center,
		explicit-sign = +,
		table-format = +1.1(2)e+2]
		S[table-number-alignment = center,
		table-format = 1.2(2)]
		S[table-format = 1.2(2)e+2,
		table-comparator = true]
		}
    \toprule
    Molecule & {Date\tablefootmark{\emph{a}}} &
    {Transition\tablefootmark{\emph{b}}} &
    Frequency\tablefootmark{\emph{c}} & {Backend} &
    {\( \sigma_{T_\mathrm{mB}} \)} &
    {$\int T_\textrm{mB}\, dv$\tablefootmark{\emph{d}}} &
    {$\Delta v$\tablefootmark{\emph{e}}} &
    {$\vexp$} &
    {$Q$\tablefootmark{\emph{f}}}\\
    & {(UT)} & & (GHz) & & {(\si{\kelvin})} & {(\Kkms)} & {(\ms)} &
    {(\kms)} & {(\s)}\\
    \midrule
    \ce{HCN} & 7.98 & 3.2 & 265.886180 & AOSA & 0.033 & 3.607\pm0.070 & -1.6\pm0.4e2 &0.93\pm0.09 & 3.04\pm0.06e26\\
\ce{HCN} & 7.98 & 3.2 & 265.886180 & AOSB & 0.028 & 3.556\pm0.062 & -5.7\pm0.7e2 &0.93\pm0.09 & 2.99\pm0.05e26\\
\ce{HCN} & 7.98 & 3.2 & 265.886180 & CTSB & 0.165 & 3.374\pm0.061 & -1.9\pm0.1e2 &0.73\pm0.07 & 2.84\pm0.05e26\\
\ce{H2CO} & 8.12 & 3_{12}.2_{11} & 225.697775 & AOSA & 0.024 & 0.208\pm0.049 & -4.8\pm3.0e2 &1.02\pm0.10 & 3.50\pm0.82e26\\
\ce{H2CO} & 8.12 & 3_{12}.2_{11} & 225.697775 & AOSB & 0.026 & 0.199\pm0.045 & -7.0\pm1.7e2 &0.75\pm0.08 & 3.36\pm0.75e26\\
\ce{H2CO} & 8.12 & 3_{12}.2_{11} & 225.697775 & CTSB & 0.166 & <0.193 & & & <3.26e+26\\
\ce{CO} & 9.01 & 2.1 & 230.538000 & AOSA & 0.020 & 0.108\pm0.032 & -2\pm15e1 &0.33\pm0.03 & 7.8\pm2.3e27\\
\ce{CO} & 9.01 & 2.1 & 230.538000 & AOSB & 0.020 & 0.108\pm0.033 & -5.0\pm2.8e2 &0.75\pm0.07 & 7.8\pm2.4e27\\
\ce{CO} & 9.01 & 2.1 & 230.538000 & CTSB & 0.095 & 0.160\pm0.036 & -1.0\pm1.7e2 &0.73\pm0.07 & 1.15\pm0.26e28\\
\ce{CH3OH} & 11.03 & 5_0.4_0~\text{E} & 241.700159 & AOSA & 0.016 & 0.181\pm0.024 & -1.2\pm0.1e3 &1.13\pm0.11 & 4.89\pm0.66e27\\
\ce{CH3OH} & 11.03 & 5_0.4_0~\text{E} & 241.700159 & AOSB & 0.017 & 0.174\pm0.026 & -2.1\pm0.2e3 &1.15\pm0.11 & 4.69\pm0.71e27\\
\ce{CH3OH} & 11.03 & 5_{-1}.4_{-1}~\text{E} & 241.767234 & AOSA & 0.016 & 0.163\pm0.023 & -1.2\pm0.1e3 &1.13\pm0.11 & 3.88\pm0.54e27\\
\ce{CH3OH} & 11.03 & 5_{-1}.4_{-1}~\text{E} & 241.767234 & AOSB & 0.017 & 0.163\pm0.024 & -1.3\pm0.2e3 &0.62\pm0.06 & 3.89\pm0.58e27\\
\ce{CH3OH} & 11.03 & 5_0.4_0~\text{A}^+ & 241.791352 & AOSA & 0.016 & 0.274\pm0.026 & -7\pm12e1 &1.15\pm0.12 & 5.94\pm0.57e27\\
\ce{CH3OH} & 11.03 & 5_0.4_0~\text{A}^+ & 241.791352 & AOSB & 0.017 & 0.161\pm0.026 & -1.1\pm0.2e3 &0.77\pm0.08 & 3.49\pm0.57e27\\
\ce{CH3OH} & 11.03 & \mathbf{5_3}.\mathbf{4_3~\text{A}^+} & 241.832718 & AOSA & 0.016 & 0.108\pm0.023 &  & & \\
\ce{CH3OH} & 11.03 & \mathbf{5_3}.\mathbf{4_3~\text{A}^+} & 241.832718 & AOSB & 0.017 & 0.099\pm0.022 &  & & \\
\ce{CH3OH} & 11.03 & \mathbf{5_3}.\mathbf{4_3~\text{A}^+} & 241.832718 & CTSB & 0.069 & 0.080\pm0.026 &  & & \\
\ce{CH3OH} & 11.03 & \mathbf{5_3}.\mathbf{4_3~\text{A}^-} & 241.833106 & AOSA & 0.016 & 0.108\pm0.023 &  & & \\
\ce{CH3OH} & 11.03 & \mathbf{5_3}.\mathbf{4_3~\text{A}^-} & 241.833106 & AOSB & 0.017 & 0.099\pm0.022 &  & & \\
\ce{CH3OH} & 11.03 & \mathbf{5_3}.\mathbf{4_3~\text{A}^-} & 241.833106 & CTSB & 0.069 & 0.080\pm0.026 &  & & \\
\ce{CH3OH} & 11.03 & \mathbf{5_2}.\mathbf{4_2~\text{A}^-} & 241.842284 & AOSA & 0.016 & 0.122\pm0.023 &  & & \\
\ce{CH3OH} & 11.03 & \mathbf{5_2}.\mathbf{4_2~\text{A}^-} & 241.842284 & AOSB & 0.017 & 0.123\pm0.026 &  & & \\
\ce{CH3OH} & 11.03 & \mathbf{5_2}.\mathbf{4_2~\text{A}^-} & 241.842284 & CTSB & 0.069 & 0.191\pm0.032 &  & & \\
\ce{CH3OH} & 11.03 & \mathbf{5_3}.\mathbf{4_3~\text{E}} & 241.843604 & AOSA & 0.016 & 0.122\pm0.023 &  & & \\
\ce{CH3OH} & 11.03 & \mathbf{5_3}.\mathbf{4_3~\text{E}} & 241.843604 & AOSB & 0.017 & 0.123\pm0.026 &  & & \\
\ce{CH3OH} & 11.03 & \mathbf{5_3}.\mathbf{4_3~\text{E}} & 241.843604 & CTSB & 0.069 & 0.191\pm0.032 &  & & \\
\ce{CH3OH} & 11.03 & 5_1.4_1~\text{E} & 241.879025 & AOSA & 0.016 & 0.082\pm0.023 & 2.9\pm3.1e2 &0.88\pm0.09 & 2.99\pm0.82e27\\
\ce{CH3OH} & 11.03 & 5_1.4_1~\text{E} & 241.879025 & AOSB & 0.017 & 0.077\pm0.024 & -9.0\pm3.2e2 &0.60\pm0.06 & 2.80\pm0.89e27\\
\ce{CH3OH} & 11.03 & 5_1.4_1~\text{E} & 241.879025 & CTSB & 0.069 & 0.204\pm0.032 & -1.1\pm0.2e3 &1.59\pm0.16 & 7.4\pm1.2e27\\
\ce{CH3OH} & 11.03 & \mathbf{5_{-2}}.\mathbf{4_{-2}~\text{E}} & 241.904147 & AOSA & 0.016 & 0.191\pm0.024 &  & & \\
\ce{CH3OH} & 11.03 & \mathbf{5_{-2}}.\mathbf{4_{-2}~\text{E}} & 241.904147 & AOSB & 0.017 & 0.255\pm0.026 &  & & \\
\ce{CH3OH} & 11.03 & \mathbf{5_{-2}}.\mathbf{4_{-2}~\text{E}} & 241.904147 & CTSB & 0.069 & 0.280\pm0.032 &  & & \\
\ce{CH3OH} & 11.03 & \mathbf{5_2}.\mathbf{4_2~\text{E}} & 241.904643 & AOSA & 0.016 & 0.191\pm0.024 &  & & \\
\ce{CH3OH} & 11.03 & \mathbf{5_2}.\mathbf{4_2~\text{E}} & 241.904643 & AOSB & 0.017 & 0.255\pm0.026 &  & & \\
\ce{CH3OH} & 11.03 & \mathbf{5_2}.\mathbf{4_2~\text{E}} & 241.904643 & CTSB & 0.069 & 0.280\pm0.032 &  & & \\
\ce{CS} & 11.03 & 5.4 & 244.935643 & AOSA & 0.013 & 0.644\pm0.023 & 2.0\pm0.5e2 &0.96\pm0.10 & 2.00\pm0.07e26\\
\ce{CS} & 11.03 & 5.4 & 244.935643 & AOSB & 0.013 & 0.688\pm0.029 & 4.9\pm0.9e2 &0.92\pm0.09 & 2.13\pm0.09e26\\
\ce{CS} & 11.03 & 5.4 & 244.935643 & CTSB & 0.068 & 0.676\pm0.025 & 8.1\pm2.5e1 &0.75\pm0.08 & 2.10\pm0.08e26\\
\ce{HNC} & 11.95 & 3.2 & 271.981142 & AOSA & 0.016 & 0.222\pm0.029 & -7\pm17e1 &0.82\pm0.08 & 1.66\pm0.22e25\\
\ce{HNC} & 11.95 & 3.2 & 271.981142 & AOSB & 0.016 & 0.233\pm0.028 & 6\pm12e1 &1.04\pm0.10 & 1.74\pm0.21e25\\

    \bottomrule
  \end{tabular}
  \tablefoot{ This table is available in machine-readable format at the CDS.
  \tablefoottext{\emph{a}}{Mid-time of the observations recorded as
    decimal days in May 2004.}
  \tablefoottext{\emph{b}}{\ce{CH3OH} transitions marked in boldface
    indicate blended lines.}
  \tablefoottext{\emph{c}}{The line frequencies were obtained from the latest
  online edition of the JPL Molecular Spectroscopy Catalog
  \citep{1998JQSRT..60..883P}.}
  \tablefoottext{\emph{d}}{Line intensities and production rates of
    \ce{H2CO} measured with the CTSB are derived 3-$\sigma$ upper limits.}
  \tablefoottext{\emph{e}}{
    The velocity offsets are computed with respect to
    the optocenters of the complete components for blended lines.}
  \tablefoottext{\emph{f}}{Production rates derived for a gas kinetic
    temperature of \rotnea, expansion velocity of \vnea, electron
    density scaling factor of $\xne = 0.2$, and a pointing offset of
    2\arcsec.}
  }
\end{table*}

\begin{table*}
  \caption{
    Observed flux densities, velocity shifts and production
    rates in comet \linear{} observed with the SMT in 8--11 May 2004.
  }
  \label{tbl:linear:prodrates}
  \centering
  \begin{tabular}{c S
		S[table-number-alignment = center,
		output-decimal-marker = {\text{--}},
		parse-numbers = false]
		c c
		S[table-number-alignment = center,
		table-format = 1.3]
		S[table-number-alignment = center,
		table-figures-integer = 1,
		table-figures-uncertainty = 3,
		table-comparator = true,
		table-space-text-post = \textsuperscript{\emph{f}}]
		S[table-number-alignment = center,
		explicit-sign = +,
		table-format = +1.1(2)e+2]
		S[table-number-alignment = center,
		table-format = 1.2(2)]
		S[table-format = 1.2(2)e+2,
		table-comparator = true]
		}
    \toprule
    Molecule & {Date\tablefootmark{\emph{a}}} &
    {Transition\tablefootmark{\emph{b}}} &
    Frequency\tablefootmark{\emph{c}} & {Backend} &
    {\( \sigma_{T_\mathrm{mB}} \)} &
    {$\int T_\textrm{mB}\, dv$\tablefootmark{\emph{d}}} &
    {$\Delta v$\tablefootmark{\emph{e}}} &
    {$\vexp$} &
    {$Q$\tablefootmark{\emph{f}}}\\
    & {(UT)} & & (GHz) & & {(\si{\kelvin})} & {(\Kkms)} & {(\ms)} &
    {(\kms)} & {(\s)}\\
    \midrule
    \ce{HCN} & 8.68 & 3.2 & 265.886180 & AOSA & 0.031 & 2.757\pm0.061 & -1.6\pm0.4e2 &1.05\pm0.10 & 4.37\pm0.10e26\\
\ce{HCN} & 8.68 & 3.2 & 265.886180 & AOSB & 0.031 & 2.878\pm0.061 & -5.5\pm0.5e2 &1.04\pm0.10 & 4.56\pm0.10e26\\
\ce{HCN} & 8.68 & 3.2 & 265.886180 & CTSB & 0.132 & 2.814\pm0.056 & -2.8\pm0.2e2 &0.90\pm0.09 & 4.46\pm0.09e26\\
\ce{H2CO} & 8.82 & 3_{12}.2_{11} & 225.697775 & AOSA & 0.028 & 0.458\pm0.061 & 2.4\pm1.3e2 &1.15\pm0.11 & 2.26\pm0.30e27\\
\ce{H2CO} & 8.82 & 3_{12}.2_{11} & 225.697775 & AOSB & 0.032 & 0.497\pm0.067 & -1.1\pm0.8e2 &1.21\pm0.12 & 2.45\pm0.33e27\\
\ce{H2CO} & 8.82 & 3_{12}.2_{11} & 225.697775 & CTSB & 0.181 & 0.347\pm0.070 & 1.5\pm1.5e2 &0.95\pm0.10 & 1.71\pm0.35e27\\
\ce{CO} & 9.67 & 2.1 & 230.538000 & AOSA & 0.036 & <0.133 & & & <1.62e+28\\
\ce{CO} & 9.67 & 2.1 & 230.538000 & AOSB & 0.030 & <0.112 & & & <1.36e+28\\
\ce{CO} & 9.67 & 2.1 & 230.538000 & CTSB & 0.191 & <0.217 & & & <2.63e+28\\
\ce{HCN} & 9.70 & 3.2 & 265.886180 & AOSA & 0.045 & 2.474\pm0.081 & -3.0\pm5.0e1 &0.99\pm0.10 & 3.92\pm0.13e26\\
\ce{HCN} & 9.70 & 3.2 & 265.886180 & AOSB & 0.040 & 2.471\pm0.073 & -5.3\pm0.5e2 &1.01\pm0.10 & 3.91\pm0.12e26\\
\ce{HCN} & 9.70 & 3.2 & 265.886180 & CTSB & 0.216 & 2.778\pm0.087 & -2.0\pm0.3e2 &0.89\pm0.09 & 4.40\pm0.14e26\\
\ce{CH3OH} & 10.74 & 5_0.4_0~\text{E} & 241.700159 & AOSA & 0.017 & 0.212\pm0.026 & 2.2\pm1.5e2 &0.68\pm0.07 & 1.06\pm0.13e28\\
\ce{CH3OH} & 10.74 & 5_0.4_0~\text{E} & 241.700159 & AOSB & 0.019 & 0.250\pm0.029 & -7.3\pm1.3e2 &0.99\pm0.10 & 1.25\pm0.15e28\\
\ce{CH3OH} & 10.74 & 5_{-1}.4_{-1}~\text{E} & 241.767234 & AOSA & 0.017 & 0.257\pm0.024 & 6.4\pm0.9e2 &0.98\pm0.10 & 1.13\pm0.11e28\\
\ce{CH3OH} & 10.74 & 5_{-1}.4_{-1}~\text{E} & 241.767234 & AOSB & 0.019 & 0.376\pm0.031 & -5\pm11e1 &1.11\pm0.11 & 1.66\pm0.14e28\\
\ce{CH3OH} & 10.74 & 5_0.4_0~\text{A}^+ & 241.791352 & AOSA & 0.017 & 0.337\pm0.028 & 4.1\pm1.1e2 &1.12\pm0.11 & 1.36\pm0.11e28\\
\ce{CH3OH} & 10.74 & 5_0.4_0~\text{A}^+ & 241.791352 & AOSB & 0.019 & 0.310\pm0.029 & -1.5\pm1.1e2 &0.93\pm0.09 & 1.24\pm0.12e28\\
\ce{CH3OH} & 10.74 & \mathbf{5_3}.\mathbf{4_3~\text{A}^+} & 241.832718 & AOSA & 0.017 & 0.281\pm0.026 &  & & \\
\ce{CH3OH} & 10.74 & \mathbf{5_3}.\mathbf{4_3~\text{A}^+} & 241.832718 & AOSB & 0.019 & 0.194\pm0.027 &  & & \\
\ce{CH3OH} & 10.74 & \mathbf{5_3}.\mathbf{4_3~\text{A}^+} & 241.832718 & CTSB & 0.092 & 0.296\pm0.035 &  & & \\
\ce{CH3OH} & 10.74 & \mathbf{5_3}.\mathbf{4_3~\text{A}^-} & 241.833106 & AOSA & 0.017 & 0.281\pm0.026 &  & & \\
\ce{CH3OH} & 10.74 & \mathbf{5_3}.\mathbf{4_3~\text{A}^-} & 241.833106 & AOSB & 0.019 & 0.194\pm0.027 &  & & \\
\ce{CH3OH} & 10.74 & \mathbf{5_3}.\mathbf{4_3~\text{A}^-} & 241.833106 & CTSB & 0.092 & 0.296\pm0.035 &  & & \\
\ce{CH3OH} & 10.74 & \mathbf{5_2}.\mathbf{4_2~\text{A}^-} & 241.842284 & AOSA & 0.017 & 0.304\pm0.031 &  & & \\
\ce{CH3OH} & 10.74 & \mathbf{5_2}.\mathbf{4_2~\text{A}^-} & 241.842284 & AOSB & 0.019 & 0.380\pm0.031 &  & & \\
\ce{CH3OH} & 10.74 & \mathbf{5_2}.\mathbf{4_2~\text{A}^-} & 241.842284 & CTSB & 0.092 & 0.242\pm0.044 &  & & \\
\ce{CH3OH} & 10.74 & \mathbf{5_3}.\mathbf{4_3~\text{E}} & 241.843604 & AOSA & 0.017 & 0.304\pm0.031 &  & & \\
\ce{CH3OH} & 10.74 & \mathbf{5_3}.\mathbf{4_3~\text{E}} & 241.843604 & AOSB & 0.019 & 0.380\pm0.031 &  & & \\
\ce{CH3OH} & 10.74 & \mathbf{5_3}.\mathbf{4_3~\text{E}} & 241.843604 & CTSB & 0.092 & 0.242\pm0.044 &  & & \\
\ce{CH3OH} & 10.74 & 5_1.4_1~\text{E} & 241.879025 & AOSA & 0.017 & 0.374\pm0.028 & 2.4\pm1.0e2 &1.15\pm0.12 & 2.53\pm0.19e28\\
\ce{CH3OH} & 10.74 & 5_1.4_1~\text{E} & 241.879025 & AOSB & 0.019 & 0.353\pm0.031 & -2.3\pm1.2e2 &1.21\pm0.12 & 2.38\pm0.21e28\\
\ce{CH3OH} & 10.74 & 5_1.4_1~\text{E} & 241.879025 & CTSB & 0.092 & 0.405\pm0.041 & 7.1\pm1.0e2 &0.85\pm0.09 & 2.73\pm0.28e28\\
\ce{CH3OH} & 10.74 & 5_2.4_2~\text{A}^+ & 241.887674 & AOSA & 0.017 & 0.272\pm0.028 & 3.4\pm1.3e2 &1.48\pm0.15 & 2.42\pm0.25e28\\
\ce{CH3OH} & 10.74 & 5_2.4_2~\text{A}^+ & 241.887674 & AOSB & 0.019 & 0.224\pm0.031 & 2.5\pm1.8e2 &1.12\pm0.11 & 2.00\pm0.28e28\\
\ce{CH3OH} & 10.74 & 5_2.4_2~\text{A}^+ & 241.887674 & CTSB & 0.092 & 0.269\pm0.043 & 6.4\pm1.9e2 & & 2.40\pm0.38e28\\
\ce{CH3OH} & 10.74 & \mathbf{5_{-2}}.\mathbf{4_{-2}~\text{E}} & 241.904147 & AOSA & 0.017 & 0.403\pm0.026 &  & & \\
\ce{CH3OH} & 10.74 & \mathbf{5_{-2}}.\mathbf{4_{-2}~\text{E}} & 241.904147 & AOSB & 0.019 & 0.471\pm0.029 &  & & \\
\ce{CH3OH} & 10.74 & \mathbf{5_{-2}}.\mathbf{4_{-2}~\text{E}} & 241.904147 & CTSB & 0.092 & 0.418\pm0.047 &  & & \\
\ce{CH3OH} & 10.74 & \mathbf{5_2}.\mathbf{4_2~\text{E}} & 241.904643 & AOSA & 0.017 & 0.403\pm0.026 &  & & \\
\ce{CH3OH} & 10.74 & \mathbf{5_2}.\mathbf{4_2~\text{E}} & 241.904643 & AOSB & 0.019 & 0.471\pm0.029 &  & & \\
\ce{CH3OH} & 10.74 & \mathbf{5_2}.\mathbf{4_2~\text{E}} & 241.904643 & CTSB & 0.092 & 0.418\pm0.047 &  & & \\
\ce{CS} & 10.74 & 5.4 & 244.935643 & AOSA & 0.021 & 0.909\pm0.039 & 3.2\pm0.5e2 &0.98\pm0.10 & 5.54\pm0.24e26\\
\ce{CS} & 10.74 & 5.4 & 244.935643 & AOSB & 0.017 & 1.071\pm0.039 & 6.5\pm0.6e2 &1.06\pm0.11 & 6.53\pm0.24e26\\
\ce{CS} & 10.74 & 5.4 & 244.935643 & CTSB & 0.092 & 0.975\pm0.037 & 1.4\pm0.3e2 &0.81\pm0.08 & 5.95\pm0.23e26\\
\ce{HCN} & 11.60 & 3.2 & 265.886180 & AOSA & 0.034 & 3.154\pm0.062 & -2.7\pm3.6e1 &1.05\pm0.10 & 4.99\pm0.10e26\\
\ce{HCN} & 11.60 & 3.2 & 265.886180 & AOSB & 0.030 & 3.285\pm0.053 & -4.7\pm0.3e2 &1.07\pm0.11 & 5.20\pm0.08e26\\
\ce{HCN} & 11.60 & 3.2 & 265.886180 & CTSB & 0.129 & 3.184\pm0.052 & -1.7\pm0.1e2 &0.88\pm0.09 & 5.04\pm0.08e26\\
\ce{HNC} & 11.69 & 3.2 & 271.981142 & AOSA & 0.013 & 0.315\pm0.023 & -1.3\pm0.7e2 &0.95\pm0.09 & 4.36\pm0.32e25\\
\ce{HNC} & 11.69 & 3.2 & 271.981142 & AOSB & 0.013 & 0.365\pm0.023 & 3.0\pm0.7e2 &0.96\pm0.10 & 5.05\pm0.32e25\\
\ce{HNC} & 11.69 & 3.2 & 271.981142 & CTSB & 0.063 & 0.396\pm0.023 & 4.2\pm4.0e1 &0.85\pm0.09 & 5.48\pm0.32e25\\

    \bottomrule
  \end{tabular}
  \tablefoot{ This table is available in machine-readable format at the CDS.
  \tablefoottext{\emph{a}}{Mid-time of the observations recorded as
    decimal days in May 2004.}
  \tablefoottext{\emph{b}}{\ce{CH3OH} transitions marked in boldface
    indicate blended lines.}
  \tablefoottext{\emph{c}}{The line frequencies were obtained from the latest
  online edition of the JPL Molecular Spectroscopy Catalog
  \citep{1998JQSRT..60..883P}.}
  \tablefoottext{\emph{d}}{Line intensities and production rates of
    \ce{CO} are derived 3-$\sigma$ upper limits.}
  \tablefoottext{\emph{e}}{
    The velocity offsets are computed with respect to
    the optocenters of the complete components for blended lines.}
  \tablefoottext{\emph{f}}{Production rates computed for
    $T_\mathrm{kin}$ = \rotlin, $\vexp$ = \vlin, $\xne = 0.2$, and a
    pointing offset of 2\arcsec.}
  }
\end{table*}

\subsubsection{HCN}

The abundance of HCN relative to water has been observed to have a range
between 0.08--0.25\% in several comets at submillimeter wavelengths for
a wide range of heliocentric distances with most comets having a value
around 0.1\%.  For this reason, HCN has been generally used as a proxy to
determine water production rates \citep{2002EM&P...90..323B}.  To
calculate the HCN production rates, we have considered the effects of
radiative pumping of the fundamental vibrational levels by infrared
radiation from the Sun.  Nonetheless, the transition between the
collision-dominated region in the inner coma and fluorescent equilibrium
in the outer regions of the coma, where absorption of solar infrared
radiation results in the excitation of vibrational levels, has been
determined to be 82\arcsec--119\arcsec for comet \lin{} and
59\arcsec--86\arcsec in comet \nea{} \citep[see discussion
in][]{2005ApJ...630..623F}.  Therefore, a beam size of about 30\arcsec{}
in our HCN observations probes a region where molecular excitation is
dominated by collisions with water and electrons, and the infrared
pumping effects are largely negligible \citep[c.f.][]{2009AJ....137.4837H}.

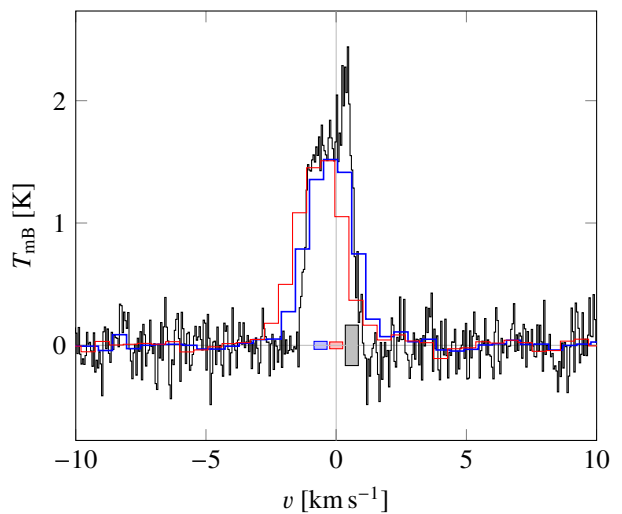
\begin{figure}
  \centering
  \begin{tikzpicture}
    \begin{axis}[spectrum style, clip=true]
      \rmshcnbars{\qhcnAOSArms}{\qhcnAOSBrms}{\qhcnCTSBrms}
      \addplot table [x index=0,y index=1] {Q4_HCN_7.98_CTSB.dat};
      \addplot table [x index=0,y index=1] {Q4_HCN_7.98_AOSA.dat};
      \addplot table [x index=0,y index=1] {Q4_HCN_7.98_AOSB.dat};
    \end{axis}
  \end{tikzpicture}
  \caption{Averaged spectrum of the \ce{HCN} (3--2) line at 265.886~GHz
  toward comet \neat{} observed on UT 7.98 May 2004 with the
  AOSA (thick blue line), AOSB (thin red line), and CTSB (black line)
  spectrometers.  The vertical axis is the calibrated main beam
  brightness temperature and the horizontal axis is the Doppler velocity
  in the comet nucleus rest frame.  The rms noise value on the
  brightness temperature measured with respect to the zero reference
  level is shown by the blue, red, and gray bars close to the origin for
  the AOSA, AOSB, and CTSB backends, respectively.  Note the difference
  in the native spectral resolutions.}
  \label{fig:neat:hcn}
\end{figure}

The HCN (3--2) line at 265.886~GHz was
observed in comet \nea{} on UT 7.98 May 2004 with a total integration
time of 120 minutes.  Figure~\ref{fig:neat:hcn} shows the averaged
\ce{HCN} rotational spectrum toward \nea.  We derived an HCN production
rate of \qqhcn{}~\s\, which corresponds to a mixing ratio with
respect to \ce{H2O} of \qqhcnqho{} by using the $Q_\ce{H2O}$ predicted by
the Gaussian fit to the daily averaged measurements obtained with the
{\it Odin} satellite \citep{2009A&A...501..359B}.

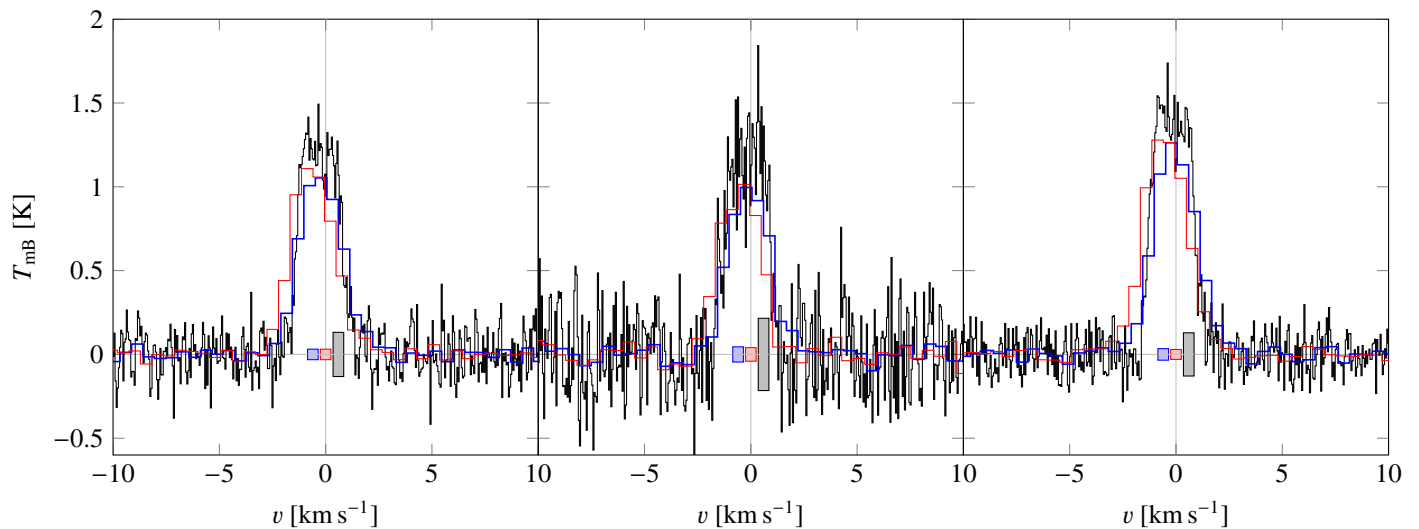
\begin{figure*}
  \centering
  \pgfplotsset{stack style/.style=
    {width=.39\hsize, height=.4\hsize, ymin=-0.6, ymax=2,clip=true
    }}
  \begin{tikzpicture}
    \begin{axis}[spectrum style,stack style,name=left]
      \rmshcnbars{\thcnaAOSArms}{\thcnaAOSBrms}{\thcnaCTSBrms}
      \addplot table [x index=0,y index=1] {T7_HCN_8.68_CTSB.dat};
      \addplot table [x index=0,y index=1] {T7_HCN_8.68_AOSA.dat};
      \addplot table [x index=0,y index=1] {T7_HCN_8.68_AOSB.dat};
    \end{axis}
    \begin{axis}[spectrum style,stack style,name=center,
      at={(left.east)},anchor=west,
      yticklabel=\empty, ylabel=\empty,
      xtick={-5,0,5,10},
    ]
      \rmshcnbars{\thcnbAOSArms}{\thcnbAOSBrms}{\thcnbCTSBrms}
      \addplot table [x index=0,y index=1] {T7_HCN_9.70_CTSB.dat};
      \addplot table [x index=0,y index=1] {T7_HCN_9.70_AOSA.dat};
      \addplot table [x index=0,y index=1] {T7_HCN_9.70_AOSB.dat};
    \end{axis}
    \begin{axis}[spectrum style,stack style,
      at={(center.east)},anchor=west,
      yticklabel=\empty, ylabel=\empty,
      xtick={-5,0,5,10},
    ]
      \rmshcnbars{\thcncAOSArms}{\thcncAOSBrms}{\thcncCTSBrms}
      \addplot table [x index=0,y index=1] {T7_HCN_11.60_CTSB.dat};
      \addplot table [x index=0,y index=1] {T7_HCN_11.60_AOSA.dat};
      \addplot table [x index=0,y index=1] {T7_HCN_11.60_AOSB.dat};
    \end{axis}
  \end{tikzpicture}
  \caption{Averaged spectrum of the \ce{HCN} (3--2) line at 265.886~GHz
  toward comet \linear{} observed on UT 8.68 (\emph{left panel}), 9.70
  (\emph{second panel}), and 11.60 (\emph{right panel}) May 2004 with
  integration times of 120, 60, and 28 minutes respectively using the
  AOSA (thick blue line), AOSB (thin red line), and CTSB (black line)
  spectrometers.  The vertical axis is the calibrated main beam
  brightness temperature and the horizontal axis is the Doppler velocity
  in the cometocentric frame.  The rms noise level measured with respect
  to the zero reference level is shown by the blue, red, and gray bars
  near the plots origin for AOSA, AOSB, and CTSB, respectively.
  }
  \label{fig:linear:hcn}
\end{figure*}

The HCN (3--2) line was observed in comet \lin{} on three different
epochs (8.68, 9.70 and 11.60 May 2004 UT) with integration times of 120,
60, and 28 minutes, respectively.  Figure~\ref{fig:linear:hcn} shows the
averaged AOSA, AOSB, and CTSB spectra toward \lin\ from those dates.
The mean HCN production rate from those observations is \qthcn{}~\s,
giving a mixing ratio with respect to \ce{H2O} of \qthcnqho, with a
variation of 20\% around the mean value over the observing period. This
mixing ratio is lower than the value of \num{3.3(11)e-3} that is
obtained by \citet{2005ApJ...630..623F} using the
Berkeley-Illinois-Maryland Association (BIMA) array and closer to the
standard value for the relative abundance of $Q_\ce{HCN}/Q_\ce{H2O} \sim
\num{1e-3}$ that is inferred from simultaneous observations of \lin{} at
the BIMA and Owens Valley Radio Observatory (OVRO) arrays
\citep{2009AJ....137.4837H}.

\subsubsection{\ce{H2CO}}\label{sec:h2co}

Formaldehyde (\ce{H2CO}) has been found to be ubiquitous in dense
molecular clouds and has been observed in several comets, since its
first detection in comet 1P/Halley
\citep{1987Sci...237..626M,1989AJ.....97..246S}.  The density profile of
\ce{H2CO} in cometary atmospheres is believed to originate in the
photodissociation of a distributed source in the coma
\citep{1997EM&P...78....5B}.  Figure~\ref{fig:neat:h2co} shows the
observed \ce{H2CO} rotational transition at 225.698~GHz in comet \nea{}.
There is a tentative detection of the emission line at the 3-$\sigma$
detection limit using a window of (-1, 1) \kms{} around the transition
rest frequency in the CTSB spectrum, while there are 4-$\sigma$ emission
features in both the AOSA and AOSB spectra.  The mean value of
$Q_\ce{H2CO}$ derived from the AOSA and AOSB is \qqhco{}, corresponding
to a $Q_\ce{H2CO}/Q_\ce{H2O}$ ratio of \qqhcoqho{}. On the other hand,
the \ce{H2CO} production rate is \qqhcod{}~\s in \nea{} with
$Q_\ce{H2CO}/Q_\ce{H2O}$ of \qqhcoqhod{}, which is about twice as large,
assuming a parent distributed source in the coma.

We show the averaged spectra of \ce{H2CO} in comet \lin{} in
Fig.~\ref{fig:linear:h2co}.  The emission line is detected toward this
object by all the spectrometers with a 7- to 8-$\sigma$ significance.
The mean value of the derived formaldehyde production rate from the
three spectrometers is \qthco{}, which corresponds to a $Q_\ce{H2CO}/Q_\ce{H2O}$
ratio of \qthcoqho{}.  This value agrees well with the mean
$Q_\ce{H2CO}/Q_\ce{H2O}$ mixing ratio of \num{7.9\pm0.9e-3} derived from
observations of \lin{} with the NASA Infrared Telescope Facility on 5, 7,
and 9 May 2004 \citep{2006ApJ...650..470D} and is substantially larger
than the mean relative abundance of $Q_\ce{H2CO}/Q_\ce{H2O}$ =
\num{1.3e-3} that is derived from three different rotational lines observed from
15.60 to 25.01 May with the 12-m ARO telescope
\citep{2006ApJ...649.1169M}.  The production rate for a distributed
source is \qthcod{}~\s, and the mixing ratio relative to water is
\qthcoqhod{} in \lin{}, which is about twice the average value of
\num{4.3e-3} that is obtained from observations with the 12-m ARO telescope
with the assumption that \ce{H2CO} is released from small refractory
particles in the coma \citep{2006ApJ...649.1169M}.

\begin{figure}
  \centering
  \begin{tikzpicture}
    \begin{axis}[spectrum style]
      \rmsbars{\qhcoAOSArms}{\qhcoAOSBrms}{\qhcoCTSBrms}
      \addplot table [x index=0,y index=1] {Q4_H2CO_8.12_CTSB.dat};
      \addplot table [x index=0,y index=1] {Q4_H2CO_8.12_AOSA.dat};
      \addplot table [x index=0,y index=1] {Q4_H2CO_8.12_AOSB.dat};
    \end{axis}
  \end{tikzpicture}
  \caption{Averaged spectrum of the \ce{H2CO} ($3_{12}$--$2_{11}$) line
at 225.698~GHz toward comet \neat{} observed on UT 8.12 May 2004 with
the AOSA (thick blue line), AOSB (thin red line), and CTSB (black line).
The vertical axis is the calibrated main beam brightness temperature and
the horizontal axis is the Doppler velocity in the comet rest frame.
  The 1-$\sigma$ rms noise
  is shown by the blue, red, and gray bars to the right of the figure
  for AOSA, AOSB, and CTSB, respectively.
  }
  \label{fig:neat:h2co}
\end{figure}
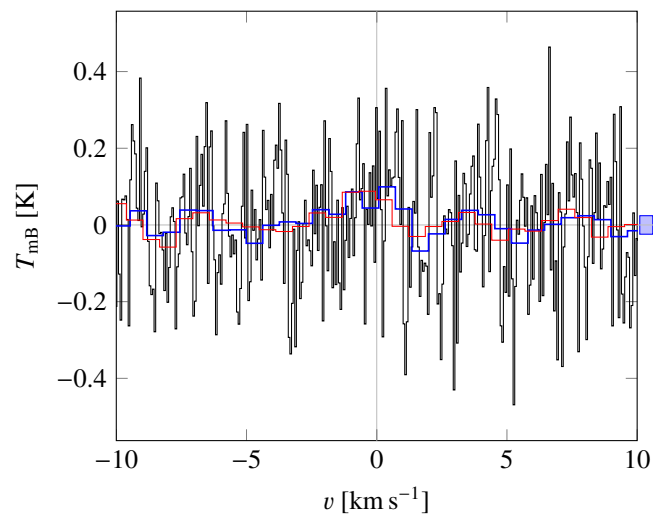

\begin{figure}
  \centering
  \begin{tikzpicture}
    \begin{axis}[spectrum style, ymin=-0.5]
      \rmsbars{\thcoAOSArms}{\thcoAOSBrms}{\thcoCTSBrms}
      \addplot table [x index=0,y index=1] {T7_H2CO_8.82_CTSB.dat};
      \addplot table [x index=0,y index=1] {T7_H2CO_8.82_AOSA.dat};
      \addplot table [x index=0,y index=1] {T7_H2CO_8.82_AOSB.dat};
    \end{axis}
  \end{tikzpicture}
  \caption{Averaged spectrum of the \ce{H2CO} ($3_{12}$--$2_{11}$) line
at 225.698~GHz toward comet \linear{} observed on 8.82 UT May 2004 with
the AOSA (thick blue line), AOSB (thin red line), and CTS (black line).
The vertical axis is the calibrated main beam brightness temperature and
the horizontal axis is the Doppler velocity in the comet rest frame.
  The rms noise level measured with respect to the zero reference level
  is shown by the blue, red, and gray bars to the right of the figure
  for AOSA, AOSB, and CTSB, respectively.
}
  \label{fig:linear:h2co}
\end{figure}

\subsubsection{CO}

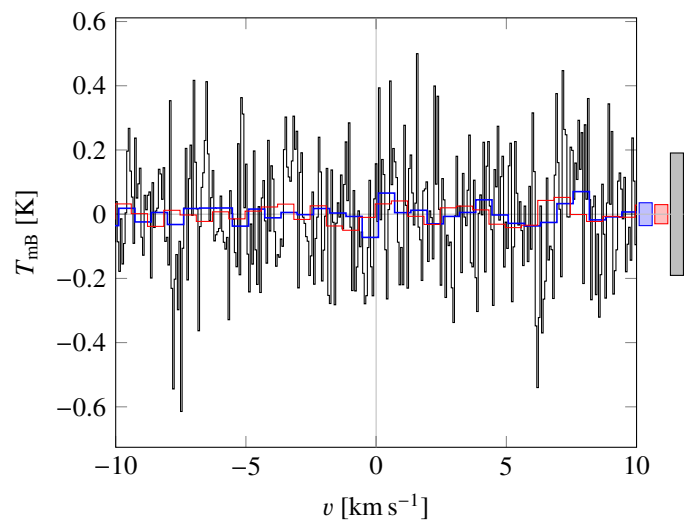
\begin{figure}
  \centering
  \begin{tikzpicture}
    \begin{axis}[spectrum style]
      \rmsbars{\tcoAOSArms}{\tcoAOSBrms}{\tcoCTSBrms}
      \addplot table [x index=0,y index=1] {T7_12CO_9.67_CTSB.dat};
      \addplot table [x index=0,y index=1] {T7_12CO_9.67_AOSA.dat};
      \addplot table [x index=0,y index=1] {T7_12CO_9.67_AOSB.dat};
    \end{axis}
  \end{tikzpicture}
  \caption{Averaged spectrum of the \ce{CO} (2--1) line at 230.538~GHz
toward comet \linear{} observed on UT 9.67 May 2004 with the AOSA (thick
blue line), AOSB (thin red line), and CTSB (black line) spectrometers.  The
vertical axis is the calibrated main beam brightness temperature and the
horizontal axis is the Doppler velocity in the comet rest frame.
  The rms noise level measured with respect to the zero reference level
  is shown by the blue, red, and gray bars to the right of the figure
  for AOSA, AOSB, and CTSB, respectively.
}
  \label{fig:linear:co}
\end{figure}

Carbon monoxide (CO) is the main driver of cometary activity at large
heliocentric distances and has been detected in the radio with
comparatively high abundances relative to water of 0.15--23\% for comets
that are close to the Sun.  Figures~\ref{fig:neat:co} and
\ref{fig:linear:co} show the averaged \ce{CO} spectrum of the (2--1)
rotational transition in comets \nea\ and \lin, respectively.  There is
a tentative 4-$\sigma$ detection of an emission feature in the \nea{}
spectrum using the three spectrometers at the frequency of the CO line,
which has a production rate of \qqco{} \s, and no detection in the
\lin{} spectrum.  An upper limit was obtained for the \ce{CO} production
rate of \qtco{}\ \s{} in comet \lin{} from the 3-$\sigma$ values of the
integrated main beam brightness temperature.  These values correspond to
the mixing ratios $Q_\ce{CO}/Q_\ce{H2O}$ of \qqcoqho{} and \qtcoqho{}
for \nea{} and \lin{}, respectively.

The CO abundance relative to \ce{H2O} is relatively low in both comets.
There is marginal statistical evidence from radio observations of low CO
abundance in dynamically new comets \citep[see
e.g.][]{2002EM&P...90..323B}. This may be caused by partial evaporation
of CO from the surface layer in previous visits through the outer part
of the Solar System. While \nea{} and \lin{} are widely believed to have
visited the inner Solar System for the first time during their 2004
passage, it is possible that they have passed by its outer parts several
times.  Recent dynamical models suggest that \lin{} is a dynamically new
comet with its previous perihelion passage at a distance larger than 400
AU from the Sun, while \nea{} has visited the inner part of the Solar
System during a previous apparition at around a 6--7 AU heliocentric
distance and could therefore be considered as dynamically old
\citep{2010MNRAS.404.1886K,2012A&A...544A.119K}.  Since CO is the most
volatile of the observed molecules, it may have evaporated during past
perihelion passages even at large heliocentric distances.

\begin{figure}
  \centering
  \begin{tikzpicture}
    \begin{axis}[spectrum style]
      \rmsbars{\qcoAOSArms}{\qcoAOSBrms}{\qcoCTSBrms}
      \addplot table [x index=0,y index=1] {Q4_12CO_9.01_CTSB.dat};
      \addplot table [x index=0,y index=1] {Q4_12CO_9.01_AOSA.dat};
      \addplot table [x index=0,y index=1] {Q4_12CO_9.01_AOSB.dat};
    \end{axis}
  \end{tikzpicture}
  \caption{Averaged spectrum of the \ce{CO} (2--1) line at 230.538~GHz
  toward comet \neat{} observed on UT 9.01 May 2004 with the
  AOSA (thick blue line), AOSB (thin red line), and CTSB (black line).
  The vertical axis is the calibrated main beam brightness temperature
  and the horizontal axis is the Doppler velocity in the comet rest
  frame.  The rms noise level measured with respect to the zero
  reference level is shown by the blue, red, and gray bars to the right
  of the figure for AOSA, AOSB, and CTSB, respectively.
  }
  \label{fig:neat:co}
\end{figure}

\subsubsection{CS}\label{sec:cs}

The carbon monosulfide radical (CS) is a daughter species that is
believed to be formed in cometary coma from photodissociation of carbon
disulfide (\ce{CS2}), which has a short lifetime
\citep{2001AJ....121.1147S}.  CS has been observed in comets at UV
\citep{2004come.book..425F} and several rotational lines at radio
wavelengths \citep{2004come.book..391B}.  The CS ($J$ = 5--4) transition
was detected at the SMT in the upper sideband of the receiver
simultaneously with the \ce{CH3OH} observations in the lower sideband in
both \nea{} and \lin.  To obtain the radial density profile and derive
the production rates, we used a photodissociation rate of
$\beta_\mathrm{0,CS} = 2.5 \times 10^{-5} \s$ at heliocentric distance
of 1~AU estimated from spectroscopic observations
\citep{2007A&A...475.1131B,2011A&A...528A.142B}.  Scaled to the
appropriate $\rh$, this photodissociation rate corresponds to
scale-lengths of \SI{29000}{km} and \SI{19000}{km} for \nea{} and
\lin{}.  These scale-lengths assume expansion velocities of \vnea{} and
\vlin{} obtained from the width of the HCN lines in comets \nea{} and
\lin{}, respectively (see Table~\ref{tbl:beta}).

\begin{figure}
  \centering
  \begin{tikzpicture}
    \begin{axis}[spectrum style, clip=true]
      \rmshcnbars{\qcsAOSArms}{\qcsAOSBrms}{\qcsCTSBrms}
      \addplot table [x index=0,y index=1] {Q4_CS_11.03_CTSB.dat};
      \addplot table [x index=0,y index=1] {Q4_CS_11.03_AOSA.dat};
      \addplot table [x index=0,y index=1] {Q4_CS_11.03_AOSB.dat};
    \end{axis}
  \end{tikzpicture}
  \caption{Averaged spectrum of the \ce{CS} (5--4) line at 244.936~GHz
  toward comet \neat{} observed on UT 11.03 May 2004 with the
  AOSA (thick blue line), AOSB (thin red line), and CTSB (black line).
  The vertical axis is the calibrated main beam brightness temperature
  and the horizontal axis is the Doppler velocity in the comet rest
  frame.
  The rms noise level measured with respect to the zero reference level
  is shown by the blue, red, and gray bars close to the origin of the
  figure for AOSA, AOSB, and CTSB, respectively.
  }
  \label{fig:neat:cs}
\end{figure}

Figure~\ref{fig:neat:cs} shows the averaged \ce{CS} rotational
transition at 244.936~GHz in comet \nea{} with a derived production rate
of \qqcs{} \s and a mixing ratio $Q_\ce{CS}/Q_\ce{H2O}$ of \qqcsqho{},
assuming direct release from the nucleus. A daughter distribution
profile yields a CS production rate of \qqcsd{} \s with a mixing ratio
relative to \ce{H2O} of \qqcsqhod{}.

\begin{figure}
  \centering
  \begin{tikzpicture}
    \begin{axis}[spectrum style, clip=true]
      \rmshcnbars{\tcsAOSArms}{\tcsAOSBrms}{\tcsCTSBrms}
      \addplot table [x index=0,y index=1] {T7_CS_10.74_CTSB.dat};
      \addplot table [x index=0,y index=1] {T7_CS_10.74_AOSA.dat};
      \addplot table [x index=0,y index=1] {T7_CS_10.74_AOSB.dat};
    \end{axis}
  \end{tikzpicture}
  \caption{Averaged spectrum of the \ce{CS} (5--4) line at 244.936~GHz
toward comet \linear{} observed on UT 10.74 May 2004 with the
AOSA (thick blue line), AOSB (thin red line), and CTSB (black line).  The
vertical axis is the calibrated main beam brightness temperature and the
horizontal axis is the Doppler velocity in the comet rest frame.
  The rms noise level measured with respect to the zero reference level
  is shown by the blue, red, and gray bars near the origin of the figure
  for AOSA, AOSB, and CTSB, respectively.
}
  \label{fig:linear:cs}
\end{figure}
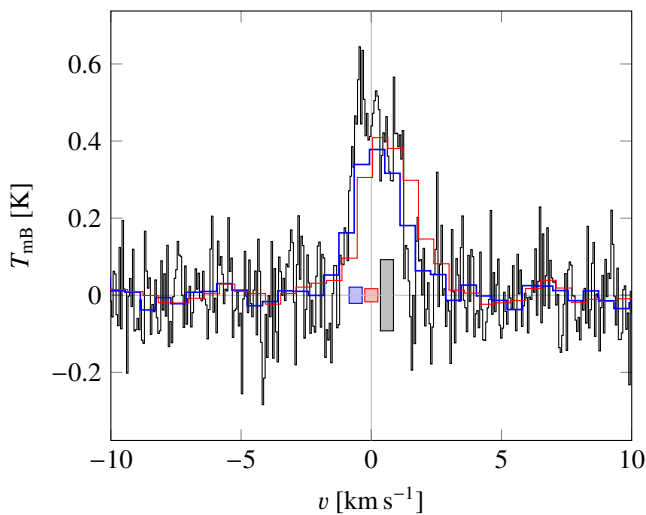

We show the same \ce{CS} transition toward comet \lin{} in
Fig.~\ref{fig:linear:cs}. A production rate of \qtcs{} \s is derived for
this object assuming that it is a parent species, that corresponds to a
relative abundance with respect to water of \qtcsqho{}.  Assuming
release from an extended coma source distribution, the CS production
rate is \qtcsd{} \s with $Q_\ce{CS}/Q_\ce{H2O}$ of \qtcsqhod{}.  Since
the beam size is considerably larger than the scale-length of the parent
species at the distance of both objects, the production rates are not
significantly modified by using a distributed source in the coma for
this species.

\subsubsection{\ce{CH3OH}}
\label{sec:ch3oh}

Methanol is a parent species that has been detected in several comets
since the initial observations of rotational transitions in comet C/1989
X1 (Austin) by \citet{1990ESASP.315..143B}.  \ce{CH3OH} has been found to
have variable production rates with a wide spread of abundances relative
to water, ranging between about 0.6\% to 6.2\% with some dependence on
heliocentric distance \citep[see][]{2004come.book..391B}.  Methanol
rotational lines often appear in multiplets at millimeter and
submillimeter wavelengths, allowing the estimation of the rotational
temperature and excitation conditions in the coma.  The averaged
\ce{CH3OH} spectra toward comets \nea{} and \lin{} measured with the
AOSA, AOSB, and CTSB backends are shown in Figs.~\ref{fig:neat:ch3oh} and
\ref{fig:linear:ch3oh}, respectively.  We derived \ce{CH3OH} production
rates of \qqchoh{} and \qtchoh{} \s{} for \nea{} and \lin{},
respectively. These values are comparable to the production rates
determined using the column densities obtained from the intercept of the
rotational diagram at $E_\mathrm{u} = 0$ as shown in
Figs.~\ref{fig:neat:rotdiagram} and \ref{fig:linear:rotdiagram}.  The
corresponding $Q_\ce{CH3OH}/Q_\ce{H2O}$ mixing ratios of \qqchohqho{}
and \qtchohqho{} in \nea{} and \lin{} represent typical and high-end
relative \ce{CH3OH} abundances in comparison to other Oort cloud
comets, respectively.

\subsubsection{HNC}

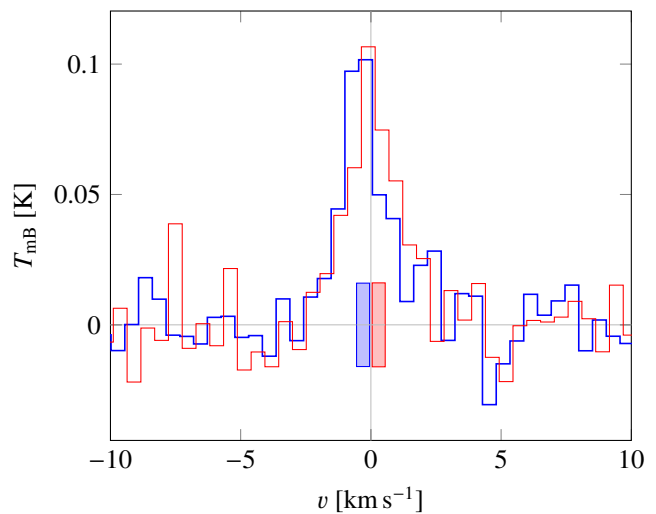
\begin{figure}
  \centering
  \begin{tikzpicture}
    \begin{axis}[spectrum style,
      cycle list name=aoslist, clip=true
      ]
      \rmshncbars{\qhncAOSArms}{\qhncAOSBrms}
      \addplot table [x index=0,y index=1] {Q4_HNC_11.95_AOSA.dat};
      \addplot table [x index=0,y index=1] {Q4_HNC_11.95_AOSB.dat};
    \end{axis}
  \end{tikzpicture}
  \caption{Averaged spectrum of the \ce{HNC} (3--2) line at 271.981~GHz
  toward comet \neat{} observed on UT 11.95 May 2004 with the
  AOSA (thick blue line) and AOSB (thin red line). The
  vertical axis is the calibrated main beam brightness temperature and
  the horizontal axis is the Doppler velocity in the comet rest frame.
  The rms noise level measured with respect to the zero reference level
  is shown by the blue and red bars near the origin of the figure
  for the AOSA and AOSB backends, respectively.
  }
  \label{fig:neat:hnc}
\end{figure}

Hydrogen isocyanide (HNC) is a metastable isomer of HCN that has been
observed in the interstellar medium. It is a relatively abundant species
in cometary atmospheres that may be produced by chemical reactions from
parent volatile species in the coma.  Detected $Q_\ce{HNC}/Q_\ce{HCN}$
mixing ratios in a sample of comets range from 0.03 to 0.3
\citep{2008ApJ...675..931L}.  Since we do not have precise estimates on
the distribution of HNC in comets \nea{} and \lin{}, we assumed the
Haser formula for a parent molecule to compute the production rates.

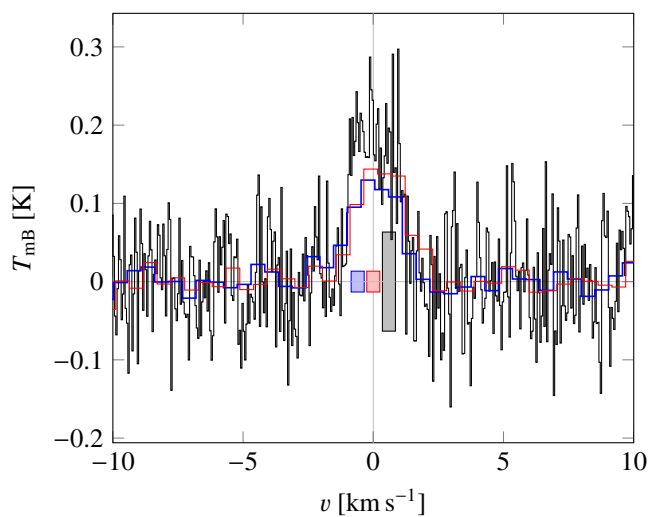
\begin{figure}
  \centering
  \begin{tikzpicture}
    \begin{axis}[spectrum style, clip=true]
      \rmshcnbars{\thncAOSArms}{\thncAOSBrms}{\thncCTSBrms}
      \addplot table [x index=0,y index=1] {T7_HNC_11.69_CTSB.dat};
      \addplot table [x index=0,y index=1] {T7_HNC_11.69_AOSA.dat};
      \addplot table [x index=0,y index=1] {T7_HNC_11.69_AOSB.dat};
    \end{axis}
  \end{tikzpicture}
  \caption{Averaged spectrum of the \ce{HNC} (3--2) line at 271.981~GHz
toward comet \linear{} observed on 11.69 UT May 2004 with the AOSA
(thick blue line), AOSB (thin red line), and CTSB (black line).  The
vertical axis is the calibrated main beam brightness temperature and the
horizontal axis is the Doppler velocity in the comet rest frame.
  The rms noise level measured with respect to the zero reference level
  is shown by the blue, red, and gray bars close to the origin of the
  figure for AOSA, AOSB, and CTSB, respectively.
}
  \label{fig:linear:hnc}
\end{figure}

Figure~\ref{fig:neat:hnc} shows the \ce{HNC} rotational transition
(3--2) in comet \nea{} detected with the AOSA and AOSB spectrometers.
This transition was not detected in the CTSB data due to an apparent
problem in the patching of the receiver.  In Fig.~\ref{fig:linear:hnc},
we show the same \ce{HNC} transition observed toward \lin{} with the
AOSA, AOSB, and CTSB. Production rates of $Q_\ce{HNC}$ = \qqhnc{} and
\qthnc{} \s{} were estimated for \nea{} and \lin{}, corresponding to
mixing ratios $Q_\ce{HNC}/Q_\ce{HCN}$ of \qqhncqhcn{} and \qthncqhcn{}
and $Q_\ce{HNC}/Q_\ce{H2O}$ of \qqhncqho{} and \qthncqho{},
respectively.  These values are intermediate in comparison to the
$Q_\ce{HNC}/Q_\ce{HCN}$ mixing ratios of 0.02--0.29 observed in
other Oort cloud comets \citep{2008ApJ...675..931L,2011IAUS..280..261B},
with a higher relative abundance in \lin.

\subsection{Comparison to other comets}

\begin{table*}
  \caption{Comparison of production rates of detected species relative
  to HCN and \ce{H2O} in comet \neat{} with 1-$\sigma$ statistical
  uncertainties.}
  \label{tbl:neat:mixing}
  \centering
  \begin{tabular}{c
      S[table-format = 1.2(2)e+2,
	table-comparator = true]
      S[table-format = 1.2(2)e+1,
	table-comparator = true]
      S[table-format = 1.2(2)e+1,
	table-comparator = true]
      S[table-format = 1.1(2)e+1,
	table-comparator = true]
      c}
    \toprule
    Molecule & {$Q$\tablefootmark{\emph{a}}} & {$Q/Q_\ce{HCN}$} &
    {$Q/Q_\ce{H2O}$\tablefootmark{\emph{b}}} &
    {$Q/Q_\ce{H2O}$\tablefootmark{\emph{c}}} \\
    & {(\s)} \\
    \midrule
    HCN&2.96\pm0.05e26&1&1.12\pm0.02e-3&7.4\pm3.9e-4&\\
\ce{H2CO}&3.43\pm0.79e26&1.16\pm0.27e0&1.30\pm0.30e-3&1.0e-3&\\
\ce{CO}&9.0\pm2.4e27&3.05\pm0.83e1&3.45\pm0.93e-2&&\\
\ce{CS}&2.08\pm0.08e26&7.02\pm0.30e-1&8.12\pm0.31e-4&<1.4e-3&\\
\ce{CH3OH}&4.45\pm0.72e27&1.50\pm0.25e1&1.74\pm0.28e-2&2.4\pm1.4e-2&\\
\ce{HNC}&1.70\pm0.21e25&5.75\pm0.73e-2&6.72\pm0.85e-5&&\\

    \bottomrule
  \end{tabular}
  \tablefoot{
  \tablefoottext{\emph{a}}{Production rates measured at the SMT on 7--12
    May 2004 derived from an excitation and radiative transfer model
    assuming a pointing offset of 2\arcsec{}.}
  \tablefoottext{\emph{b}}{A water production rate of $Q_\ce{H2O} =
    2.7 \times 10^{29}\ \s$ was obtained from the fit to the daily
    averaged production rates measured by the {\it Odin} satellite
    \citep{2009A&A...501..359B}.}
  \tablefoottext{\emph{c}}{Mixing ratios obtained between 20--24 May
    2004 with the BIMA array and on 15 and 26 May 2004 with the 12-m ARO
    telescope for a parent source distribution
    \citep{2005ApJ...630..623F,2006ApJ...643..567R,2006ApJ...649.1169M}.
    }
  }
\end{table*}

From the recent observations of volatile abundances during the {\it Deep
Impact} encounter with Tempel 1, the composition of the nucleus ices to
depths of tens of meters is found to be essentially the same as in the
fraction released as gas from the surface \citep{2008EP&S...60...61A}.
Thus, cometary nuclei are regarded to be well mixed and the outgassing
abundances to represent the bulk composition of the nucleus, although
strong inhomogeneities have been found in some comets \citep[e.g.
103P/Hartley2,][]{2012ApJ...756...80D}.  The cometary diversity has been
studied by narrowband photometric observations in the visible that
suggest distinct classes, according to the carbon abundance
\citep{1995Icar..118..223A}.  However, no evident correlations have been
found between the relative abundances and the dynamical class for a
larger sample of comets that have been observed at submillimeter and
radio wavelengths with the James Clerk Maxwell Telescope (JCMT), the
Caltech Submillimeter Observatory (CSO), and the IRAM telescope
\citep[see][and references
therein]{2002EM&P...90..323B,2009EM&P..105..267C}.

Tables~\ref{tbl:neat:mixing} and \ref{tbl:linear:mixing} show a summary
of the mixing ratios of the detected volatiles in \nea{} and \lin{} with
respect to \ce{HCN} and \ce{H2O} with statistical uncertainties
(1-$\sigma$ rms noise from the integrated intensities) and a comparison
to other measurements using various facilities.  The production rate
values and the inferred mixing ratios depend on the model input
parameters and collisional excitation rates, as well as on the radiative
transfer model that is used to calculate the level populations and line
emission.  The uncertainties in the production rates can amount up to
50\% due to systematic errors in the adopted model, and up to a factor
of 2 due the uncertainties in the collisional excitation rates
\citep[see][for a detailed assesment of the systematic
errors]{2009AJ....137.4837H}.  To compute the mixing ratios, the water
production in \nea\ was obtained from the Gaussian fit to the daily
averaged production rates in \citet{2009A&A...501..359B}.  We adopt a
water production rate of $Q_\ce{H2O} = 3 \times 10^{29}\ \s$ for comet
\lin{} at the time of our observations.  Several species including HCN,
\ce{H2CO}, \ce{CH3OH}, CS, and HNC were observed in \nea{} and \lin{}
with the BIMA array near Hat Creek, California and the 12-m ARO
telescope on Kitt Peak, Arizona
\citep{2005ApJ...630..623F,2006ApJ...643..567R,2006ApJ...649.1169M,2008ApJ...689..613R}.
Their derived mixing ratios with quoted errors or upper limits are shown
in the rightmost columns in Tables~\ref{tbl:neat:mixing} and
\ref{tbl:linear:mixing}.

\begin{figure}
  \begin{tikzpicture}
    \begin{loglogaxis}[
	axis equal=true, grid=both,
	legend pos=north west,
	cycle list name=exotic,
	xlabel={$Q/Q_\ce{HCN}$ in \nea{}},
	ylabel={$Q/Q_\ce{HCN}$ in \lin{}},
	xmax=200,
	]
      \addplot+[comp style]
      coordinates { (\qcohcn, \tcohcn) +- (\qcohcne, 0.) };
      \draw[->,color=teal!80!black] (axis cs:\qcohcn, \tcohcn) --
			    (axis cs:\qcohcn, 26.);
      \addplot+[comp style]
      coordinates { (\qchohhcn, \tchohhcn) +- (\qchohhcne, \tchohhcne) };
      \addplot+[mark=triangle*,comp style]
      coordinates { (\qhcohcn, \thcohcn) +- (\qhcohcne, \thcohcne) };
      \addplot+[comp style]
      coordinates { (\qcshcn, \tcshcn) +- (\qcshcne, \tcshcne) };
      \addplot+[comp style]
      coordinates { (\qhnchcn, \thnchcn) +- (\qhnchcne, \thnchcne) };
      \legend{CO, \ce{CH3OH}, \ce{H2CO}, CS, HNC}
      \addplot[dashed,domain=0.04:3e2,update limits=false] {x};
      \fill[teal!80!black,path fading=fade out]
	(axis cs:\COlow, \COlow) rectangle (axis cs:\COup, \COup);
      \fill[orange!80!black,path fading=fade out]
	(axis cs:\CHOHlow, \CHOHlow) rectangle (axis cs:\CHOHup, \CHOHup);
      \fill[cyan!80!black,path fading=fade out]
	(axis cs:\HCOlow, \HCOlow) rectangle (axis cs:\HCOup, \HCOup);
      \fill[red!70!white,path fading=fade out]
	(axis cs:\CSlow, \CSlow) rectangle (axis cs:\CSup, \CSup);
      \fill[lime,path fading=fade out]
	(axis cs:\HNClow, \HNClow) rectangle (axis cs:\HNCup, \HNCup);
    \end{loglogaxis}
  \end{tikzpicture}
  \caption{Comparison of the CO, \ce{CH3OH}, \ce{H2CO}, CS, and HNC
    mixing ratios with respect to \ce{HCN} derived from the SMT
    observations assuming direct release from the nucleus in comets
    \neat{} and \linear{} plotted in logarithmic scale. Error bars are
    1-$\sigma$ statistical uncertainties. The shaded areas indicate
    typical values of the relative $Q$/$Q_\ce{HCN}$ abundances measured
    in a sample of comets \citep{2009EM&P..105..267C}.}
  \label{fig:comparison}
\end{figure}
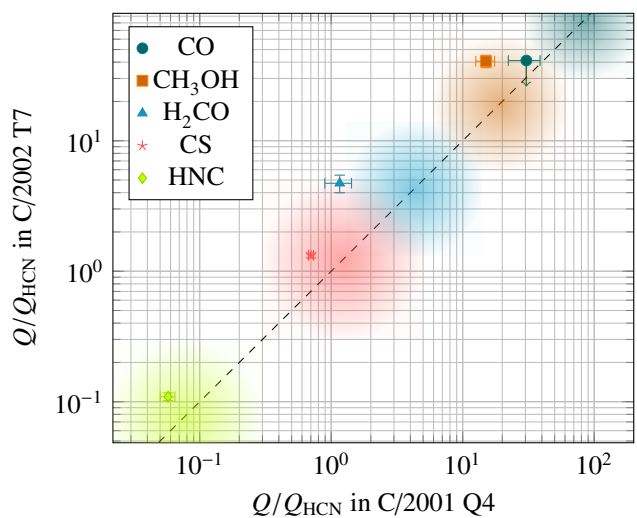

\begin{table*}
  \caption{Comparison of production rates of detected species relative
  to HCN and \ce{H2O} in comet \linear{} with 1-$\sigma$ statistical
  uncertainties.}
  \label{tbl:linear:mixing}
  \centering
  \begin{tabular}{c
      S[table-format = 1.2(2)e+2,
	table-comparator = true]
      S[table-format = 1.2(2)e+1,
	table-comparator = true]
      S[table-format = 1.2(2)e+1,
	table-comparator = true]
      S[table-format = 1.1(2)e+1,
	table-comparator = true]
      S[table-format = 1e+1,
	table-number-alignment = center]
      }
    \toprule
    Molecule & {$Q$\tablefootmark{\emph{a}}} & {$Q/Q_\ce{HCN}$} &
    {$Q/Q_\ce{H2O}$\tablefootmark{\emph{b}}} &
    {$Q/Q_\ce{H2O}$\tablefootmark{\emph{c}}} &
    {$Q/Q_\ce{H2O}$\tablefootmark{\emph{d}}} \\
    & {(\s)} \\
    \midrule
    HCN&4.54\pm0.10e26&1&1.51\pm0.03e-3&3.3\pm1.1e-3&1e-3\\
\ce{H2CO}&2.14\pm0.33e27&4.72\pm0.73e0&7.1\pm1.1e-3&1.3e-3&\\
\ce{CO}&<1.87e28&<4.12e1&<6.23e-2&&\\
\ce{CS}&6.01\pm0.23e26&1.32\pm0.06e0&2.00\pm0.08e-3&7.5\pm2.5e-3&\\
\ce{CH3OH}&1.85\pm0.20e28&4.07\pm0.44e1&6.16\pm0.65e-2&3.8\pm0.8e-2&7e-3\\
\ce{HNC}&4.96\pm0.32e25&1.09\pm0.08e-1&1.65\pm0.11e-4&<7.5e-5&\\

    \bottomrule
  \end{tabular}
  \tablefoot{
  \tablefoottext{\emph{a}}{Weighted mean of production rates measured at the
    SMT on 8--12 May 2004 derived from an excitation and radiative
    transfer model assuming a pointing offset of 2\arcsec{}.}
  \tablefoottext{\emph{b}}{A water production rate of $Q_\ce{H2O} =
    3 \times 10^{29}\ \s$ was adopted to calculate the abundances, close
    to the measurement by the {\it Odin} satellite on UT 25.9 May 2004
    \citep{2007P&SS...55.1058B}.}
  \tablefoottext{\emph{c}}{Mixing ratios obtained between 11 and 15 May
    2004 with the BIMA array and on 15 and 25 May 2004 with the 12-m ARO
    telescope for a parent source distribution
    \citep{2005ApJ...630..623F,2006ApJ...643..567R,2006ApJ...649.1169M}.}
  \tablefoottext{\emph{d}}{Mixing ratios obtained in the period 11--15 May
    at the OVRO and BIMA arrays \citep{2009AJ....137.4837H}.}
  }
\end{table*}

The observed mixing ratios in \nea\ and \lin\ are mostly in the usual
range for dynamically new comets and indicate that both objects are
chemically similar, although relative abundances in \nea{} are
systematically lower.  Figure~\ref{fig:comparison} shows a comparison of
the mixing ratios in both objects normalized to HCN that are listed in
Tables~\ref{tbl:neat:mixing} and \ref{tbl:linear:mixing} with the range
of measurements from other comets \citep{2009EM&P..105..267C}.  The
$Q/Q_\ce{HCN}$ ratio is more reliable than the relative abundance
relative to \ce{H2O}, since HCN was detected in our observations at
around the same time as the other species using approximately the same
beam size and all the data were analyzed homogeneously.  The same
molecular species (namely \volatiles{}) were observed in the dynamically
old comet \machholz{} in January 2005 at the SMT
\citep{2012A&A...545A...2D}. The mixing ratios relative to HNC in
\machholz{} are lower than the abundances in \lin{} and generally lie
closer to the \nea{} values, except for the HNC abundance.  Comet
\machholz{} seems to be remarkably rich in HNC and its
$Q_\ce{HNC}/Q_\ce{HCN}$ ratio is about a factor of three greater than the
values derived for \nea{} and \lin{}.

Both \nea{} and \lin{} have been widely considered in the literature as
Oort cloud comets that visit the inner Solar System for the first time
based on their orbital parameters.  However, according to recent
dynamical models that include the effects of non-gravitational
acceleration on the orbital elements, \lin{} is found to be a
dynamically new comet with its previous perihelion passage at a distance
larger than 400 AU from the Sun, while \nea{} has visited the inner part
of the Solar System during a past apparition at around 6--7 AU
heliocentric distance and is therefore dynamically old
\citep{2010MNRAS.404.1886K,2012A&A...544A.119K}. We observe higher
relative abundances in \lin{} than in \nea{}, except for CO, which is
not detected in \lin{}; the upper limit agrees with this object being
largely depleted in CO.  \ce{H2CO}, CS, and CO relative abundances in
comet \nea{} lie in the lower end of the observed abundances in other
comets, while HNC and \ce{CH3OH} have typical values.  This relative
depletion is consistent with dynamical simulations that indicate that
\nea{} have visited the inner part of the Solar System and may have been
thermally processed.

\subsection{Short-term periodic variations in the HCN production rate}
\label{sec:variability}

In addition to a seasonal variation with a peak emission close to
perihelion, periodic variations in the outgassing activity of comets
have been associated with the rotation of the nucleus.  Using various
techniques, the rotation periods of cometary nuclei have been determined
by measuring this periodic variability of the outgassing \citep[see][for
a comprehensive review]{2004come.book..281S}.  Rotation periods have
been computed in previous works from the periodic variability of the HCN
production rate in comets 9P/Tempel 1 \citep{2007Icar..187..253B},
73P-C/Schwassmann-Wachmann 3 \citep{2010A&A...510A..55D}, 2P/Encke
\citep{2011Icar..215..153J} and 103P/Hartley 2
\citep{2011ApJ...734L...4D}.

\subsubsection{Comet \neat{}}

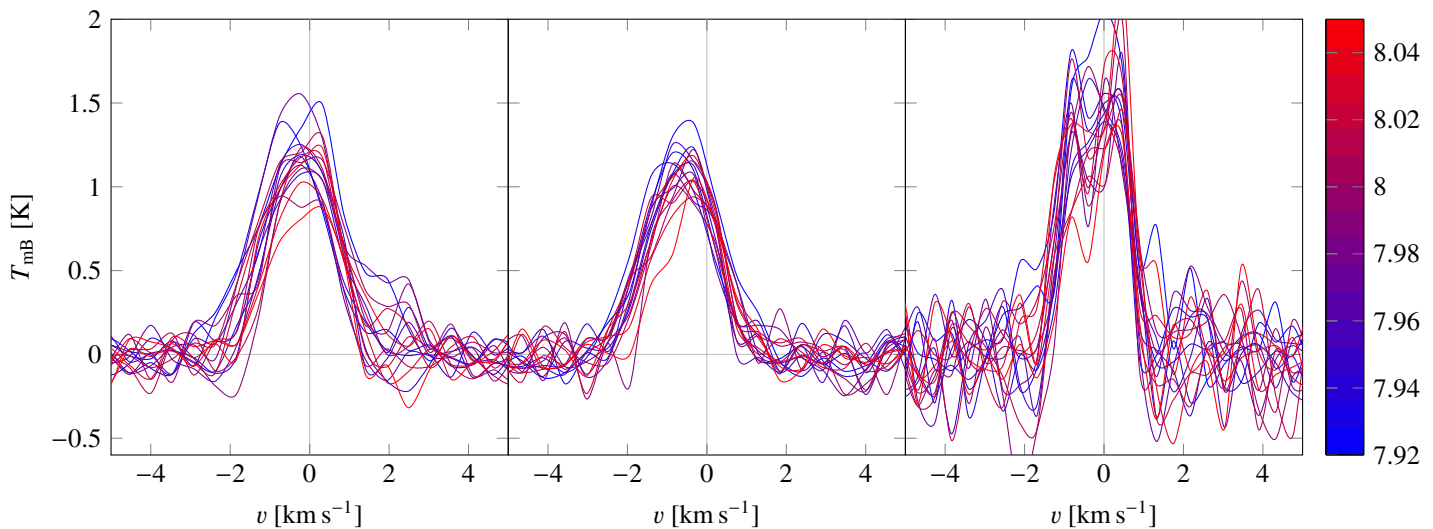
\begin{figure*}
  \centering
  \pgfplotsset{stack style/.style=
    {width=.37\hsize, height=.4\hsize, ymin=-0.6, ymax=2.,
    xmin=-5, xmax=5,
    extra x ticks       = 0,
    extra y ticks       = 0,
    extra x tick labels = ,
    extra y tick labels = ,
    extra tick style  = { grid = major },
    }}
  \begin{tikzpicture}
    \begin{axis}[
      stack style,
      name=aosa,
      smooth,
      no markers,
      xlabel={$v$ [\kms]},
      ylabel={$T_\mathrm{mB}$ [\si{\kelvin}]},
      ]
      \pgfplotsinvokeforeach{02, 03, 04, 05, 06, 07, 08, 09, 11,
      12, 13, 14, 15, 16, 17, 18, 19} {
	\addplot [
	  execute at begin plot visualization={%
	    \pgfplotscolormapdefinemappedcolor{\numexpr#1*58-110\relax}
	  },
	  mapped color,
	]
	table [x index=0,y index=1] {Q4_AOSA_spec_#1.txt};
      }
    \end{axis}
    \begin{axis}[
      stack style,
      name=aosb,
      at={(aosa.east)},anchor=west,
      smooth,
      no markers,
      xlabel={$v$ [\kms]},
      yticklabel=\empty,
      ]
      \pgfplotsinvokeforeach{02, 03, 04, 05, 06, 07, 08, 09, 11,
      12, 13, 14, 15, 16, 17, 18, 19} {
	\addplot [
	  execute at begin plot visualization={%
	    \pgfplotscolormapdefinemappedcolor{\numexpr#1*58-110\relax}
	  },
	  mapped color,
	]
	table [x index=0,y index=1] {Q4_spec_#1.txt};
      }
    \end{axis}
    \begin{axis}[
      colorbar sampled ,
      colorbar style={samples=17,
      point meta min=7.92, point meta max=8.05},
      stack style,
      name=ctsb,
      at={(aosb.east)},anchor=west,
      smooth,
      no markers,
      xlabel={$v$ [\kms]},
      yticklabel=\empty,
      ]
      \pgfplotsinvokeforeach{02, 03, 04, 05, 06, 07, 08, 09, 11,
      12, 13, 14, 15, 16, 17, 18, 19} {
	\addplot [
	  execute at begin plot visualization={%
	    \pgfplotscolormapdefinemappedcolor{\numexpr#1*58-110\relax}
	  },
	  mapped color,
	]
	table [x index=0,y index=1] {Q4_CTSB_spec_#1.txt};
      }
    \end{axis}
  \end{tikzpicture}
  \caption{Variability of the \ce{HCN} (3--2) line over time
  in comet \neat{} obtained by the AOSA (\emph{left panel}), AOSB
  (\emph{middle panel}), and CTSB (\emph{right panel}) spectrometers from UT
  7.92--8.05 May 2004.  All the spectra have been interpolated smoothly
  between data points for clarity. The CTSB spectra have been resampled
  to a 380~kHz resolution per channel with a rectangular window
  function.  Each individual spectrum represents a 6-minute scan
  excluding the spectra, where the system temperature of the receiver
  deviates more that 1-$\sigma$ from the mean system temperature. The
  sampled color scale corresponds to the time of the observation and
  runs from blue to red.
  }
  \label{fig:neat:variability}
\end{figure*}

\begin{table}
  \caption{Line variability of the HCN (3--2) transition in \neat{}.}
  \centering
  \label{tbl:neat:variability}
  \begin{tabular}{c c
		S[table-number-alignment = center,
		table-format = 1.2(2)]
		S[table-number-alignment = center,
		explicit-sign = +,
		table-format = +1.1(2)e+1]
		c}
    \toprule
    Date & Backend & {$\int T_\textrm{mB}\, dv$} &
    {$\Delta v$} & {$T_\textrm{sys}$} \\
    (UT) & & {(\Kkms)} & {(\ms)} & {(\si{\kelvin})} \\
    \midrule
    7.910&AOSA&3.98\pm0.29e0&-7.3\pm8.8e1&2682\\
7.910&AOSB&3.24\pm0.22e0&-4.7\pm0.8e2&2672\\
7.910&CTSB&3.93\pm0.70e0&-1.6\pm2.1e2&2763\\
7.915&AOSA&4.95\pm0.71e0&-4.6\pm1.8e2&2952\\
7.915&AOSB&4.99\pm0.30e0&-6.4\pm0.8e2&2929\\
7.915&CTSB&5.89\pm0.64e0&-2.6\pm1.3e2&3066\\
7.921&AOSA&4.71\pm0.14e0&-3.7\pm3.5e1&1121\\
7.921&AOSB&4.67\pm0.10e0&-3.9\pm0.3e2&1115\\
7.921&CTSB&5.32\pm0.30e0&-6.6\pm6.5e1&1157\\

    \bottomrule
  \end{tabular}
  \tablefoot{ This table is available in machine-readable format at the
    CDS. Part of the table is shown here for guidance regarding its format.
  }
\end{table}

Figure~\ref{fig:neat:variability} shows the change over time of the line
intensity and velocity in the HCN (3--2) spectra in \nea{} for each of
the 6-minute scans obtained with the AOSA, AOSB, and CTSB.  The line
intensity and velocity shift results are summarized in
Table~\ref{tbl:neat:variability}.  The emission line is detected in each
single scan with a median S/N ratio of $\sim 12$ in all the
spectrometers.  The line profiles are asymmetric with a larger variation
in the blueshifted wing, which is larger than the rms noise in the
emission-free part of the spectrum. There is a decrease in the emission
toward the end of the observing period, which is visible in the spectral
channels corresponding to the blueshifted wing of the AOSA and AOSB
spectra (left and middle panels in Fig.~\ref{fig:neat:variability}).
Since the phase angle is about 90\degr\ at the time of the observations,
this variation can be interpreted as the modulation of an active region
in the surface that initially is outgassing roughly toward the observer
in the direction along the line of sight, and the source becomes less
irradiated by the Sun during the nucleus rotation cycle.

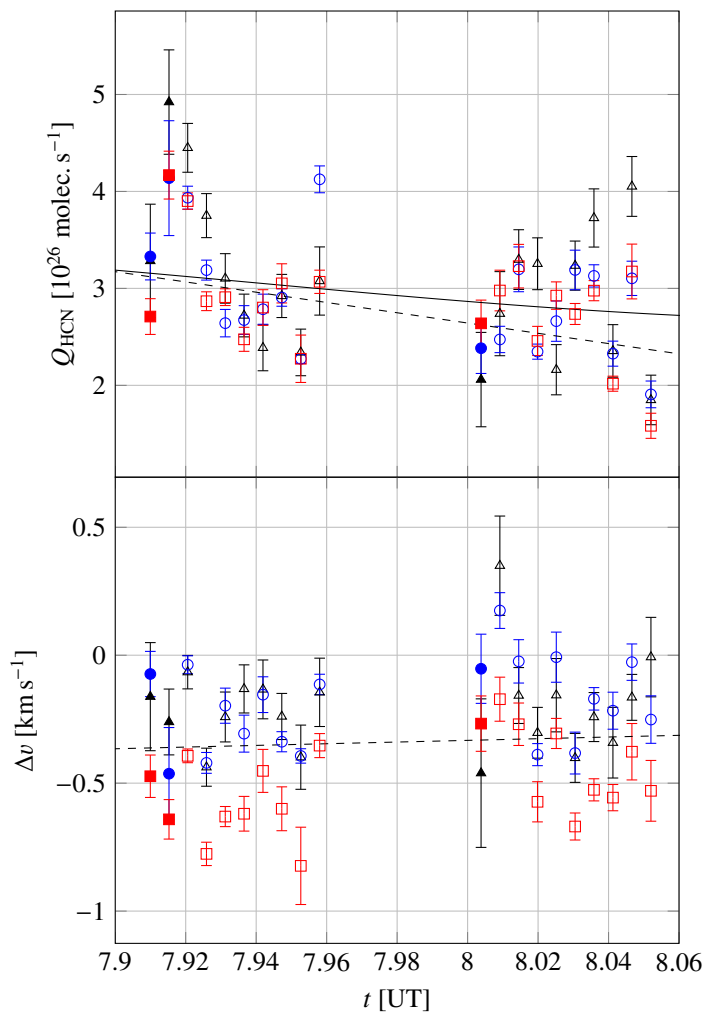
\begin{figure}
  \centering
  \begin{tikzpicture}
    \begin{axis}[
      name=prodrates,
      grid=major,
      cycle list name=rotdiaglist,
      ylabel={$Q_\ce{HCN}$ [\num{e26} \s]},
      xticklabel=\empty,
      width=\hsize, xmin=7.9, xmax=8.06
      ]
      \addplot+ table [x index=0,y index=1,y error index=2] {Q4_CTSB_good.txt};
      \addplot+ table [x index=0,y index=1,y error index=2] {Q4_AOSA_good.txt};
      \addplot+ table [x index=0,y index=1,y error index=2] {Q4_AOSB_good.txt};
      \addplot+ table [x index=0,y index=1,y error index=2] {Q4_CTSB_bad.txt};
      \addplot+ table [x index=0,y index=1,y error index=2] {Q4_AOSA_bad.txt};
      \addplot+ table [x index=0,y index=1,y error index=2] {Q4_AOSB_bad.txt};
      \addplot[mark=none] table [x index=0,y index=1] {Q4_biver.txt};
      \addplot[mark=none, dashed] table [x index=0,y index=1] {Q4_..S[AB]_linear.txt};
    \end{axis}
    \begin{axis}[
      at={(prodrates.south)},anchor=north,
      grid=major,
      cycle list name=rotdiaglist,
      xlabel={$t$ [UT]},
      ylabel={$\Delta v$ [\kms]},
      width=\hsize,
      xmin=7.9, xmax=8.06
      ]
      \addplot+ table [x index=0,y index=1,y error index=2] {vel_Q4_CTSB_good.txt};
      \addplot+ table [x index=0,y index=1,y error index=2] {vel_Q4_AOSA_good.txt};
      \addplot+ table [x index=0,y index=1,y error index=2] {vel_Q4_AOSB_good.txt};
      \addplot+ table [x index=0,y index=1,y error index=2] {vel_Q4_CTSB_bad.txt};
      \addplot+ table [x index=0,y index=1,y error index=2] {vel_Q4_AOSA_bad.txt};
      \addplot+ table [x index=0,y index=1,y error index=2] {vel_Q4_AOSB_bad.txt};
      \addplot[mark=none, dashed] table [x index=0,y index=1]
	{Q4_dv_..S[AB]_linear.txt};
    \end{axis}
  \end{tikzpicture}
  \caption{\ce{HCN} production rates (\emph{upper panel}) and line velocity
  offsets (\emph{lower panel}) as a function of time in comet \neat{} derived
  from the AOSA (blue circles), AOSB (red squares), and CTSB (black
  triangles) spectrometers which include 1-$\sigma$ uncertainties. The unit
  of time in the horizontal axis is UT decimal days in May 2004.  The
  time resolution is roughly 6 min with a small overhead from the
  telescope movement (integration of 3 min at the on-source and
  off-source positions in the position-switching observing mode with a
  relative azimuth offset of 0\fdg5 between both positions using 15 s
  samples).  Filled symbols indicate spectra that were obtained with a
  system temperature that was more than 1-$\sigma$ away from the mean
  system temperature during the adjacent scans.  In the \emph{upper panel}, the
  solid line shows the best fit to the \ce{H2O} apparent outgassing
  rates with a period of 0.8139 days obtained by fitting a sine function
  with one harmonic \citep{2009A&A...501..359B} scaled by the mean
  $Q_\ce{HCN}/Q_\ce{H2O}$ during our observations.  The dashed lines
  show linear fits to the data from the three spectrometers using a
  weighted linear least-squares method.
  }
  \label{fig:neat:qhcn}
\end{figure}

Figure~\ref{fig:neat:qhcn} shows the time variability of the HCN
production rate and line velocity offset.  The apparent \ce{H2O}
production rate is described by fitting a sine function with a period of
0.8139 days and one harmonic with half of the period
\citep{2009A&A...501..359B}:
\begin{equation}
  \begin{split}
    Q_\ce{H2O} = \num{2.709e29} + \num{4.45e28} \times \sin \left( 2\pi
    \frac{t/\mathrm{days}+15.7898}{0.8139}\right)\\ + \num{1.46e28}\sin
    \left( \pi \frac{t/\mathrm{days}+16.105}{0.8139}\right) \ \s,
  \end{split}
\end{equation}
where $t$ is the UT time in decimal days relative to perihelion. We have
scaled this production rate by the mean $Q_\ce{HCN}/Q_\ce{H2O}$ ratio
during the observing period (solid line in the upper panel of
Fig.~\ref{fig:neat:qhcn}). The weighted linear fits to the line-intensity
and velocity-shift data are shown by the dashed lines that exclude scans 
with a system temperature that deviates more than 1-$\sigma$ from
neighboring scans. These are indicated by solid symbols in
Fig.~\ref{fig:neat:qhcn}.  Our HCN observations over four hours are
consistent with the derived $\sim$ 0.816 day (19.58 h) rotation period
and are in phase with the fit to the apparent \ce{H2O} production rate
calculated by \citet{2009A&A...501..359B}.  However, we are not able to
calculate the rotation period from the variability of the HCN production
curve in comet \nea{} due to the insufficient temporal coverage of our
observations. This variation could also be possibly introduced by
instrument fluctuations or changes in the main beam efficiency.

The HCN line velocity shift measurements with the three backends range
roughly between 0 to -0.5 \kms with the AOSB velocity offsets lying
closer to the lower value (see Fig.~\ref{fig:neat:qhcn}).  Given the
observing geometry with a solar phase angle $\sim$ 90\degr, blueshifted
lines are expected with outgassing toward the observer if an active area
on the nucleus surface is irradiated by the Sun.  The small velocity
offset variation of about 15\% toward zero values, according to the
weighted linear fit shown by the dashed line in
Fig.~\ref{fig:neat:qhcn}, is consistent with a jet that is rotating away
from the direction toward the telescope during our observations.  This
effect could also explain the simultaneous decrease in line intensity as
the total HCN column density within the beam drops off when the
preferential outflow direction moves aside from the line-of-sight.

\subsubsection{Comet \linear{}}

\begin{table}
  \caption{Line variability of the HCN (3--2) transition in \linear{}.}
  \centering
  \label{tbl:linear:variability}
  \begin{tabular}{c c
		S[table-number-alignment = center,
		table-format = 1.2(2)]
		S[table-number-alignment = center,
		explicit-sign = +,
		table-format = +1.1(2)e+1]
		c}
    \toprule
    Date & Backend & {$\int T_\textrm{mB}\, dv$} &
    {$\Delta v$} & {$T_\textrm{sys}$} \\
    (UT) & & {(\Kkms)} & {(\ms)} & {(\si{\kelvin})} \\
    \midrule
    8.623&AOSA&2.79\pm0.23e0&-2.5\pm1.0e2&1132\\
8.623&AOSB&2.62\pm0.19e0&-4.8\pm0.9e2&1124\\
8.623&CTSB&2.68\pm0.40e0&-2.8\pm1.7e2&1160\\
8.626&AOSA&2.76\pm0.14e0&-3.3\pm0.7e2&1115\\
8.626&AOSB&2.36\pm0.16e0&-9.6\pm1.1e2&1115\\
8.626&CTSB&4.57\pm0.34e0&-2.5\pm0.9e2&1147\\
8.630&AOSA&3.02\pm0.11e0&-2.3\pm0.4e2&1100\\
8.630&AOSB&2.91\pm0.08e0&-4.4\pm0.3e2&1095\\
8.630&CTSB&3.12\pm0.41e0&-4.4\pm1.7e2&1129\\

    \bottomrule
  \end{tabular}
  \tablefoot{ This table is available in machine-readable format at the
    CDS. Part of the table is shown here for guidance regarding its format.
  }
\end{table}

The HCN production rate derived from the $J$ = 2--1 transition varied
over the 4 days of observations of comet \lin{} between 8 and 12 May 2004.
Figure~\ref{fig:linear:qhcn} shows the variability of the HCN outgassing
rate and line velocity shift for each individual 4-minute scan obtained
with the AOSA, AOSB, and CTSB spectrometers as a function of time. The
individual intensity and velocity measurements are reported in
Table~\ref{tbl:linear:variability}. The median S/N ratio of the scans
obtained with the three spectrometers during the observing period is 10.
There is an increase of about 20\% in the HCN production rate of
\qthcnc{} \s{} measured around UT 11.60 May when compared to the previous
measurements on UT 8.68 and 9.70 May of \qthcna{} and \qthcnb{} \s{},
respectively. There is also considerably less scatter in the
observations from UT 11.60 May, as shown in Fig.~\ref{fig:linear:qhcn}.
However, our observations did not achieve complete phase coverage, and
it is not possible to determine the rotation period of the nucleus from
this dataset.  It is possible that this brightness variability is due to
instrumental effects.  Small outbursts or activity changes are also
expected as comets approach their perihelion passage, which could
explain partly the observed fluctuations in the HCN production rate
\citep[see e.g.][]{2005Sci...310..258A}.

Line velocity offsets with respect to the comet rest frame in \lin{} are
shown as a function of time in the lower panel of
Fig.~\ref{fig:linear:qhcn} for AOSA, AOSB and CTSB.  Overall, the values
are negative within the uncertainties and the line velocity position is
shifted toward the blue in the AOSB spectra when compared to the other
two backends.  The magnitude of the difference in the average velocity
offset values between the AOSA and AOSB is between 0.2--0.3 \kms{} at
various times.  It is likely that this discrepancy is caused by the
frequency scale in this range of the AOS backends being slightly
inaccurate.  The correspondence between channel number and frequency is
not perfectly linear for acousto-optical spectrometers, and the
correction may be significant and not well constrained in certain parts
of the band. The average velocity offset obtained with the CTSB lies
within the AOSA and AOSB measurements, generally being closer to the
former of the two.

There are significant fluctuations in the line offsets but no clear
correlation with the variability in the HCN production rate is observed.
The line positions are measured with higher accuracy on UT 11.60 May,
corresponding with a mean value of -0.2 \kms{} as derived from the three
spectrometers. Considering a phase angle of $\sim$ 118\degr{}
at the time of the observations, negative line velocities suggest that
there is preferential outgassing in the plane perpendicular to the Sun
direction or roughly in the anti-Sunward direction.

\pgfplotsset{
  cycle list name=rotdiaglist,
  height=.4\hsize,
  xlabel style={xshift=4cm}
}
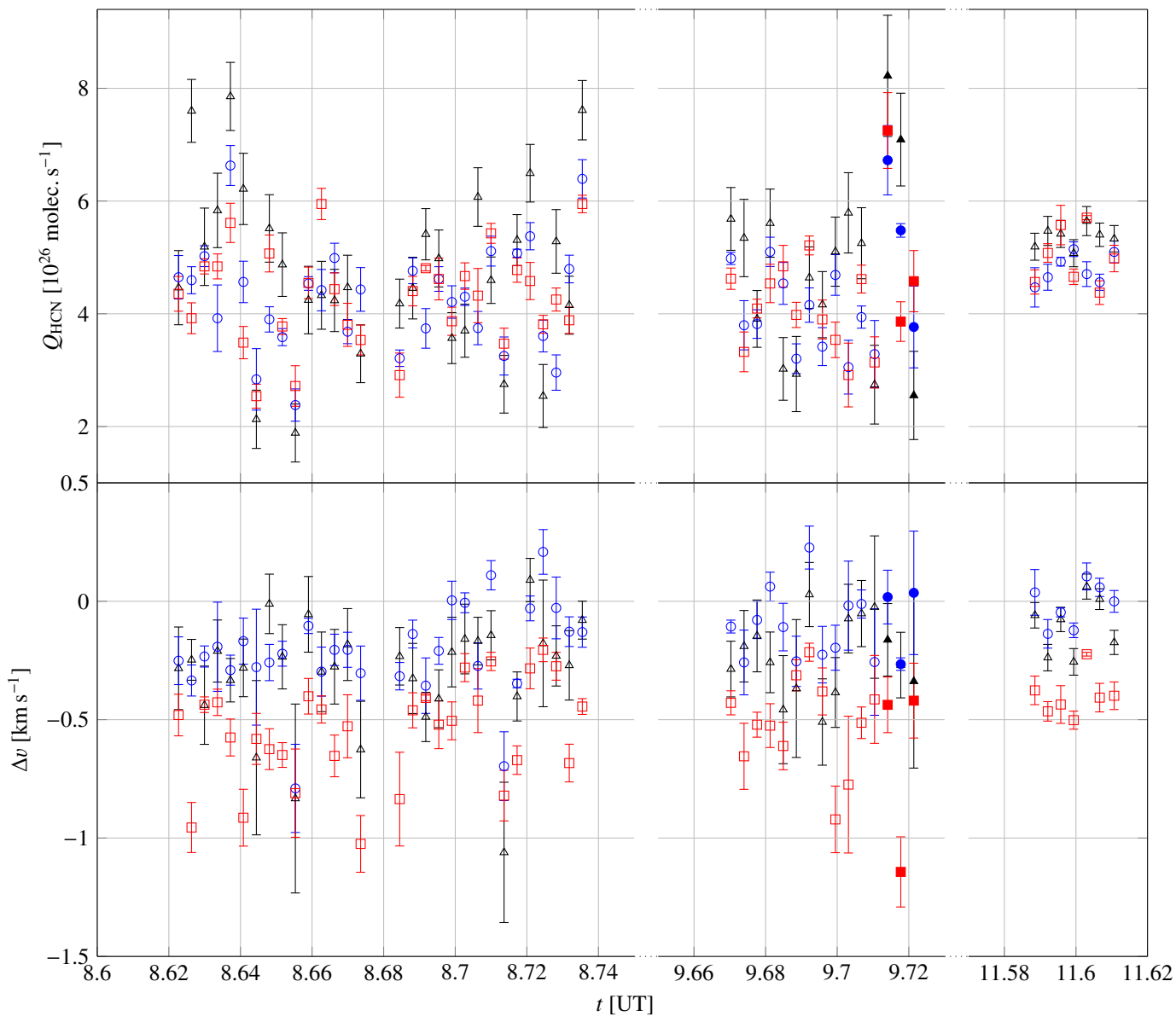
\begin{figure*}
  \centering
  \begin{tikzpicture}
    \begin{axis}[name=a,
	    grid=major,
	    ymin=1, ymax=9.4,
	    xmin=8.6,xmax=8.75,
	    axis y line*=left,
	    width=.45\hsize,
	    scale only axis,
	    xticklabel=\empty,
	    ylabel={$Q_\ce{HCN}$ [\num{e26} \s]},
	    ]
      \addplot+ table [x index=0,y index=1,y error index=2] {T7_CTSB_good.txt};
      \addplot+ table [x index=0,y index=1,y error index=2] {T7_AOSA_good.txt};
      \addplot+ table [x index=0,y index=1,y error index=2] {T7_AOSB_good.txt};
    \end{axis}
    \begin{axis}[name=b,
	    grid=major,
	    xshift=.47\hsize,
	    xticklabel=\empty,
	    xmin=9.65,xmax=9.73,
	    ymin=1, ymax=9.4,
	    width=.24\hsize,
	    axis y line=none,
	    scale only axis,
	    ]
      \addplot+ table [x index=0,y index=1,y error index=2] {T7_CTSB_good.txt};
      \addplot+ table [x index=0,y index=1,y error index=2] {T7_AOSA_good.txt};
      \addplot+ table [x index=0,y index=1,y error index=2] {T7_AOSB_good.txt};
      \addplot+ table [x index=0,y index=1,y error index=2] {T7_CTSB_bad.txt};
      \addplot+ table [x index=0,y index=1,y error index=2] {T7_AOSA_bad.txt};
      \addplot+ table [x index=0,y index=1,y error index=2] {T7_AOSB_bad.txt};
      \addplot [/pgfplots/every axis grid, no markers] coordinates {(9.65,2) (9.73,2)};
      \addplot [/pgfplots/every axis grid, no markers] coordinates {(9.65,4) (9.73,4)};
      \addplot [/pgfplots/every axis grid, no markers] coordinates {(9.65,6) (9.73,6)};
      \addplot [/pgfplots/every axis grid, no markers] coordinates {(9.65,8) (9.73,8)};
    \end{axis}
    \begin{axis}[name=c,
	    grid=major,
	    xshift=.73\hsize,
	    xmin=11.57,xmax=11.62,
	    ymin=1, ymax=9.4,
	    axis y line*=right,
	    width=.15\hsize,
	    scale only axis,
	    xticklabel=\empty,
	    yticklabel=\empty,
	    ]
      \addplot+ table [x index=0,y index=1,y error index=2] {T7_CTSB_good.txt};
      \addplot+ table [x index=0,y index=1,y error index=2] {T7_AOSA_good.txt};
      \addplot+ table [x index=0,y index=1,y error index=2] {T7_AOSB_good.txt};
    \end{axis}
    \begin{axis}[name=d,
	    grid=major,
	    at=(a.south), anchor=north,
	    xmin=8.6,xmax=8.75,
	    ymin=-1.5, ymax=.5,
	    axis y line*=left,
	    width=.45\hsize,
	    scale only axis,
	    xlabel={$t$ [UT]},
	    ylabel={$\Delta v$ [\kms]},
	    ]
      \addplot+ table [x index=0,y index=1,y error index=2] {vel_T7_CTSB_good.txt};
      \addplot+ table [x index=0,y index=1,y error index=2] {vel_T7_AOSA_good.txt};
      \addplot+ table [x index=0,y index=1,y error index=2] {vel_T7_AOSB_good.txt};
    \end{axis}
    \begin{axis}[name=e,
	    grid=major,
	    at=(b.south), anchor=north,
	    xmin=9.65,xmax=9.73,
	    ymin=-1.5, ymax=.5,
	    width=.24\hsize,
	    axis y line=none,
	    scale only axis,
	    ]
      \addplot+ table [x index=0,y index=1,y error index=2] {vel_T7_CTSB_good.txt};
      \addplot+ table [x index=0,y index=1,y error index=2] {vel_T7_AOSA_good.txt};
      \addplot+ table [x index=0,y index=1,y error index=2] {vel_T7_AOSB_good.txt};
      \addplot+ table [x index=0,y index=1,y error index=2] {vel_T7_CTSB_bad.txt};
      \addplot+ table [x index=0,y index=1,y error index=2] {vel_T7_AOSA_bad.txt};
      \addplot+ table [x index=0,y index=1,y error index=2] {vel_T7_AOSB_bad.txt};
      \addplot [/pgfplots/every axis grid, no markers] coordinates {(9.65,-1) (9.73,-1)};
      \addplot [/pgfplots/every axis grid, no markers] coordinates {(9.65,-0.5) (9.73,-0.5)};
      \addplot [/pgfplots/every axis grid, no markers] coordinates {(9.65,0) (9.73,0)};
    \end{axis}
    \begin{axis}[name=f,
	    grid=major,
	    at=(c.south), anchor=north,
	    xmin=11.57,xmax=11.62,
	    ymin=-1.5, ymax=.5,
	    axis y line*=right,
	    width=.15\hsize,
	    scale only axis,
	    yticklabel=\empty,
	    ]
      \addplot+ table [x index=0,y index=1,y error index=2] {vel_T7_CTSB_good.txt};
      \addplot+ table [x index=0,y index=1,y error index=2] {vel_T7_AOSA_good.txt};
      \addplot+ table [x index=0,y index=1,y error index=2] {vel_T7_AOSB_good.txt};
    \end{axis}
    \draw [dotted] (a.south east) -- (b.south west);
    \draw [dotted] (a.north east) -- (b.north west);
    \draw [dotted] (b.south east) -- (c.south west);
    \draw [dotted] (b.north east) -- (c.north west);
    \draw [dotted] (d.south east) -- (e.south west);
    \draw [dotted] (e.south east) -- (f.south west);
  \end{tikzpicture}
  \caption{\ce{HCN} production rates (\emph{upper panel}) and line velocity
    shifts (\emph{lower panel}) derived from the AOSA (blue circles), AOSB (red
    squares), and CTSB (black triangles) spectrometers as a function of
    time in comet \linear{} with 1-$\sigma$ uncertainties. The unit of
    time in the horizontal axis is UT decimal days in May 2004 and the
    time resolution is about 4 min (integration of 2 min at the
    on-source and off-source positions with 15 s samples plus a small
    overhead from telescope movement in the position-switching observing
    mode).  Gaps in the time coverage of our observations are indicated
    by the dotted lines in the horizontal axes.  Filled symbols
    designate spectra that were obtained with a system temperature that
    was more than 1-$\sigma$ away from the mean system temperature
    during the adjacent scans.
  }
  \label{fig:linear:qhcn}
\end{figure*}

\section{Summary and discussion}\label{sec:discussion}

The chemical composition of comets has been investigated using
millimeter/submillimeter spectroscopic molecular observations
\citep[see][for a review]{2009EM&P..105..267C}.  We have presented the
results of our observing campaign to survey several molecular species in
the bright Oort cloud comets \nea{} and \lin{} in May 2004 using
the SMT.  These observations were carried out in the position-switching
observing mode and allowed us to measure the production rates or derive
upper limits of the \volatiles{} abundances in these objects using a
radiative transfer code.  Both comets were likely to be on their first
observed apparition through the inner Solar System, although it has been
proposed that \nea{} is probably a dynamically old comet based on
dynamical simulations \citep{2010MNRAS.404.1886K,2012A&A...544A.119K}.

We derived an HCN production rate of \qqhcn{}~\s\ for comet \neat{} that
was obtained on UT 7.98 May 2004 which gave a mixing ratio with respect
to \ce{H2O} of \qqhcnqho, and an HCN production rate for comet \linear{}
of \qthcn{}~\s\ which gave a $Q_\ce{HCN}$/$Q_\ce{H2O}$ mixing ratio of
\qthcnqho.  These relative abundances are close to the value of 0.1\%
observed in several comets for a wide range of heliocentric distances.
Methanol rotational temperatures derived from multiple lines using the
rotational diagram technique are consistent with those of other
dynamically new comets observed at similar heliocentric distances
\citep[e.g.][]{1999AJ....118.1850B}.

Outgassing variations induced by the nucleus rotation were expected to
appear from non-sphericity of the nucleus or the presence of active
region areas on the surface.  The HCN production rates in comet \nea{}
showed clear evidence of a uniform brightness decrease in the linear fit
over a period of about four hours, which could be a periodic phenomenon
that is consistent with the rotation period of the nucleus derived by {\it Odin}
observations \citep{2009A&A...501..359B}, and of a simultaneous increase
in the line velocity shift.  There is a time variability of the HCN
production rate in comet \lin{} of about 20\% around the mean value, but
it is possible that they are caused by instrumental fluctuations and our
observations do not provide sufficient phase coverage to constrain the
nucleus rotation period.

The results presented here show that millimeter observations are an
effective tool to investigate the chemical composition of cometary
atmospheres.  Accounting for the uncertainties in the excitation
parameters in our model, the errors in the mixing ratios of the observed
species are large, but our observations suggest that the
production rates are in the usual range for dynamically new comets with
lower relative abundances in comet \nea{}.  Therefore, these
measurements will complement the observations of \nea{} and
\lin{} conducted with other facilities
\citep[e.g.][]{2005ApJ...630..623F,2006ApJ...650..470D,2006ApJ...649.1169M,2008ApJ...689..613R,2009AJ....137.4837H},
and will contribute to future statistical studies of the chemical
composition of comets.

As noted in the previous section, there are no obvious correlations
between the relative abundances of the observed molecular species at
millimeter/submillimeter wavelengths and the dynamical class of the
comets from the comparison of a large sample of comets belonging to
different families \citep{2002EM&P...90..323B,2009EM&P..105..267C}.
Hence, more observations are needed to find a robust systematic
correlation between the observed relative abundances and the comet
origin in the protosolar disk.  Future progress in the field of radio
observations of cometary atmospheres will build on ground-based
observations with the large millimeter/submillimeter array (ALMA) and
CCAT telescope, that will significantly increase the current sample of
isotopic ratio measurements in comets \citep{2013ApJ...774L...3L}, and
on in situ measurements with the microwave instrument for the Rosetta
orbiter (MIRO) onboard the Rosetta spacecraft.

\begin{acknowledgements}
Results presented in this paper are based on observations at the SMT
which is operated by the Arizona Radio Observatory (ARO), Steward
Observatory, University of Arizona.  We are grateful to the ARO staff
for their expertise and support during these observations, and to
M.~Drahus for valuable comments that improved the manuscript.  We
gratefully acknowledge useful discussions about the spectroscopic
analysis package \texttt{pyspeckit} with Adam~G.~Ginsburg.  The
anonymous referee is thanked for useful and positive feedback.  MdVB
acknowledges support from the Special Priority Program 1488 (PlanetMag,
\url{http://www.planetmag.de}) of the German Science Foundation and
partial support from grants NSF AST-1108686 and NASA NNX12AH91H.
\end{acknowledgements}

\bibliographystyle{aa}
\bibliography{ads,preprints}

\begin{thebibliography}{81}
\expandafter\ifx\csname natexlab\endcsname\relax\def\natexlab#1{#1}\fi

\bibitem[{{A'Hearn} {et~al.}(2008){A'Hearn}, {Belton}, {Collins}, {Farnham},
  {Feaga}, {Groussin}, {Lisse}, {Meech}, {Schultz}, \&
  {Sunshine}}]{2008EP&S...60...61A}
{A'Hearn}, M.~F., {Belton}, M.~J.~S., {Collins}, S.~M., {et~al.} 2008, Earth,
  Planets, and Space, 60, 61

\bibitem[{{A'Hearn} {et~al.}(2005){A'Hearn}, {Belton}, {Delamere}, {Kissel},
  {Klaasen}, {McFadden}, {Meech}, {Melosh}, {Schultz}, {Sunshine}, {Thomas},
  {Veverka}, {Yeomans}, {Baca}, {Busko}, {Crockett}, {Collins}, {Desnoyer},
  {Eberhardy}, {Ernst}, {Farnham}, {Feaga}, {Groussin}, {Hampton}, {Ipatov},
  {Li}, {Lindler}, {Lisse}, {Mastrodemos}, {Owen}, {Richardson}, {Wellnitz}, \&
  {White}}]{2005Sci...310..258A}
{A'Hearn}, M.~F., {Belton}, M.~J.~S., {Delamere}, W.~A., {et~al.} 2005,
  Science, 310, 258

\bibitem[{{A'Hearn} {et~al.}(2012){A'Hearn}, {Feaga}, {Keller}, {Kawakita},
  {Hampton}, {Kissel}, {Klaasen}, {McFadden}, {Meech}, {Schultz}, {Sunshine},
  {Thomas}, {Veverka}, {Yeomans}, {Besse}, {Bodewits}, {Farnham}, {Groussin},
  {Kelley}, {Lisse}, {Merlin}, {Protopapa}, \&
  {Wellnitz}}]{2012ApJ...758...29A}
{A'Hearn}, M.~F., {Feaga}, L.~M., {Keller}, H.~U., {et~al.} 2012, \apj, 758, 29

\bibitem[{{A'Hearn} {et~al.}(1995){A'Hearn}, {Millis}, {Schleicher}, {Osip}, \&
  {Birch}}]{1995Icar..118..223A}
{A'Hearn}, M.~F., {Millis}, R.~L., {Schleicher}, D.~G., {Osip}, D.~J., \&
  {Birch}, P.~V. 1995, \icarus, 118, 223

\bibitem[{{Baars} \& {Martin}(1996)}]{1996RvMA....9..111B}
{Baars}, J.~W.~M. \& {Martin}, R.~N. 1996, in Reviews in Modern Astronomy, ed.
  {R.~E.~Schielicke}, Vol.~9, 111--126

\bibitem[{{Baars} {et~al.}(1999){Baars}, {Martin}, {Mangum}, {McMullin}, \&
  {Peters}}]{1999PASP..111..627B}
{Baars}, J.~W.~M., {Martin}, R.~N., {Mangum}, J.~G., {McMullin}, J.~P., \&
  {Peters}, W.~L. 1999, \pasp, 111, 627

\bibitem[{{Bensch} \& {Bergin}(2004)}]{2004ApJ...615..531B}
{Bensch}, F. \& {Bergin}, E.~A. 2004, \apj, 615, 531

\bibitem[{{Birtwhistle} \& {Spahr}(2002)}]{2002IAUC.8003....1B}
{Birtwhistle}, P. \& {Spahr}, T.~B. 2002, \iaucirc, 8003, 1

\bibitem[{{Biver}(1997)}]{1997PhDT........51B}
{Biver}, N. 1997, PhD thesis, Univ.~Paris 7-Diderot

\bibitem[{{Biver} {et~al.}(2007{\natexlab{a}}){Biver}, {Bockel{\'e}e-Morvan},
  {Boissier}, {Crovisier}, {Colom}, {Lecacheux}, {Moreno}, {Paubert}, {Lis},
  {Sumner}, {Frisk}, {Hjalmarson}, {Olberg}, {Winnberg}, {Flor{\'e}n},
  {Sandqvist}, \& {Kwok}}]{2007Icar..187..253B}
{Biver}, N., {Bockel{\'e}e-Morvan}, D., {Boissier}, J., {et~al.}
  2007{\natexlab{a}}, \icarus, 187, 253

\bibitem[{{Biver} {et~al.}(1997){Biver}, {Bockel{\'e}e-Morvan}, {Colom},
  {Crovisier}, {Germain}, {Lellouch}, {Davies}, {Dent}, {Moreno}, {Paubert},
  {Wink}, {Despois}, {Lis}, {Mehringer}, {Benford}, {Gardner}, {Phillips},
  {Gunnarsson}, {Rickman}, {Winnberg}, {Bergman}, {Johansson}, \&
  {Rauer}}]{1997EM&P...78....5B}
{Biver}, N., {Bockel{\'e}e-Morvan}, D., {Colom}, P., {et~al.} 1997, Earth Moon
  and Planets, 78, 5

\bibitem[{{Biver} {et~al.}(2009){Biver}, {Bockel{\'e}e-Morvan}, {Colom},
  {Crovisier}, {Lecacheux}, {Frisk}, {Hjalmarson}, {Olberg}, \&
  {Sandqvist}}]{2009A&A...501..359B}
{Biver}, N., {Bockel{\'e}e-Morvan}, D., {Colom}, P., {et~al.} 2009, \aap, 501,
  359

\bibitem[{{Biver} {et~al.}(2011){Biver}, {Bockel{\'e}e-Morvan}, {Colom},
  {Crovisier}, {Paubert}, {Weiss}, \& {Wiesemeyer}}]{2011A&A...528A.142B}
{Biver}, N., {Bockel{\'e}e-Morvan}, D., {Colom}, P., {et~al.} 2011, \aap, 528,
  A142

\bibitem[{{Biver} {et~al.}(2002){Biver}, {Bockel{\'e}e-Morvan}, {Crovisier},
  {Colom}, {Henry}, {Moreno}, {Paubert}, {Despois}, \&
  {Lis}}]{2002EM&P...90..323B}
{Biver}, N., {Bockel{\'e}e-Morvan}, D., {Crovisier}, J., {et~al.} 2002, Earth
  Moon and Planets, 90, 323

\bibitem[{{Biver} {et~al.}(1999){Biver}, {Bockel{\'e}e-Morvan}, {Crovisier},
  {Davies}, {Matthews}, {Wink}, {Rauer}, {Colom}, {Dent}, {Despois}, {Moreno},
  {Paubert}, {Jewitt}, \& {Senay}}]{1999AJ....118.1850B}
{Biver}, N., {Bockel{\'e}e-Morvan}, D., {Crovisier}, J., {et~al.} 1999, \aj,
  118, 1850

\bibitem[{{Biver} {et~al.}(2007{\natexlab{b}}){Biver}, {Bockel{\'e}e-Morvan},
  {Crovisier}, {Lecacheux}, {Frisk}, {Hjalmarson}, {Olberg}, {Flor{\'e}n},
  {Sandqvist}, \& {Kwok}}]{2007P&SS...55.1058B}
{Biver}, N., {Bockel{\'e}e-Morvan}, D., {Crovisier}, J., {et~al.}
  2007{\natexlab{b}}, \planss, 55, 1058

\bibitem[{{Biver} {et~al.}(2012){Biver}, {Crovisier}, {Bockel{\'e}e-Morvan},
  {Szutowicz}, {Lis}, {Hartogh}, {de Val-Borro}, {Moreno}, {Boissier},
  {Kidger}, {K{\"u}ppers}, {Paubert}, {Dello Russo}, {Vervack}, \&
  {Weaver}}]{2012A&A...539A..68B}
{Biver}, N., {Crovisier}, J., {Bockel{\'e}e-Morvan}, D., {et~al.} 2012, \aap,
  539, A68

\bibitem[{{Bockel{\'e}e-Morvan}(1987)}]{1987A&A...181..169B}
{Bockel{\'e}e-Morvan}, D. 1987, \aap, 181, 169

\bibitem[{{Bockel{\'e}e-Morvan}(2011)}]{2011IAUS..280..261B}
{Bockel{\'e}e-Morvan}, D. 2011, in IAU Symposium, Vol. 280, 261--274

\bibitem[{{Bockel{\'e}e-Morvan} {et~al.}(2012){Bockel{\'e}e-Morvan}, {Biver},
  {Swinyard}, {de Val-Borro}, {Crovisier}, {Hartogh}, {Lis}, {Moreno},
  {Szutowicz}, {Lellouch}, {Emprechtinger}, {Blake}, {Courtin}, {Jarchow},
  {Kidger}, {K{\"u}ppers}, {Rengel}, {Davis}, {Fulton}, {Naylor}, {Sidher}, \&
  {Walker}}]{2012A&A...544L..15B}
{Bockel{\'e}e-Morvan}, D., {Biver}, N., {Swinyard}, B., {et~al.} 2012, \aap,
  544, L15

\bibitem[{{Bockel{\'e}e-Morvan} {et~al.}(1994){Bockel{\'e}e-Morvan},
  {Crovisier}, {Colom}, \& {Despois}}]{1994A&A...287..647B}
{Bockel{\'e}e-Morvan}, D., {Crovisier}, J., {Colom}, P., \& {Despois}, D. 1994,
  \aap, 287, 647

\bibitem[{{Bockel{\'e}e-Morvan} {et~al.}(1990){Bockel{\'e}e-Morvan},
  {Crovisier}, {Colom}, {Despois}, \& {Paubert}}]{1990ESASP.315..143B}
{Bockel{\'e}e-Morvan}, D., {Crovisier}, J., {Colom}, P., {Despois}, D., \&
  {Paubert}, G. 1990, in ESA Special Publication, ed. {B.~Battrick}, Vol. 315,
  143--148

\bibitem[{{Bockel{\'e}e-Morvan} {et~al.}(2004){Bockel{\'e}e-Morvan},
  {Crovisier}, {Mumma}, \& {Weaver}}]{2004come.book..391B}
{Bockel{\'e}e-Morvan}, D., {Crovisier}, J., {Mumma}, M.~J., \& {Weaver}, H.~A.
  2004, in Comets II, ed. M.~C. {Festou}, H.~U. {Keller}, \& H.~A. {Weaver}
  (Univ. Arizona Press), 391--423

\bibitem[{{Bockel{\'e}e-Morvan} {et~al.}(2010){Bockel{\'e}e-Morvan}, {Hartogh},
  {Crovisier}, {Vandenbussche}, {Swinyard}, {Biver}, {Lis}, {Jarchow},
  {Moreno}, {Hutsem{\'e}kers}, {Jehin}, {K{\"u}ppers}, {Lara}, {Lellouch},
  {Manfroid}, {de Val-Borro}, {Szutowicz}, {Banaszkiewicz}, {Bensch}, {Blecka},
  {Emprechtinger}, {Encrenaz}, {Fulton}, {Kidger}, {Rengel}, {Waelkens},
  {Bergin}, {Blake}, {Blommaert}, {Cernicharo}, {Decin}, {Encrenaz}, {de
  Graauw}, {Leeks}, {Medvedev}, {Naylor}, {Schieder}, \&
  {Thomas}}]{2010A&A...518L.149B}
{Bockel{\'e}e-Morvan}, D., {Hartogh}, P., {Crovisier}, J., {et~al.} 2010, \aap,
  518, L149

\bibitem[{{Boissier} {et~al.}(2007){Boissier}, {Bockel{\'e}e-Morvan}, {Biver},
  {Crovisier}, {Despois}, {Marsden}, \& {Moreno}}]{2007A&A...475.1131B}
{Boissier}, J., {Bockel{\'e}e-Morvan}, D., {Biver}, N., {et~al.} 2007, \aap,
  475, 1131

\bibitem[{{Cavali{\'e}} {et~al.}(2013){Cavali{\'e}}, {Feuchtgruber},
  {Lellouch}, {de Val-Borro}, {Jarchow}, {Moreno}, {Hartogh}, {Orton},
  {Greathouse}, {Billebaud}, {Dobrijevic}, {Lara}, {Gonz{\'a}lez}, \&
  {Sagawa}}]{2013A&A...553A..21C}
{Cavali{\'e}}, T., {Feuchtgruber}, H., {Lellouch}, E., {et~al.} 2013, \aap,
  553, A21

\bibitem[{{Combi} {et~al.}(2004){Combi}, {Harris}, \&
  {Smyth}}]{2004come.book..523C}
{Combi}, M.~R., {Harris}, W.~M., \& {Smyth}, W.~H. 2004, in Comets II, ed.
  {Festou, M.~C., Keller, H.~U., \& Weaver, H.~A.} (Univ. Arizona Press),
  523--552

\bibitem[{{Combi} {et~al.}(2009){Combi}, {M{\"a}kinen}, {Bertaux}, {Lee}, \&
  {Qu{\'e}merais}}]{2009AJ....137.4734C}
{Combi}, M.~R., {M{\"a}kinen}, J.~T.~T., {Bertaux}, J.-L., {Lee}, Y., \&
  {Qu{\'e}merais}, E. 2009, \aj, 137, 4734

\bibitem[{{Crovisier}(1984)}]{1984A&A...130..361C}
{Crovisier}, J. 1984, \aap, 130, 361

\bibitem[{{Crovisier}(1994)}]{1994JGR....99.3777C}
{Crovisier}, J. 1994, \jgr, 99, 3777

\bibitem[{{Crovisier} {et~al.}(2009){Crovisier}, {Biver},
  {Bockel{\'e}e-Morvan}, {Boissier}, {Colom}, \& {Lis}}]{2009EM&P..105..267C}
{Crovisier}, J., {Biver}, N., {Bockel{\'e}e-Morvan}, D., {et~al.} 2009, Earth
  Moon and Planets, 105, 267

\bibitem[{{Crovisier} \& {Encrenaz}(1983)}]{1983A&A...126..170C}
{Crovisier}, J. \& {Encrenaz}, T. 1983, \aap, 126, 170

\bibitem[{{de Val-Borro} {et~al.}(2010){de Val-Borro}, {Hartogh}, {Crovisier},
  {Bockel{\'e}e-Morvan}, {Biver}, {Lis}, {Moreno}, {Jarchow}, {Rengel},
  {Szutowicz}, {Banaszkiewicz}, {Bensch}, {B{\l}{\c e}cka}, {Emprechtinger},
  {Encrenaz}, {Jehin}, {K{\"u}ppers}, {Lara}, {Lellouch}, {Swinyard},
  {Vandenbussche}, {Bergin}, {Blake}, {Blommaert}, {Cernicharo}, {Decin},
  {Encrenaz}, {de Graauw}, {Hutsem{\'e}kers}, {Kidger}, {Manfroid}, {Medvedev},
  {Naylor}, {Schieder}, {Stam}, {Thomas}, {Waelkens}, {Szczerba}, {Saraceno},
  {di Giorgio}, {Philipp}, {Klein}, {Ossenkopf}, {Zaal}, \&
  {Shipman}}]{2010A&A...521L..50D}
{de Val-Borro}, M., {Hartogh}, P., {Crovisier}, J., {et~al.} 2010, \aap, 521,
  L50

\bibitem[{{de Val-Borro} {et~al.}(2012{\natexlab{a}}){de Val-Borro}, {Hartogh},
  {Jarchow}, {Rengel}, {Villanueva}, {K{\"u}ppers}, {Biver},
  {Bockel{\'e}e-Morvan}, \& {Crovisier}}]{2012A&A...545A...2D}
{de Val-Borro}, M., {Hartogh}, P., {Jarchow}, C., {et~al.} 2012{\natexlab{a}},
  \aap, 545, A2

\bibitem[{{de Val-Borro} {et~al.}(2011){de Val-Borro}, {Jarchow}, {Hartogh},
  {Villanueva}, \& {K\"uppers}}]{2011AOGS}
{de Val-Borro}, M., {Jarchow}, C., {Hartogh}, P., {Villanueva}, G., \&
  {K\"uppers}, M. 2011, Advances in Geosciences, 25, 149

\bibitem[{{de Val-Borro} {et~al.}(2012{\natexlab{b}}){de Val-Borro}, {Rezac},
  {Hartogh}, {Biver}, {Bockel{\'e}e-Morvan}, {Crovisier}, {K{\"u}ppers}, {Lis},
  {Szutowicz}, {Blake}, {Emprechtinger}, {Jarchow}, {Jehin}, {Kidger}, {Lara},
  {Lellouch}, {Moreno}, \& {Rengel}}]{2012A&A...546L...4D}
{de Val-Borro}, M., {Rezac}, L., {Hartogh}, P., {et~al.} 2012{\natexlab{b}},
  \aap, 546, L4

\bibitem[{{DiSanti} {et~al.}(2006){DiSanti}, {Bonev}, {Magee-Sauer}, {Dello
  Russo}, {Mumma}, {Reuter}, \& {Villanueva}}]{2006ApJ...650..470D}
{DiSanti}, M.~A., {Bonev}, B.~P., {Magee-Sauer}, K., {et~al.} 2006, \apj, 650,
  470

\bibitem[{{Drahus} {et~al.}(2011){Drahus}, {Jewitt}, {Guilbert-Lepoutre},
  {Waniak}, {Hoge}, {Lis}, {Yoshida}, {Peng}, \&
  {Sievers}}]{2011ApJ...734L...4D}
{Drahus}, M., {Jewitt}, D., {Guilbert-Lepoutre}, A., {et~al.} 2011, \apjl, 734,
  L4+

\bibitem[{{Drahus} {et~al.}(2012){Drahus}, {Jewitt}, {Guilbert-Lepoutre},
  {Waniak}, \& {Sievers}}]{2012ApJ...756...80D}
{Drahus}, M., {Jewitt}, D., {Guilbert-Lepoutre}, A., {Waniak}, W., \&
  {Sievers}, A. 2012, \apj, 756, 80

\bibitem[{{Drahus} {et~al.}(2010){Drahus}, {K{\"u}ppers}, {Jarchow},
  {Paganini}, {Hartogh}, \& {Villanueva}}]{2010A&A...510A..55D}
{Drahus}, M., {K{\"u}ppers}, M., {Jarchow}, C., {et~al.} 2010, \aap, 510, A55

\bibitem[{{Feldman} {et~al.}(2004){Feldman}, {Cochran}, \&
  {Combi}}]{2004come.book..425F}
{Feldman}, P.~D., {Cochran}, A.~L., \& {Combi}, M.~R. 2004, in Comets II, ed.
  {Festou, M.~C., Keller, H.~U., \& Weaver, H.~A.} (Univ. Arizona Press),
  425--447

\bibitem[{{Friedel} {et~al.}(2005){Friedel}, {Remijan}, {Snyder}, {A'Hearn},
  {Blake}, {de Pater}, {Dickel}, {Forster}, {Hogerheijde}, {Kraybill},
  {Looney}, {Palmer}, \& {Wright}}]{2005ApJ...630..623F}
{Friedel}, D.~N., {Remijan}, A.~J., {Snyder}, L.~E., {et~al.} 2005, \apj, 630,
  623

\bibitem[{{Ginsburg} \& {Mirocha}(2011)}]{2011ascl.soft09001G}
{Ginsburg}, A. \& {Mirocha}, J. 2011, {PySpecKit: Python Spectroscopic
  Toolkit}, astrophysics Source Code Library

\bibitem[{{Giorgini} {et~al.}(1996){Giorgini}, {Yeomans}, {Chamberlin},
  {Chodas}, {Jacobson}, {Keesey}, {Lieske}, {Ostro}, {Standish}, \&
  {Wimberly}}]{1996DPS....28.2504G}
{Giorgini}, J.~D., {Yeomans}, D.~K., {Chamberlin}, A.~B., {et~al.} 1996, in
  Bulletin of the American Astronomical Society, Vol.~28, AAS/Division for
  Planetary Sciences Meeting Abstracts \#28, 1158

\bibitem[{{Hartogh}(1997{\natexlab{a}})}]{1997SPIE.3220..115H}
{Hartogh}, P. 1997{\natexlab{a}}, in Society of Photo-Optical Instrumentation
  Engineers (SPIE) Conference Series, ed. J.~D. {Haigh}, Vol. 3220, 115--124

\bibitem[{{Hartogh}(1997{\natexlab{b}})}]{1997SPIE.3221..328H}
{Hartogh}, P. 1997{\natexlab{b}}, in Society of Photo-Optical Instrumentation
  Engineers (SPIE) Conference Series, ed. H.~{Fujisada}, Vol. 3221, 328--339

\bibitem[{{Hartogh} {et~al.}(2010){Hartogh}, {Crovisier}, {de Val-Borro},
  {Bockel{\'e}e-Morvan}, {Biver}, {Lis}, {Moreno}, {Jarchow}, {Rengel},
  {Emprechtinger}, {Szutowicz}, {Banaszkiewicz}, {Bensch}, {Blecka},
  {Cavali{\'e}}, {Encrenaz}, {Jehin}, {K{\"u}ppers}, {Lara}, {Lellouch},
  {Swinyard}, {Vandenbussche}, {Bergin}, {Blake}, {Blommaert}, {Cernicharo},
  {Decin}, {Encrenaz}, {de Graauw}, {Hutsemekers}, {Kidger}, {Manfroid},
  {Medvedev}, {Naylor}, {Schieder}, {Thomas}, {Waelkens}, {Roelfsema},
  {Dieleman}, {G{\"u}sten}, {Klein}, {Kasemann}, {Caris}, {Olberg}, \&
  {Benz}}]{2010A&A...518L.150H}
{Hartogh}, P., {Crovisier}, J., {de Val-Borro}, M., {et~al.} 2010, \aap, 518,
  L150

\bibitem[{{Hartogh} \& {Hartmann}(1990)}]{1990MeScT...1..592H}
{Hartogh}, P. \& {Hartmann}, G.~K. 1990, Measurement Science and Technology, 1,
  592

\bibitem[{{Hartogh} {et~al.}(2011){Hartogh}, {Lis}, {Bockel{\'e}e-Morvan}, {de
  Val-Borro}, {Biver}, {K{\"u}ppers}, {Emprechtinger}, {Bergin}, {Crovisier},
  {Rengel}, {Moreno}, {Szutowicz}, \& {Blake}}]{2011Natur.478..218H}
{Hartogh}, P., {Lis}, D.~C., {Bockel{\'e}e-Morvan}, D., {et~al.} 2011, \nat,
  478, 218

\bibitem[{{Haser}(1957)}]{1957BSRSL..43..740H}
{Haser}, L. 1957, Bulletin de la Societe Royale des Sciences de Liege, 43, 740

\bibitem[{{Hogerheijde} {et~al.}(2009){Hogerheijde}, {Qi}, {de Pater}, {Blake},
  {Friedel}, {Forster}, {Palmer}, {Remijan}, {Snyder}, \&
  {Wright}}]{2009AJ....137.4837H}
{Hogerheijde}, M.~R., {Qi}, C., {de Pater}, I., {et~al.} 2009, \aj, 137, 4837

\bibitem[{{Hogerheijde} \& {van der Tak}(2000)}]{2000A&A...362..697H}
{Hogerheijde}, M.~R. \& {van der Tak}, F.~F.~S. 2000, \aap, 362, 697

\bibitem[{{Jockers} {et~al.}(2011){Jockers}, {Szutowicz}, {Villanueva},
  {Bonev}, \& {Hartogh}}]{2011Icar..215..153J}
{Jockers}, K., {Szutowicz}, S., {Villanueva}, G., {Bonev}, T., \& {Hartogh}, P.
  2011, \icarus, 215, 153

\bibitem[{{Kawakita} {et~al.}(2005){Kawakita}, {Watanabe}, {Furusho}, {Fuse},
  \& {Boice}}]{2005ApJ...623L..49K}
{Kawakita}, H., {Watanabe}, J.-i., {Furusho}, R., {Fuse}, T., \& {Boice}, D.~C.
  2005, \apjl, 623, L49

\bibitem[{{Kr{\'o}likowska} \& {Dybczy{\'n}ski}(2010)}]{2010MNRAS.404.1886K}
{Kr{\'o}likowska}, M. \& {Dybczy{\'n}ski}, P.~A. 2010, \mnras, 404, 1886

\bibitem[{{Kr{\'o}likowska} {et~al.}(2012){Kr{\'o}likowska}, {Dybczy{\'n}ski},
  \& {Sitarski}}]{2012A&A...544A.119K}
{Kr{\'o}likowska}, M., {Dybczy{\'n}ski}, P.~A., \& {Sitarski}, G. 2012, \aap,
  544, A119

\bibitem[{{K{\"u}ppers} {et~al.}(2004){K{\"u}ppers}, {Hartogh}, \&
  {Villanueva}}]{2004DPS....36.2505K}
{K{\"u}ppers}, M., {Hartogh}, P., \& {Villanueva}, G. 2004, in BAAS, Vol.~36,
  AAS/Division for Planetary Sciences Meeting Abstracts \#36, 1125

\bibitem[{{Lecacheux} {et~al.}(2004){Lecacheux}, {Biver}, {Crovisier}, \&
  {Bockelee-Morvan}}]{2004IAUC.8304....2L}
{Lecacheux}, A., {Biver}, N., {Crovisier}, J., \& {Bockelee-Morvan}, D. 2004,
  \iaucirc, 8304, 2

\bibitem[{{Lecacheux} {et~al.}(2003){Lecacheux}, {Biver}, {Crovisier},
  {Bockel{\'e}e-Morvan}, {Baron}, {Booth}, {Encrenaz}, {Flor{\'e}n}, {Frisk},
  {Hjalmarson}, {Kwok}, {Mattila}, {Nordh}, {Olberg}, {Olofsson}, {Rickman},
  {Sandqvist}, {von Sch{\'e}ele}, {Serra}, {Torchinsky}, {Volk}, \&
  {Winnberg}}]{2003A&A...402L..55L}
{Lecacheux}, A., {Biver}, N., {Crovisier}, J., {et~al.} 2003, \aap, 402, L55

\bibitem[{{Levison} {et~al.}(2010){Levison}, {Duncan}, {Brasser}, \&
  {Kaufmann}}]{2010Sci...329..187L}
{Levison}, H.~F., {Duncan}, M.~J., {Brasser}, R., \& {Kaufmann}, D.~E. 2010,
  Science, 329, 187

\bibitem[{{Lis} {et~al.}(2013){Lis}, {Biver}, {Bockel{\'e}e-Morvan}, {Hartogh},
  {Bergin}, {Blake}, {Crovisier}, {de Val-Borro}, {Jehin}, {K{\"u}ppers},
  {Manfroid}, {Moreno}, {Rengel}, \& {Szutowicz}}]{2013ApJ...774L...3L}
{Lis}, D.~C., {Biver}, N., {Bockel{\'e}e-Morvan}, D., {et~al.} 2013, \apjl,
  774, L3

\bibitem[{{Lis} {et~al.}(2008){Lis}, {Bockel{\'e}e-Morvan}, {Boissier},
  {Crovisier}, {Biver}, \& {Charnley}}]{2008ApJ...675..931L}
{Lis}, D.~C., {Bockel{\'e}e-Morvan}, D., {Boissier}, J., {et~al.} 2008, \apj,
  675, 931

\bibitem[{{Mekhtiev} {et~al.}(1999){Mekhtiev}, {Godfrey}, \&
  {Hougen}}]{1999JMoSp.194..171M}
{Mekhtiev}, M.~A., {Godfrey}, P.~D., \& {Hougen}, J.~T. 1999, Journal of
  Molecular Spectroscopy, 194, 171

\bibitem[{{Milam} {et~al.}(2006){Milam}, {Remijan}, {Womack}, {Abrell},
  {Ziurys}, {Wyckoff}, {Apponi}, {Friedel}, {Snyder}, {Veal}, {Palmer},
  {Woodney}, {A'Hearn}, {Forster}, {Wright}, {de Pater}, {Choi}, \&
  {Gesmundo}}]{2006ApJ...649.1169M}
{Milam}, S.~N., {Remijan}, A.~J., {Womack}, M., {et~al.} 2006, \apj, 649, 1169

\bibitem[{{Mitchell} {et~al.}(1987){Mitchell}, {Lin}, {Anderson}, {Carlson},
  {Curtis}, {Korth}, {Reme}, {Sauvaud}, {D'Uston}, \&
  {Mendis}}]{1987Sci...237..626M}
{Mitchell}, D.~L., {Lin}, R.~P., {Anderson}, K.~A., {et~al.} 1987, Science,
  237, 626

\bibitem[{Oliphant(2006)}]{oliphant-2006-guide}
Oliphant, T.~E. 2006, A Guide to NumPy, Vol.~1 (Trelgol Publishing USA)

\bibitem[{{O'Rourke} {et~al.}(2013){O'Rourke}, {Snodgrass}, {de Val-Borro},
  {Biver}, {Bockel{\'e}e-Morvan}, {Hsieh}, {Teyssier}, {Fernandez}, {Kueppers},
  {Micheli}, \& {Hartogh}}]{2013ApJ...774L..13O}
{O'Rourke}, L., {Snodgrass}, C., {de Val-Borro}, M., {et~al.} 2013, \apjl, 774,
  L13

\bibitem[{{Paganini} {et~al.}(2010){Paganini}, {Villanueva}, {Lara}, {Lin},
  {K{\"u}ppers}, {Hartogh}, \& {Faure}}]{2010ApJ...715.1258P}
{Paganini}, L., {Villanueva}, G.~L., {Lara}, L.~M., {et~al.} 2010, \apj, 715,
  1258

\bibitem[{{Pickett} {et~al.}(1998){Pickett}, {Poynter}, {Cohen}, {Delitsky},
  {Pearson}, \& {M{\"u}ller}}]{1998JQSRT..60..883P}
{Pickett}, H.~M., {Poynter}, R.~L., {Cohen}, E.~A., {et~al.} 1998, \jqsrt, 60,
  883

\bibitem[{{Pravdo} {et~al.}(2001){Pravdo}, {Helin}, {Lawrence}, {Ticha},
  {Tichy}, {Jelinek}, {Shelus}, \& {Marsden}}]{2001IAUC.7695....1P}
{Pravdo}, S.~H., {Helin}, E.~F., {Lawrence}, K.~J., {et~al.} 2001, \iaucirc,
  7695, 1

\bibitem[{{Remijan} {et~al.}(2006){Remijan}, {Friedel}, {de Pater},
  {Hogerheijde}, {Snyder}, {A'Hearn}, {Blake}, {Dickel}, {Forster}, {Kraybill},
  {Looney}, {Palmer}, \& {Wright}}]{2006ApJ...643..567R}
{Remijan}, A.~J., {Friedel}, D.~N., {de Pater}, I., {et~al.} 2006, \apj, 643,
  567

\bibitem[{{Remijan} {et~al.}(2008){Remijan}, {Milam}, {Womack}, {Apponi},
  {Ziurys}, {Wyckoff}, {A'Hearn}, {de Pater}, {Forster}, {Friedel}, {Palmer},
  {Snyder}, {Veal}, {Woodney}, \& {Wright}}]{2008ApJ...689..613R}
{Remijan}, A.~J., {Milam}, S.~N., {Womack}, M., {et~al.} 2008, \apj, 689, 613

\bibitem[{{Rickman}(2010)}]{2010LNP...790..341R}
{Rickman}, H. 2010, in Lecture Notes in Physics, Berlin Springer Verlag, ed.
  J.~{Souchay} \& R.~{Dvorak}, Vol. 790, 341--399

\bibitem[{{Samarasinha} {et~al.}(2004){Samarasinha}, {Mueller}, {Belton}, \&
  {Jorda}}]{2004come.book..281S}
{Samarasinha}, N.~H., {Mueller}, B.~E.~A., {Belton}, M.~J.~S., \& {Jorda}, L.
  2004, in Comets II, ed. {Festou, M.~C., Keller, H.~U., \& Weaver, H.~A.}
  (Univ. Arizona Press), 281--299

\bibitem[{{S{\'a}nchez-Lavega} {et~al.}(2010){S{\'a}nchez-Lavega}, {Wesley},
  {Orton}, {Hueso}, {Perez-Hoyos}, {Fletcher}, {Yanamandra-Fisher},
  {Legarreta}, {de Pater}, {Hammel}, {Simon-Miller}, {Gomez-Forrellad},
  {Ortiz}, {Garc{\'{\i}}a-Melendo}, {Puetter}, \&
  {Chodas}}]{2010ApJ...715L.155S}
{S{\'a}nchez-Lavega}, A., {Wesley}, A., {Orton}, G., {et~al.} 2010, \apjl, 715,
  L155

\bibitem[{{Snyder} {et~al.}(1989){Snyder}, {Palmer}, \& {de
  Pater}}]{1989AJ.....97..246S}
{Snyder}, L.~E., {Palmer}, P., \& {de Pater}, I. 1989, \aj, 97, 246

\bibitem[{{Snyder} {et~al.}(2001){Snyder}, {Veal}, {Woodney}, {Wright},
  {Palmer}, {A'Hearn}, {Kuan}, {de Pater}, \& {Forster}}]{2001AJ....121.1147S}
{Snyder}, L.~E., {Veal}, J.~M., {Woodney}, L.~M., {et~al.} 2001, \aj, 121, 1147

\bibitem[{{Ulich} \& {Haas}(1976)}]{1976ApJS...30..247U}
{Ulich}, B.~L. \& {Haas}, R.~W. 1976, \apjs, 30, 247

\bibitem[{{Villanueva} \& {Hartogh}(2004)}]{2004ExA....18...77V}
{Villanueva}, G. \& {Hartogh}, P. 2004, Experimental Astronomy, 18, 77

\bibitem[{{Villanueva} {et~al.}(2006){Villanueva}, {Hartogh}, \&
  {Reindl}}]{2006ITMTT..54.1415V}
{Villanueva}, G.~L., {Hartogh}, P., \& {Reindl}, L.~M. 2006, IEEE Transactions
  on Microwave Theory Techniques, 54, 1415

\bibitem[{{Villanueva} {et~al.}(2005){Villanueva}, {K\"uppers}, \&
  {Hartogh}}]{2005DPS....37.1108V}
{Villanueva}, G.~L., {K\"uppers}, M., \& {Hartogh}, P. 2005, in BAAS, Vol.~37,
  633

\end{thebibliography}

\end{document}